\documentclass[onecolumn,authoryear]{els-mrw} 

\usepackage{amsmath,amssymb,amsfonts,amsthm,makeidx,graphicx,dsfont}
\usepackage{txfonts}
\usepackage{helvet}
\usepackage{color}
\usepackage{xcolor}
\usepackage{hyperref}
\usepackage{caption}
\usepackage{booktabs,subcaption}
\usepackage{braket}
\hypersetup{colorlinks=true,citecolor=blue,linkcolor=blue,urlcolor=blue}
\usepackage{multirow}
\usepackage{orcidlink}
\usepackage[bottom]{footmisc}
\usepackage{tabularx}

\def\beq{\begin{equation}}
\def\eeq{\end{equation}}

\newcommand{\ord}[1]{\mathcal{O}\left(#1\right)}
\newcommand{\ops}{\mathcal{O}}
\newcommand{\TeV}{\ensuremath{\mathrm{TeV}}}
\newcommand{\MeV}{\ensuremath{\mathrm{MeV}}}

\newcommand{\vpj }{\mbox{${H^\dag\, \raisebox{1.mm}{${}^\leftrightarrow$}\hspace{-3mm} D_\mu\,H}$}}
\newcommand{\vpjt}{\mbox{${H^\dag \tau_i\,\raisebox{1.mm}{${}^\leftrightarrow$}\hspace{-3mm} D_\mu\,H}$}}
\begin{document}

\chapter{Searching for Charged Lepton Flavor Violation}\label{chap1}

\author[1]{Lorenzo Calibbi\,\orcidlink{0000-0002-9322-8076}}%
\author[2,3]{Kaori Fuyuto\,\orcidlink{0000-0002-6924-0665}}%
\author[4,5]{Matthias Heinz\,\orcidlink{0000-0002-6363-0056}}%
\author[6]{Jure Zupan\,\orcidlink{0000-0001-9445-537X}}%

\address[1]{\orgname{Nankai University,} \orgdiv{School of Physics,} \orgaddress{Tianjin 300071, China}}
\address[2]{\orgname{High Energy Accelerator Research Organization},\\ \orgdiv{\,~KEK Theory Center, Institute of Particle and Nuclear Studies, } \orgaddress{Tsukuba 305–0801, Japan}}
\address[3]{\orgname{RIKEN iTHEMS}, \orgdiv{RIKEN-Berkeley Center},  \orgaddress{Wako, Saitama 351-0198, Japan}}
\address[4]{\orgname{Oak Ridge National Laboratory}, \orgdiv{National Center for Computational Sciences}, \orgaddress{Oak Ridge, TN 37831, USA}}
\address[5]{\orgname{Oak Ridge National Laboratory}, \orgdiv{Physics Division}, \orgaddress{Oak Ridge, TN 37831, USA}}
\address[6]{\orgname{University of Cincinnati}, \orgdiv{Department of Physics}, \orgaddress{Cincinnati, Ohio 45221, USA}}

\maketitle

\begin{abstract}[Abstract]

Searches for Charged Lepton Flavor Violation (CLFV) are powerful tools to probe physics beyond the Standard Model (BSM). This chapter provides a comprehensive survey of CLFV theory, focusing primarily on low-energy muon and tau transitions while outlining complementary searches from high-energy colliders. Grounded in the Effective Field Theory (EFT) framework, we systematically review key observables including radiative and purely leptonic decays of muon and tau as well as $\mu\to e$ conversion in nuclei and $\tau$ hadronic decay channels. CLFV involving light new physics particles is also discussed. This review captures the modern synergy between EFTs, novel light-mediator scenarios, and hadronic and nuclear physics in the ongoing hunt for BSM signals.     
\end{abstract}

\begin{glossary}[Key words]
Charged Lepton Flavor Violation; Effective Field Theory; Standard Model EFT (SMEFT); rare muon transitions; rare $\tau$ decays; physics beyond the Standard Model; axion-like particles;
neutrino mass mechanisms
\end{glossary}

\begin{glossary}[Key points]
\begin{itemize}
   \item CLFV processes are highly suppressed in the Standard Model so that any observation of $\mu\to e\gamma$, $\mu\to 3e$, $\mu\to e$ conversion, or of analogous CLFV  $\tau$ decays, would constitute an unambiguous signal of physics beyond the Standard Model.
    \item SMEFT, including renormalization group running of the operators from the new-physics scale $\Lambda$ down to the electroweak scale and matching onto LEFT below it, provides the model-independent language connecting dipole and four-fermion contact interactions to the full range of low- and high-energy CLFV observables discussed in this chapter.
    \item For anarchic flavor structure rare muon transitions constitute the most powerful CLFV probe, already testing new-physics scales of $\Lambda\approx10^3$--$10^4\,\mathrm{TeV}$; the next-generation program at MEG~II, Mu3e, Mu2e, and COMET is set to improve current sensitivities on $\mu\to e\gamma$, $\mu\to 3e$, and $\mu\to e$ conversion by one to four orders of magnitude over the coming decade.
    \item $\mu \to e$ conversion in nuclei provides a complementary probe of CLFV through the sensitivity to interactions with quarks and gluons in the nucleus.
    Modeling it requires a tower of effective field theories including accurate computations of the lepton wave functions and the nuclear structure functions.
    \item If CLFV is mediated by light axion-like particles (ALPs), it arises from dimension-five operators inducing muon decays into ALPs, whereas CLFV from heavy new physics originates from dimension-six operators. As a result, searches for CLFV involving ALPs probe substantially higher energy scales than $\mu\to e\gamma$, $\mu\to 3e$, and $\mu\to e$ conversion.    
    \item CLFV $\tau$ decays extend the CLFV program to third-generation couplings and hadronic final states, while light new-physics mediators such as ALPs and dark photons open complementary high-multiplicity signatures ($\tau\to\ell V$, $\tau\to5\mu$, and beyond) that will become accessible at Belle~II, the proposed STCF, and future Tera-$Z$ factories.
    \item High-energy searches for CLFV in $Z$ and Higgs boson decays, and in non-resonant $e^+e^-\to\tau\mu$ production at future colliders, are governed by the same SMEFT operators as the low-energy transitions; they provide a genuinely complementary, and in the $\tau\ell$ channels competitive, probe of the underlying flavor structure of new physics.
\end{itemize}
\end{glossary}

\begin{glossary}[Glossary]

\term{Axion-like particle.} A light pseudo-Nambu-Goldstone boson of a spontaneously broken (approximate) global symmetry, which can couple derivatively to charged leptons, with possibly flavor-violating couplings. 

\term{Charged lepton flavor violation.} Transitions among charged leptons of different flavor, such as \ $\mu\to e\gamma$, $\mu\to e $ conversion, $\tau\to 3\mu$. These violate individual lepton family numbers and are suppressed in the SM by neutrino masses. 

\term{Dark photon.} The gauge boson of a spontaneously broken dark $U(1)_d$ symmetry, kinetically mixed with hypercharge. When endowed with flavor-violating couplings, it can mediate exotic high-multiplicity CLFV cascades such as $\mu\to 5e$ or $\tau\to 3\mu\,2\pi$.

\term{Low-energy Effective Field Theory.} An effective field theory obtained from SMEFT after integrating out the electroweak-scale degrees of freedom ($W$, $Z$, Higgs, top quark). 

\term{Seesaw mechanism (Type I, II, III).} Three canonical extensions of the Standard Model that generate small Majorana neutrino masses by introducing, respectively, heavy right-handed neutrinos (Type I), an $SU(2)_L$ triplet scalar (Type II), or heavy triplet fermions (Type III). 

\term{Standard Model Effective Field Theory.}  An effective field theory built out of the SM fields, encoding  in higher-dimensional operators the effects of heavier degrees of freedom, which were integrated out.

\term{Super-Tau-Charm factory.} A proposed high-luminosity $e^+e^-$ collider operating at center-of-mass energies of a few GeV, in the $\tau$ and open charm threshold region ($\sim2-7$\,GeV).

\term{Tera-$Z$ factory.} A future high-luminosity $e^+e^-$ collider (such as FCC-ee or CEPC) operating at or near the $Z$-pole and expected to produce order $10^{12}$ $Z$ bosons.

\end{glossary}

\begin{glossary}[Nomenclature]
\begin{tabular}{@{}lp{34pc}@{}}
ALP & Axion Like Particle\\
BSM & Physics beyond the Standard Model \\
CEPC & Circular Electron Positron Collider \\
ChPT & Chiral Perturbation Theory \\
CLFV & Charged Lepton Flavor Violation\\
EFT & Effective Field Theory\\
EIC & Electron Ion Collider\\
FCC-ee & Future Circular Collider ($e^+ e^-$)\\
LEFT & Low-energy  Effective Field Theory\\
LHC & Large Hadron Collider \\
LFV &  Lepton Flavor Violation\\
NP & New Physics\\
SM & Standard Model\\
SMEFT & Standard Model Effective Field Theory\\
STCF & Super Tau Charm Factory\\
WC & Wilson Coefficient 
\end{tabular}
\end{glossary}

\section{Introduction}\label{chap1:sec1}

The Standard Model (SM) of particle physics successfully passed nearly every experimental test conducted to date. However, it fails to explain several significant phenomena observed in our Universe, such as the baryon asymmetry (the prevalence of matter over antimatter), the existence of dark matter, and, in its minimal form, the origin of neutrino masses. These unresolved issues suggest that there must be physics beyond the Standard Model (BSM). Low-energy processes that do not occur or are heavily suppressed within the SM can serve as powerful probes of BSM physics. A prime example of such probes are  Charged Lepton Flavor Violation (CLFV) processes. In the limit of massless neutrinos, individual lepton family numbers\footnote{These are commonly denoted as $L_e$, $L_\mu$, $L_\tau$, and are defined as the number of negative leptons plus neutrinos of a given flavor minus the number of the corresponding positive leptons and antineutrinos.} are exactly conserved in the SM. Hence, flavor-changing processes without neutrinos (such as $\mu\to e\gamma$, $\tau\to \mu\gamma$, $\mu \to eee$ etc.) are forbidden, and thus serve as an important test of the SM. CLFV processes have therefore been extensively investigated in experiments over a wide range of energies, from low-energy searches such as CLFV $\mu$ and $\tau$ decays to high-energy processes such as $pp\to \ell\ell^{\prime}$ at the Large Hadron Collider~(LHC) and $ep\to \ell X$ in deep-inelastic scattering.
In this review, we primarily focus on low-energy CLFV processes, which offer stronger constraints on new physics~(NP) than high-energy searches in most cases, and we discuss their implications for BSM physics; for other recent reviews see, e.g.,~\cite{Calibbi:2017uvl,Ardu:2022sbt,Davidson:2022jai,Banerjee:2022xuw,Altmannshofer:2022fvz}.

The searches for CLFV have a decades-long history, starting in 1948 with the search for $\mu \to e \gamma$ performed by Hincks and Pontecorvo using cosmic-ray muons \citep{Hincks:1948vr}, followed by first bounds on $\mu\to e$ conversion a few years later \citep{Lagarrigue:1952,Steinberger:1955hfk}.
The first calculations of $\mu\to e\gamma$ induced by neutrino mixing were performed already in the very early days of the SM \citep{Petcov:1976ff,Marciano:1977wx,Lee:1977qz,Lee:1977tib}, showing that these transitions are highly suppressed.  The experimental searches for CLFV with muons culminated in the classic bounds set by SINDRUM on $\mu\to 3e$ \citep{Bellgardt:1987du} and by SINDRUM~II on $\mu\to e$ conversion in titanium and gold
 \citep{Wintz:1998rp,SINDRUMII:2006dvw}, 
 which stood for decades. The $\mu\to e\gamma$ program was subsequently advanced by MEG  and its successor MEG~II at PSI \citep{MEG:2016leq,MEGII:2023fog,MEGII:2023ltw,MEGII:2025gzr},
 currently holding the world's best limit. The analogous searches for CLFV transitions in $\tau$ decays sector began soon after the tau's discovery, with the first bounds set on a whole list of CLFV tau decay channels by the MARK-II experiment at SLAC \citep{Hayes:1981bn}. These bounds were further improved at the $B$-factories, most recently at Belle and Belle~II \citep{Belle:2025bpu}, with $\tau\to 3\mu$ also probed by the LHC experiments. The field is now entering a new generation, with Mu2e and COMET targeting $\mu\to e$ conversion \citep{Mu2e:2014fns,COMET:2018auw}, and Mu3e targeting $\mu\to 3e$ \citep{Blondel:2013ia,Mu3e:2020gyw} at orders of magnitude improved sensitivities. Improved sensitivities for CLFVs involving taus will be reached at facilities such as Belle~II at full luminosity~\citep{Belle-II:2018jsg,Banerjee:2022vdd}, the proposed Super Tau-Charm Factory (STCF)~\citep{Achasov:2023gey}, and the future Tera-$Z$ colliders such as FCC-ee and CEPC~\citep{Dam:2018rfz,Ai:2024nmn}.
 
 %---------------- Figures ---------------------
\begin{figure}[t]
  \centering
  \includegraphics[width=4.0cm]{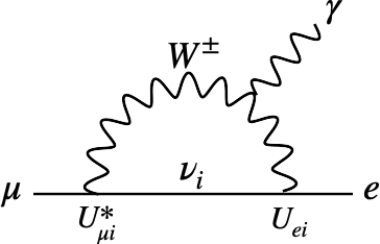}
  \vspace{0.2cm}
  \caption{Diagram inducing $\mu\to e \gamma$ in $\nu$SM (the SM extension that contains neutrino mass terms either of the Majorana or Dirac type).}
  \label{fig:Mu2EG}
\end{figure}
%------------------------------------------------
%

The CLFV processes are valuable probes of BSM physics because they fulfill two essential ingredients: {\em i)} they are highly suppressed in the minimal scenario where neutrino mass terms are introduced into the SM (often called "$\nu$SM"), and {\em ii)} they can be parametrically enhanced in generic extensions of the SM.
In $\nu$SM, the $\mu\to e \gamma$  transition is, for instance, induced at one-loop via a diagram in which neutrinos run in the loop, see Fig.~\ref{fig:Mu2EG}. If neutrinos are massless, the flavor of the intermediate neutrino $\nu_i$
is fully determined by the decaying muon, and the $W$ vertex cannot convert $\nu_i$ into an electron, i.e.~each lepton number is conserved separately. In other words, the unitary matrix $U$  that diagonalizes the neutrino masses, and that also dictates the strength of flavor transitions in the $W$ vertices, is well defined only when neutrino masses are nonzero. 
The charged-current interactions of leptons with $W$ bosons, written in the mass basis, are then given by
\begin{align}
    {\cal L}_{\rm weak}=-\frac{g_2}{\sqrt{2}}\sum_{\alpha=e,\mu,\tau}U_{\alpha i}\,\bar{\ell}_{\alpha}\gamma^{\mu}W_{\mu}^-P_L\,\nu_i++ \text{h.c.}\,,
\end{align}
where $\nu_{\alpha}=U_{\alpha i}\,\nu_{i}$, with $i=1,2,3$ labeling the neutrino mass eigenstates, $\alpha=e,\mu,\tau$ the flavor eigenstates, and $U$ the PMNS matrix rotating between neutrino flavor and mass bases ($\ell_\alpha$ are, by definition, in the charged lepton mass basis).
The resulting $\mu \to e \gamma$ decay amplitude is given by
\begin{align}
    i\mathcal{M}(\mu\to e \gamma) \propto \frac{e}{(4\pi)^2} \sum_{i=1,2,3}\,U_{\mu i}^*U_{ei}\,F_{\rm loop}\left(\frac{m^2_i}{m^2_W}\right)
    \simeq \frac{e}{(4\pi)^2}\sum_{j=2,3} U_{\mu j}^*U_{ej}\frac{\Delta m^2_{j1}}{m^2_W}\,,
\end{align}
where on the right-hand side we expanded the loop function $F_{\rm loop}$ to first nonzero order in $m^2_i/m^2_W$. The constant term vanishes due to unitarity, that is, $\sum_iU_{\mu i}^*\,U_{ei}=0$ --- an effect analogous to the GIM mechanism \citep{Glashow:1970gm} in the quark sector --- so that the amplitude is proportional to neutrino mass squared differences, $\Delta m^2_{ji}=m^2_j-m^2_i$. 

The resulting $\mu \to e\gamma$ branching ratio is \citep{Petcov:1976ff,Marciano:1977wx,Lee:1977qz, Lee:1977tib}
\begin{align}
    {\rm BR}(\mu\to e \gamma)=\frac{3\alpha}{32\pi}\,\left|~\sum_{j=2,3}U^*_{\mu i}U_{ei}\frac{\Delta m^2_{j1}}{m^2_W}~\right|^2 \approx 10^{-55}\,,
\end{align}
giving rates that are forty orders of magnitude below the sensitivities of current experiments. Similar results apply to $\mu\to 3e$ and $\mu\to e $ conversion rates, as well as the branching ratios for CLFV $\tau$ decays in $\nu$SM. Therefore, if a CLFV process is measured at any of the current or planned experiments, this would immediately imply existence of BSM physics.
%beyond the $\nu$SM.  
Since CLFV transitions are sensitive to virtual effects, searches for CLFV processes can reveal BSM particles with masses far beyond those accessible directly in the current or near-future high-energy colliders. This is true, even if the BSM only enters via loops: e.g.~, the current bound on $\mu\to e \gamma$ is already so stringent that it already probes BSM particles with masses of $\ord{100}\,{\rm TeV}$ or above, if these are contributing at one loop and have ${\mathcal O}(1)$ flavor violating couplings.

A natural target of CLFV searches is provided by models of neutrino mass generation.  These introduce new fields coupled to the SM leptons, with the goal to explain both the smallness of neutrino masses as well as the large mixing angles in the PMNS matrix, leading to correlated CLFV signatures. Explicit examples include theoretical frameworks such as the TeV-scale seesaw mechanisms (\cite{Abada:2007ux,Ibarra:2011xn,Alonso:2012ji}) and minimally flavor-violating grand unified theories (\cite{Cirigliano:2005ck}). 
%
%---------------- Figures ---------------------
\begin{figure}[t]
  \centering
  \includegraphics[width=0.6\textwidth]{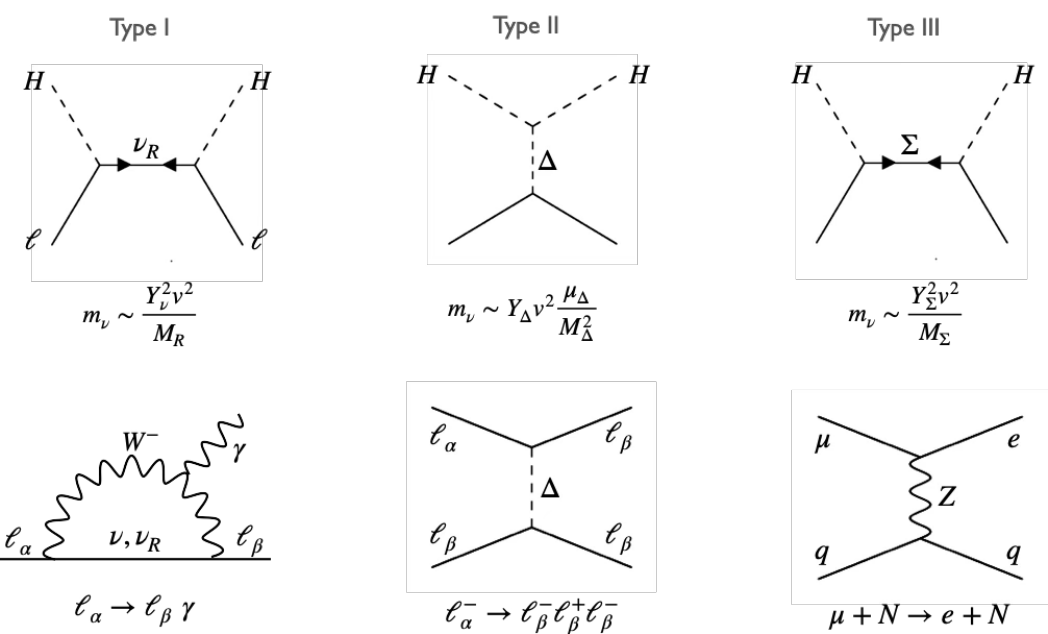}
    \vspace{0.5cm}
  \caption{Three representative neutrino mass generation mechanisms (top) and the induced CLFV diagrams (bottom). While CLFV is induced at loop level in Type I, Type II and III can generate leptonic decays, such as $\mu\to 3e$, $\tau\to 3\mu$, as well as $\mu\to e$ conversion in nuclei, already at the tree level. }
  \label{fig:NuMass}
\end{figure}
%-----------------
%
Figure~\ref{fig:NuMass} shows three representative models of neutrino mass generation, the so called Type I, II, and III seesaw mechanisms (from left to right). While Type I and Type III introduce SU(2)${}_\text{L}$ singlet fermions ($\nu_R$)~\citep{Minkowski:1977sc,Gell-Mann:1979vob,Yanagida:1979as,Mohapatra:1979ia} and triplet fermions ($\Sigma$)~\citep{Foot:1988aq}, respectively, Type II extends the SM by a SU(2)${}_\text{L}$ triplet scalar ($\Delta$)~\citep{Magg:1980ut,Lazarides:1980nt,Mohapatra:1980yp,Schechter:1980gr}. The relevant interactions are given by
\begin{align}
    {\cal L}_{\rm Type~I} &= -\frac{1}{2}\overline{\nu^c_{RI}}\,(M_R)_{IJ}\,\nu_{RJ}-\bar{L}_{\alpha}\,\tilde{H}\,(Y_{\nu})_{\alpha I}\,\nu_{RI} + \text{h.c.}\,,\\[0.3cm]
    {\cal L}_{\rm Type~ II} &=-L^T_{\alpha}\,i\tau_2\,C\,(Y_{\Delta})_{\alpha\beta}\,\Delta\,L_{\beta}+\mu_{\Delta}H^T\,i\tau_2\,\Delta^{\dagger} H+ \text{h.c.}\,,\\[0.3cm]
    {\cal L}_{\rm Type~ III} &=-\frac{1}{2}{\rm Tr}\left[\overline{\Sigma}M_{\Sigma}\Sigma^c\right]-\sqrt{2}\,L_{\alpha}\,(Y_{\Sigma})_{\alpha I}\,\Sigma_I\,\tilde{H}+ \text{h.c.}\,,
\end{align}
where $L_\alpha$ are the lepton SU(2)${}_\text{L}$ doublets and $H$ is the Higgs doublet (with $\tilde{H}=i\tau^2H^*$) and  $C$ the charge conjugation matrix.
Note that all these models result in neutrino masses $m_{\nu}$ that are of Majorana type. Interestingly, if a CLFV signal is observed, the underlying mechanism could be probed by comparing the relative rates of the observed processes \citep{Chu:2011jg, Dinh:2012bp,Alonso:2012ji}. For example, CLFV processes in Type I see-saw arise at loop level by closing neutrino and $W$ boson lines (cf.~the bottom-left diagram in Fig.~\ref{fig:NuMass}). In Type II see-saw, in contrast,  the 3-body leptonic decays $\mu\to 3e$ and $\tau\to 3\ell$ are generated already at tree level via an exchange of the SU(2)${}_\text{L}$ triplet scalar (see the bottom-middle diagram in Fig.~\ref{fig:NuMass}). The triplet fermions in Type III see-saw contains the charged states $\Sigma^{\pm}$ that mix with charged leptons due to their Yukawa interactions, inducing CLFV $Z$-boson couplings. Consequently, in Type III see-saw, both the $\ell_i\to \ell_j \ell_k \ell_\ell $ decays and the $\mu\to e $ conversion are generated already at tree level (as shown in the bottom-right diagram in Fig.~\ref{fig:NuMass}). It is then clear that studying the relative sizes of different CLFV processes can provide valuable insights into the mechanism responsible for neutrino mass generation~\citep{Hambye:2013jsa}. Note that these types of BSM extensions are also motivated by the need to explain the observed baryon asymmetry of the Universe. Possible correlations between the baryon asymmetry and CLFV processes have been explored, for example, in the framework of low-scale leptogenesis~\citep{Granelli:2022eru}, where measurements of various CLFV transitions would provide an additional handle on processes in the early Universe. 

In the following, we review how searches for CLFV serve as powerful probes of both light and heavy NP particles coupled to SM leptons. Rather than relying on specific BSM scenarios (such as the neutrino mass models discussed above), we organize the discussion within the more model-independent and flexible framework of effective field theories (EFTs), which we introduce in Sec.~\ref{sec:eft}.  
In Sec.~\ref{sec:rare:muon:trans}, we review CLFV muon processes and discuss their implications for BSM scenarios, with a particular focus on the subtleties involved in the theoretical calculation of $\mu \to e$ conversion on nuclei and on possible signals of light BSM physics in muon decays. Sec.~\ref{sec:tau} is devoted to CLFV $\tau$ decays, including purely leptonic processes, hadronic tau decays, and tau decays involving light NP particles.
 We review possible CLFV signatures at high-energy colliders in Sec.~\ref{sec:colliders}, including CLFV decays of Higgs and $Z$ bosons, while we present our conclusions in  Sec.~\ref{sec:conclusions}.

\section{Effective Field Theory framework}
\label{sec:eft}

Effective Field Theories (EFTs) enter the description of CLFV process in two ways: \emph{i}) if NP particles are heavy, they can be integrated out --- their effects at low energies are then captured fully by the Standard Model Effective Field Theory (SMEFT), for energies above electroweak breaking scale, and by the Low-Energy Effective Field Theory (LEFT) at lower energies, e.g., at energies typical for muon and tau decays (see Secs.~\ref{sec:rare:muon:trans} and~\ref{sec:tau}, respectively); and \emph{ii}) one uses EFT techniques to describe the effects of strong interactions in $\mu\to e $ conversion and in hadronic $\tau$ decays. In the following, we first introduce SMEFT and LEFT. Their extensions to the cases where $\mu$ and $\tau$ decay to new exotic light particles are denoted SMEFT${}_X$ and LEFT${}_X$, and are discussed in  Secs.~\ref{sec:mu:light:np} and~\ref{sec:tau:light:np},  respectively. The use of EFTs to describe hadronic effects is discussed in Sec.~\ref{sec:mu:e:conv} ($\mu\to e$ conversion) and Sec.~\ref{sec:tau:hadronic} ($\tau$ decays).

%---------------- Figures ---------------------
\begin{figure}[t]
  \centering
  \includegraphics[width=0.8\textwidth]{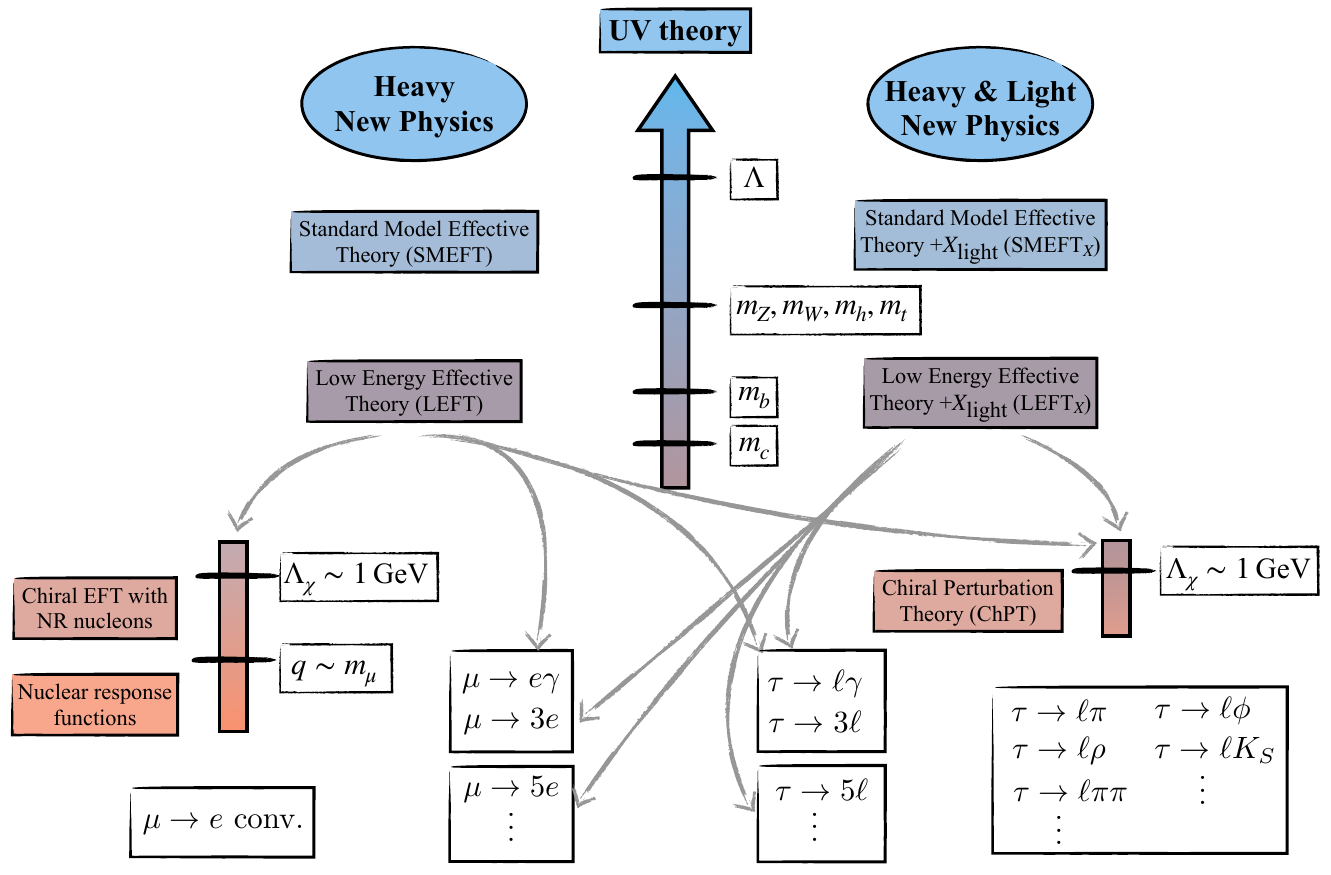}
    \vspace{0.5cm}
  \caption{A ladder of EFTs that describe CLFV processes at various energy scales. Many of the inputs to ChPT and Chiral EFT with nonrelativistic nucleons are provided by Lattice QCD (not denoted). }
  \label{fig:EFTs:ladder}
\end{figure}
%-----------------
%

A schematic depiction of the ladder of EFTs describing the CLFV processes at various scales and for different low energy transitions is given in Fig.~\ref{fig:EFTs:ladder}. For concreteness, we assumed in the schematic that the mediators inducing CLFV have a typical mass $\Lambda$ that is well above  the electroweak (EW) scale, $\Lambda \gg v_\text{EW}$.
Such heavy NP degrees of freedom can be integrated out below the scale $\Lambda$ and their effects at lower energies are then captured by higher dimension operators, of dimension $D\ge 5$. Assuming  that the low energy particle content is fully described by the SM degrees of freedom, these higher dimension operators are constructed solely out of SM fields and are invariant under the full SM gauge group, $\text{SU(3)}_c\times \text{SU(2)}_\text{L} \times \text{U(1)}_Y$. This framework is known as SMEFT~\citep{Buchmuller:1985jz,Grzadkowski:2010es}, see, e.g.,~\cite{Brivio:2017vri} for a review. The SMEFT Lagrangian is given by 
\begin{align}
\mathcal{L}_\text{SMEFT} =  \mathcal{L}_\text{SM} + 
\sum_{D,\,a} \frac{1}{\Lambda^{D-4}}\hat C^{(D)}_a Q^{(D)}_a\,,
\label{eq:smeft}
\end{align}    
where $\mathcal{L}_\text{SM}$ is the SM Lagrangian, $Q^{(D)}_a$ are operators of a given dimension $D$ and $\hat C^{(D)}_a$ the corresponding Wilson coefficients~(WC), dimensionless interaction parameters accounting for UV couplings, mixing angles, and loop factors, obtained by matching the UV-complete theory onto SMEFT. From the low energy perspective, the  SMEFT WCs are often treated as free parameters, especially when presenting general experimental bounds on BSM physics. Note that the $D=5$ SMEFT operators only give rise to the Majorana neutrino mass terms~\citep{Weinberg:1979sa}. CLFV effects first arise at the $D=6$ level, from operators involving lepton fields of different flavors, see Table~\ref{tab:smeft} for a complete basis of such operators.\footnote{In this review, we only consider CLFV operators that preserve the total lepton and baryon numbers.}

\begin{table}[t!] 
\centering
\caption{A complete list of dimension-6 SMEFT operators $Q_a^{(6)}$ (in the basis of~\cite{Grzadkowski:2010es}) that can contribute to CLFV observables~\citep{Crivellin:2013hpa}. Here,
$Q$ and $L$ denote quark and lepton $\text{SU(2)}_\text{L}$ doublets, respectively; $U,\,D$ and $E$ are the up-quark, down-quark and charged lepton singlets; $H$ is the Higgs doublet,  with $\vpj\equiv H^\dag (D_\mu H)-(D_\mu H)^\dag H$; while $B_{\mu\nu}$ and $W^i_{\mu\nu}$ are, respectively, the $\text{U(1)}_Y$ and $\text{SU(2)}_\text{L}$ field strengths. The flavor indices are omitted, and a summation over fundamental and adjoint  $\text{SU(2)}_\text{L}$ indices ($a,b=1,2$ and $i=1,2,3$) is understood ($\tau_i$ are the Pauli matrices).
\label{tab:smeft}}
\renewcommand{\arraystretch}{1.3}
\centering
\setlength{\tabcolsep}{5pt}
\begin{subtable}{0.49\textwidth}
\begin{tabular}[t]{cccc} 
 \hline
\multicolumn{2}{c}{4-lepton operators} &
\multicolumn{2}{c}{Dipole operators} \\
\hline
$Q_{\ell\ell }$ & $(\bar L \gamma_\mu L)(\bar L \gamma^\mu L)$ & 
{ $Q_{eW}$} & { $(\bar L \sigma^{\mu\nu} E) \tau_i H W_{\mu\nu}^i+ \text{h.c.}$} \\
$Q_{ee}$ & $(\bar E \gamma_\mu E)(\bar E \gamma^\mu E)$ &
{ $Q_{eB}$} & { $(\bar L \sigma^{\mu\nu} E) H B_{\mu\nu}$}+ h.c. \\
$Q_{\ell e}$ & $(\bar L \gamma_\mu L)(\bar E \gamma^\mu E)$ & & \\
\hline
\multicolumn{4}{c}{Higgs-lepton operators}\\ 
\hline
{ $Q_{H \ell }^{(1)}$} & { $i(\vpj)(\bar L\gamma^\mu L)$} &
{ $Q_{H \ell}^{(3)}$} & { $i(\vpjt)(\bar L \tau^i \gamma^\mu L)$} \\
{ $Q_{H e}$} & { $i(\vpj)(\bar E \gamma^\mu E)$} &
$Q_{eH3}$ & $(\bar L E H)(H^\dag H)+ \text{h.c.}$\\
\hline
\end{tabular}
\end{subtable}
\begin{subtable}{0.49\textwidth}
\begin{tabular}[t]{cccc} 
 \hline
\multicolumn{4}{c}{2-lepton 2-quark operators} \\
\hline 
$Q_{\ell q}^{(1)}$ & $(\bar L \gamma_\mu L)(\bar Q \gamma^\mu Q)$ & 
  $Q_{\ell u}$ & $(\bar L \gamma_\mu L)(\bar U \gamma^\mu U)$ \\
$Q_{\ell q}^{(3)}$ & $(\bar L \gamma_\mu \tau_i L)(\bar Q \gamma^\mu \tau_i Q)$ & 
$Q_{eu}$ & $(\bar E \gamma_\mu E)(\bar U \gamma^\mu U)$ \\
$Q_{eq}$ & $(\bar E \gamma^\mu E)(\bar Q \gamma_\mu Q)$ &
$Q_{\ell edq}$ & $(\bar L^a E)(\bar D Q^a)+\text{h.c.}$ \\
$Q_{\ell d}$ & $(\bar L \gamma_\mu L)(\bar D \gamma^\mu D)$ & 
$Q_{\ell equ}^{(1)}$ & $(\bar L^a E) \epsilon_{ab} (\bar Q^b U)+ \text{h.c.}$ \\
$Q_{ed}$ & $(\bar E \gamma_\mu E)(\bar D\gamma^\mu D)$ & 
$Q_{\ell equ}^{(3)}$ & $(\bar L^a \sigma_{\mu\nu} E) \epsilon_{ab} (\bar Q^b \sigma^{\mu\nu} U) +\text{h.c.}$ \\
\hline
\end{tabular}
\end{subtable}
\end{table}

At the EW scale the EW symmetry is broken, so that SMEFT is matched onto LEFT by integrating out the heavy SM fields --- $W$ and $Z$ bosons, Higgs, and the top quark~\citep{Jenkins:2017jig}. The LEFT operators  are built out of only the light SM fields and are invariant under the residual gauge symmetry $\text{SU(3)}_c\times \text{U(1)}_\text{em}$. 
The LEFT operators provide a good description of the processes that we are interested in as long as the BSM particles have masses $M  \approx \Lambda \gg m_{\ell}$ (where $m_{\ell}$ is the mass of the muon or the $\tau$), i.e., for CLFV processes LEFT can be a well behaved EFT even if $M$ lies at or below the EW scale.
Similarly to SMEFT in Eq. \eqref{eq:smeft}, the LEFT Lagrangian consists of the SM and a complete set of higher dimension operators, $O_a^{(D)}$ multiplied by WCs, $C_a^{(D)}$,
\begin{align}
\mathcal{L}_\text{LEFT} =  \mathcal{L}_\text{SM} + 
\sum_{D,\,a} {\mathcal C}^{(D)}_a {\mathcal O}^{(D)}_a\,,
\label{eq:left}
\end{align}    
where now  the WCs are defined as dimensionful quantities.
The LEFT operators relevant for CLFV observables involve vector, scalar, and tensor Lorentz structures; the complete set of CLFV $D=5,\,6$ operators is listed in Table~\ref{tab:left}.

The only $D=5$ CLFV operator is the transition dipole moment, coupling leptons to the photon. In the SMEFT to LEFT matching at tree level, it is generated from the SMEFT EW dipoles as follows,\footnote{To shorten the notation we drop from now on the operator dimension label, i.e., ${\mathcal C}_{e\gamma,\,ij}^{(5)}\to {\mathcal C}_{e\gamma,\,ij}$, etc.} 
\begin{align}
{\mathcal C}_{e\gamma,\,ij} = \frac{v_\text{EW}}{\sqrt2 \Lambda^2} \hat C_{e\gamma,\,ij}\,,
\qquad \hat C_{e\gamma,\,ij} \equiv\left(\cos\theta_W \hat C_{eB,\,ij}-\sin\theta_W \hat C_{eW,\,ij}\right)\,,
\label{eq:dipole}
\end{align}
where $v_\text{EW} \simeq 246\,{\rm GeV}$ is the EW-breaking vev of the Higgs field, $\theta_W$ is the weak mixing angle, and we neglected possible shifts to both quantities induced by other SMEFT operators, which would induce only subdominant corrections suppressed by an additional $\ord{{v_\text{EW}^2}/{\Lambda^2}}$ factor. For the full SMEFT-LEFT matching, we refer to \cite{Jenkins:2017jig}. As another example, let us consider the coefficient of the LEFT 4-lepton contact interaction ${\mathcal O}_{ee,\,ijkk}^{V\,LL}$:
\begin{align}   
{\mathcal C}^{V\,LL}_{ee,\,ijkk} = \frac{1}{\Lambda^2}\left[\hat C_{\ell\ell,\,ijkk}+(2\sin\theta_W^2-1) \left(\hat C_{H \ell,\,ij}^{(1)} + \hat C_{H\ell,\,ij}^{(3)} \right)\right]\,,
\label{eq:4lep}
\end{align}
where, besides the straightforward matching contribution from the corresponding 4-lepton SMEFT operator (the first term in the bracket above), there are also contributions arising from integrating out the $Z$ boson: $\hat C_{H \ell,\,ij}^{(1)} $  and $\hat C_{H\ell,\,ij}^{(3)}$ induce effective CLFV $Z$ vertices, and therefore tree-level $Z$-mediated diagrams result in additional contributions to the CLFV 4-fermion interactions.

\begin{table}[t!] 
\centering
\caption{Dimension-6 LEFT operators (plus the $D=5$ photon dipole) that can induce CLFV transitions~\citep{Jenkins:2017jig,Fernandez-Martinez:2024bxg}. The $e$, $u$, and $d$ denote the charged leptons, up-type quarks, and down-type quarks, respectively, while $F_{\mu\nu}$ is the electromagnetic field strength. The chirality of the fermion fields is explicitly denoted, while the flavor indices are omitted. 
\label{tab:left}}
\renewcommand{\arraystretch}{1.3}
\setlength{\tabcolsep}{7pt}
\begin{subtable}{0.33\textwidth}
\centering
\begin{tabular}[t]{cc} 
 \hline
\multicolumn{2}{c}{4-lepton operators} \\
\hline
$\ops^{V\,LL}_{ee}$ & $(\bar e_L \gamma_\mu e_L)(\bar e_L \gamma^\mu e_L)$  \\
$\ops^{V\,RR}_{ee}$ & $(\bar e_R \gamma_\mu e_R)(\bar e_R \gamma^\mu e_R)$  \\
$\ops^{V\,LR}_{ee}$ & $(\bar e_L \gamma_\mu e_L)(\bar e_R \gamma^\mu e_R)$  \\
$\ops^{S\,RR}_{ee}$ & $(\bar e_L e_R)(\bar e_L  e_R) + \text{h.c.}$  \\
\hline
\multicolumn{2}{c}{Dipole operator}\\ 
\hline
{ $\ops_{e\gamma}$} &  $(\bar e_L\sigma^{\mu\nu}  e_R)  F_{\mu\nu} + \text{h.c.}$ \\
\hline
\end{tabular}
\end{subtable}
%\hfill
\begin{subtable}{0.66\textwidth}
\begin{tabular}[t]{cccc} 
 \hline
\multicolumn{4}{c}{2-lepton 2-quark operators} \\
\hline
$\ops^{V\,LL}_{ue}$ & $(\bar u_L \gamma_\mu u_L)(\bar e_L \gamma^\mu e_L)$ &  $\ops^{V\,LL}_{de}$ & $(\bar d_L \gamma_\mu d_L)(\bar e_L \gamma^\mu e_L)$ \\
$\ops^{V\,RL}_{ue}$ & $(\bar u_R \gamma_\mu u_R)(\bar e_L \gamma^\mu e_L)$  & $\ops^{V\,RL}_{de}$ & $(\bar d_R \gamma_\mu d_R)(\bar e_L \gamma^\mu e_L)$  \\
$\ops^{V\,LR}_{ue}$ & $(\bar u_L \gamma_\mu u_L)(\bar e_R \gamma^\mu e_R)$ &  $\ops^{V\,LR}_{de}$ & $(\bar d_L \gamma_\mu d_L)(\bar e_R \gamma^\mu e_R)$ \\
$\ops^{V\,RR}_{ue}$ & $(\bar u_R \gamma_\mu u_R)(\bar e_R \gamma^\mu e_R)$  &  $\ops^{V\,RR}_{de}$ & $(\bar d_R \gamma_\mu d_R)(\bar e_R \gamma^\mu e_R)$ \\
$\ops^{S\,RR}_{ue}$ & $(\bar u_L  u_R)(\bar e_L e_R) + \text{h.c.}$ &  $\ops^{S\,RR}_{de}$ & $(\bar d_L  d_R)(\bar e_L e_R) + \text{h.c.}$ \\
$\ops^{S\,RL}_{ue}$ & $(\bar u_L  u_R)(\bar e_R e_L) + \text{h.c.}$ &  $\ops^{S\,RL}_{de}$ & $(\bar d_L  d_R)(\bar e_R e_L) + \text{h.c.}$ \\
$\ops^{T\,RR}_{ue}$ & $(\bar u_L \sigma_{\mu\nu} u_R)(\bar e_L \sigma^{\mu\nu} e_R) + \text{h.c.}$ &  $\ops^{T\,RR}_{de}$ & $(\bar d_L \sigma_{\mu\nu}  d_R)(\bar e_L \sigma^{\mu\nu}  e_R) + \text{h.c.}$ \\
\hline
\end{tabular}
\end{subtable}
\end{table}

 To constrain the SMEFT WCs using CLFV measurements, one must evolve the WCs via Renormalization Group~(RG) running from the new-physics scale $\Lambda$ down to the energy scale typical for a particular experimental  process, see Fig.~\ref{fig:EFTs:ladder}. The RG running modifies the values of the WCs (albeit, typically not very significantly), and, more importantly,  mixes operators, generating at low energies also WCs that may initially have been set to zero at scale $\Lambda$. For CLFV processes, the RG evolution typically involves three stages: \emph{i}) SMEFT running from $\Lambda$ to the EW scale~\citep{Jenkins:2013zja,Jenkins:2013wua,Alonso:2013hga}; \emph{ii}) matching SMEFT onto LEFT by removing the heavy SM particles~\citep{Jenkins:2017jig}; and \emph{iii}) QED$\times$QCD running of the LEFT operators~\citep{Jenkins:2017dyc} down to low scales such as  $\mu\approx m_\mu$ or $\mu\approx m_\tau$. Notably, QED running can induce significant operator mixing for low-energy CLFV processes~\citep{Crivellin:2017rmk}. In addition, lower energy scales can also enter the theory prediction of CLFV rates for processes involving hadrons and nuclei, such as hadronic tau decays and $\mu\to e$ conversion in nuclei, requiring the use of chiral and nuclear EFTs. This is illustrated in the bottom part of Fig.~\ref{fig:EFTs:ladder} and discussed in detail in Secs.~\ref{sec:mu:e:conv} and~\ref{sec:tau:hadronic}.

In the following, we quote limits on LEFT WCs, ignoring the RG running effects for simplicity and typically switching on a single operator at a time (although the interplay of different operators is briefly discussed). 
This provides a simple illustration of the potential for testing NP through CLFV searches. Within a generic UV-complete theory multiple operators are expected to be induced,\footnote{See, for instance,~\cite{Coy:2018bxr,Zhang:2021jdf,Li:2022ipc,Li:2023ohq}, for the matching of the above-mentioned seesaw models onto SMEFT operators.} however,  and one needs to bear in mind that correlations among WCs can potentially  weaken the limits significantly (especially if these are combined with the nuclear uncertainties in $\mu \to e$ conversion), as demonstrated in a global analysis by~\cite{Fernandez-Martinez:2024bxg}. 

\section{Muon transitions}
\label{sec:rare:muon:trans}

Due to their light mass and relatively long lifetime ($m_\mu \simeq 105.66\,{\rm MeV}$, $\tau_\mu \simeq 2.197\,{\rm \mu s}$), muons can be efficiently produced, transported in high-intensity beams, and stopped before they decay.  This motivates a rich experimental program, with muonic CLFV processes being some of the most tightly constrained processes ever searched for. Intense muon beams are produced by colliding high-power proton beams with a target, and can be stopped in thin targets. This results in large samples of muons that decay at rest, in a clean, well-controlled environment, ideal for precision searches. The MEG~II experiment at the Paul Scherrer Institute (PSI) is currently searching for $\mu^+ \to e^+\gamma$ decays with unprecedented sensitivity~\citep{MEGII:2023fog,MEGII:2025gzr}, while the upcoming Mu3e experiment~\citep{Mu3e:2020gyw}, also at PSI, aims to observe $\mu^+ \to e^+ e^+ e^-$. The COMET experiment~\citep{Kuno:2013mha} at J-PARC and the Mu2e experiment~\citep{Mu2e:2014fns} at FNAL are both going to search for coherent $\mu^- \to e^-$ conversion in the field of an atomic nucleus. The current limits and projected sensitivities for various $\mu\to e$ transitions are collected in Table~\ref{tab:muon:limits}.

As we discuss below, these processes do not measure a single quantity but rather complementary combinations of short-distance couplings, which then carry nontrivial information about the underlying new physics dynamics. 
 Assuming $\ord{1}$ new physics couplings, this program is sensitive to energies up to $\Lambda \approx 10^3 - 10^4~\mathrm{TeV}$, well beyond direct reach of any current or foreseeable collider. 
 The key point is that, in terms of LEFT operators in Table \ref{tab:left}, $\mu\to e\gamma$ is primarily sensitive to the photon dipole operator, while $\mu\to e$ conversion and $\mu\to 3e$ receive contributions from both the dipole operator and the contact 4-fermion operators. A measurement
 %, or a non-observation, 
 in more than one channel can therefore discriminate between dipole-dominated and contact-dominated scenarios; comprehensive reviews of this program can be found in \citet{Kuno:1999jp,deGouvea:2013zba,Bernstein:2013hba,Lindner:2016bgg,Calibbi:2017uvl,Ardu:2022sbt,Asadi:2025dii}. The combined model discriminating power of $\mu\to e$ observables has been discussed in detail, e.g.,~in~\cite{Davidson:2018kud,Davidson:2022nnl,Ardu:2023yyw,Ardu:2024bua}. 

\begin{table}[t]
\renewcommand{\arraystretch}{1.3}
\TBL{\caption{Current 90\%~CL limits and projected future sensitivities for the branching ratios of benchmark rare muon transitions. For $\mu\to e$ conversion the target nucleus is indicated in parentheses. The limit marked with $^\dag$ is valid for a practically massless invisible NP boson $X$ with substantial couplings to the right-handed leptons, see \cite{Calibbi:2020jvd} and Sec.~\ref{sec:mu:light:np} for details.}\label{tab:muon:limits}}
{\begin{tabular*}{\textwidth}
{@{\extracolsep{\fill}}@{}lllll@{}}
\toprule
Process & Current limit & Experiment & Future reach & Experiment\\
\colrule
$\mu^+\to e^+\gamma$            & $1.5\times10^{-13}$ & MEG~II~\citep{MEGII:2025gzr} & $ 6\times10^{-14}$ & MEG~II~\citep{MEGII:2023fog}\\[1mm]
$\mu^+\to e^+e^-e^+$            & $1.0\times10^{-12}$ & SINDRUM~\citep{Bellgardt:1987du} & $ 10^{-16}$ & Mu3e~\citep{Mu3e:2020gyw}\\[1mm]
\multirow{2}{*}{$N\,\mu^- \to N\,e^-$}    & $7\times10^{-13}~(\mathrm{Au})$ & SINDRUM~II~\citep{SINDRUMII:2006dvw}  & \multirow{2}{*}{$ 6\times10^{-17}~(\mathrm{Al})$} & COMET~\citep{Kuno:2013mha} and\\
 & $6.1\times10^{-13}~(\mathrm{Ti})$ & SINDRUM~II~\citep{Wintz:1998rp} & & Mu2e~\citep{Mu2e:2012eea}\\[1mm]
$\mu^+\to e^+ X$                & $2.6\times10^{-6}$\,$^\dag$  & Jodidio \textit{et al.}~\citep{Jodidio:1986mz} & $ 7\times 10^{-8}$ & Mu3e~\citep{Perrevoort:2018ttp}\\
\botrule
\end{tabular*}}{}
\end{table}

\subsection{$\mu\to e\gamma$ and $\mu \to 3e$}
\label{sec:mu:e:gamma}

Modern experiments search for these decay modes using positive muons, since these stop in a target without forming muonic atoms, which then enables investigating decays of free muons at rest.

The radiative decay $\mu^+\to e^+\gamma$ is directly induced by the LEFT photon dipole operator in Table~\ref{tab:left}, and is thus the cleanest probe of it. Experimentally, for a muon decaying at rest, the signal is the coincidence of a positron and a photon emitted back-to-back, each monochromatic with energy $E_e=E_\gamma=m_\mu/2\simeq 52.8~$MeV. 
For high-intensity muon beams, the main background is due to the accidental coincidence of a positron at the kinematic end-point of the energy distribution of the ordinary muon decay $\mu^+\to e^+ \nu_e \bar \nu_\mu$ (the so-called Michel spectrum) and a photon emitted from an unrelated process, e.g.,~the standard radiative decay of another muon, $\mu^+\to e^+  \nu_e \bar \nu_\mu \gamma$. 
The rate of such an accidental background grows quadratically with the instantaneous muon stopping rate. 
Together with the irreducible radiative decay, this sets the achievable resolution-driven sensitivity and favors a continuous, rather than pulsed, beam.
In order to suppress the accidental background, in addition to an excellent energy resolution, detectors need to have a high angular and timing resolution, to ensure that the positron and photon are indeed emitted back to back at the same time.
The current world best limit, $\mathrm{BR}(\mu^+\to e^+\gamma)<1.5\times10^{-13}$ at $90\%$~CL, has been set by the MEG~II experiment \citep{MEGII:2025gzr}, superseding the first-dataset result \citep{MEGII:2023ltw} and the final result of the predecessor MEG experiment, $\mathrm{BR}<4.2\times10^{-13}$~\citep{MEG:2016leq}. With the full dataset MEG~II aims at an ultimate sensitivity of approximately $6\times10^{-14}$. Due to the mentioned accidental background, increased statistics would not be beneficial without a reduction of the systematic uncertainties. Therefore, in order to push the sensitivity beyond MEG~II, future experiments will need improved energy and angular resolutions. To date, there is no approved project yet, but a possible experimental setup where thin layers of dense material convert photons into $e^+e^-$ pairs (which can be tracked with high precision) is currently under study, with the aim to reach sensitivities at the $\mathcal{O}(10^{-15})$ level~\citep{Cavoto:2017kub,Cattaneo:2025bnk}.

As mentioned above, the branching ratio of $\mu\to e\gamma$ is controlled directly by the LEFT dipole coefficients, which match to SMEFT dipoles as in Eq.~\eqref{eq:dipole}, so that an observation of $\mu\to e\gamma$ would immediately fix their magnitudes: 
\begin{align}
    \mathrm{BR}(\mu\to e\gamma) = \frac{m_\mu^3}{4\pi\,\Gamma_\mu} \left(|{\mathcal C}_{e\gamma,\,12}|^2+|{\mathcal C}_{e\gamma,\,21}|^2 \right)\,,
    \label{eq:meg}
\end{align}
where $\Gamma_\mu$ denotes the total decay width of the muon, which at leading order is $\Gamma_\mu = {m_\mu^5 G_F^2}/{192\pi^3}$, with $G_F$ the Fermi constant.
The present MEG~II limit translates into the following bound on either coefficient (setting the other one to zero):
\begin{align}
    {\mathcal C}_{e\gamma,\,12}\,,~{\mathcal C}_{e\gamma,\,21} \lesssim 2.2\times 10^{-14}~{\rm GeV}^{-1} \quad \Longrightarrow \quad \Lambda \gtrsim 8.9\times 10^7~{\rm GeV}\,,
\end{align}
where, for the last step, we used Eq.~\eqref{eq:dipole} setting $ \hat C_{e\gamma,\,ij} \equiv\left(\cos\theta_W \hat C_{eB,\,ij}-\sin\theta_W \hat C_{eW,\,ij}\right) = 1$ for illustration. Although this shows how $\mu\to e\gamma$ searches have the capability to constrain new physics up to impressively high scales, this result should not be taken at face value, interpreting it simply as a limit on the masses of NP particles. For instance, in perturbative UV-complete theories -- the supersymmetric seesaw being the textbook example~\citep{Borzumati:1986qx,Hisano:1995cp} --
 such dipole transitions are induced at one loop, leading to a typical suppression by $1/(4\pi)^2$; this factor is absent, though, in non-perturbative UV theories. The insertion of an electromagnetic coupling $e$ is always necessary, while if the required chirality flip is from the external muon line, there is an additional chiral suppression proportional to the muon Yukawa coupling, $y_\mu = \sqrt2 m_\mu/v_{\rm EW} \approx 6\times 10^{-4}$. Roughly estimating these effects by multiplying the WC by a factor $\sim y_\mu e/(4\pi)^2$, we still obtain a limit on the order of $\Lambda \gtrsim 100\,{\rm TeV}$. This shows that in order for any NP inducing $\mu\to e\gamma$ to be within reach of the LHC and/or any proposed future collider, some kind of flavor-protection is required. Furthermore, beyond its conventional sensitivity to dipole operators, $\mu\to e\gamma$ can also impose constraints on semileptonic tensor operators through non-perturbative QCD effects. Using ChPT, \cite{Dekens:2018pbu} evaluated the matching of quark-level tensor operators onto the electromagnetic dipole operator, and studied the impact of these non-perturbative QCD effects. 
The resulting hadronic contribution provides one of the strongest indirect constraints on CLFV tensor interactions, probing new physics scales far above the TeV scale.

Dipole operators also induce the three-body decay $\mu^+\to e^+e^-e^+$, which, in addition, receives contributions from the 4-lepton contact interactions.
Its experimental signature consists of three tracks that originate from a common vertex with vanishing total momentum and total energy equal to $m_\mu$. The dominant backgrounds are the irreducible internal-conversion decay $\mu\to eee\nu\bar\nu$, the relevance of which is controlled by the experimental momentum resolution; and accidental track combinations, controlled by the vertex and timing resolutions. The current limit, $\mathrm{BR}(\mu^+\to e^+e^-e^+)<1.0\times10^{-12}$, was established by SINDRUM in 1988 \citep{Bellgardt:1987du} and has stood as the world best constraint for almost four decades. The Mu3e experiment at PSI~\citep{Mu3e:2020gyw}, whose construction is now complete and data taking is starting, will improve on it dramatically by reconstructing all three tracks in a thin, high-rate silicon pixel tracker. Phase~I targets a sensitivity of $\sim 2\times10^{-15}$, while Phase~II aims at sensitivity of $10^{-16}$ \citep{Blondel:2013ia,Mu3e:2020gyw}, four orders of magnitude below the SINDRUM bound.

\begin{figure}
    \centering    
    \includegraphics[width=0.6\textwidth]{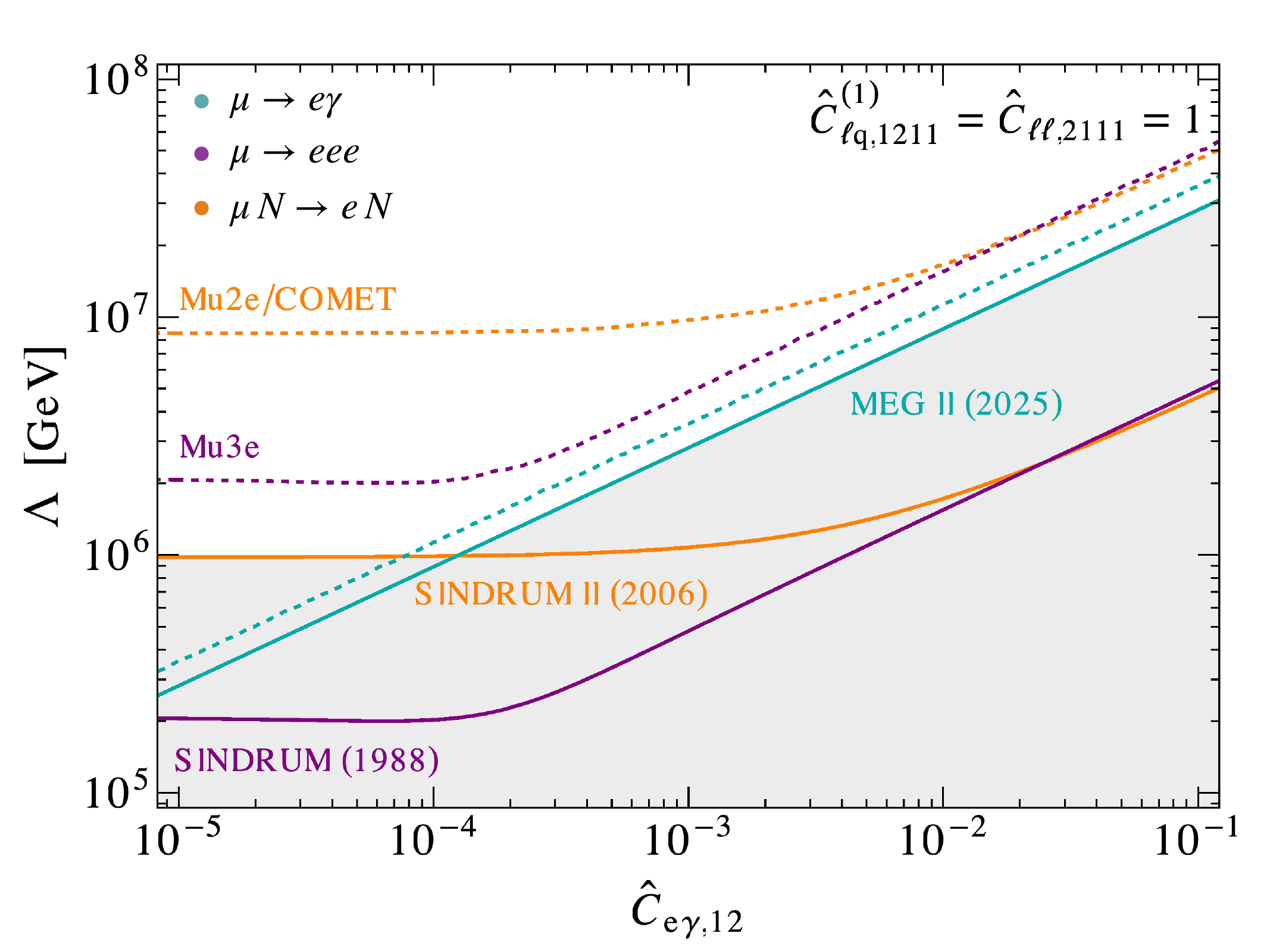}
    \caption{Sensitivity to the NP scale $\Lambda$ associated with SMEFT CLFV operators obtained from searches for $\mu \to e\gamma$ (cyan), $\mu\to 3e$ (purple), and $\mu\to e$ conversion in nuclei (orange). The solid lines represent the current lower bounds from the limits listed in Table~\ref{tab:muon:limits}, the dashed lines show the expected reach of running and upcoming experiments. The WCs of the indicated 4-fermion interactions are set to one, while the $x$-axis shows the effect of varying the dimensionless coefficient of the photon dipole operator $\hat C_{e\gamma,\,12} $, see Eq. \eqref{eq:dipole}.}
    \label{fig:mue-summary}
\end{figure}

In terms of the WCs of the dipole operator and the 4-lepton interactions in Table~\ref{tab:left}, the branching ratio is
\begin{align}
{\rm BR}(\mu^+\to e^+e^+ e^-) 
=&~\frac{m_{\mu}^5}{3 (16\pi)^3 \Gamma_{\mu}} 
\Bigg[16\big|C^{V\,LL}_{ee,\,2111}\big|^2+ 16 \big|C^{V\,RR}_{ee,\,2111}\big|^2 + 8 \big|C^{V\,LR}_{ee,\,2111}\big|^2 + 8\big|C^{V\,LR}_{ee,\,1121}\big|^2 
+ \big|C^{S\,RR}_{ee,\,2111}\big|^2 + \big|C^{S\,RR}_{ee,1121}\big|^2 + \nonumber\\
& ~\frac{256 \,e^2}{m_{\mu}^2} \left(\ln\frac{m_{\mu}^2}{m_{e}^2}-\frac{11}{4}\right) \Big(\big|C_{e\gamma,\,12}\big|^2 + \big|C_{e\gamma,\,21}\big|^2\Big)- \frac{64\, e}{m_{\mu}}\, {\rm Re}\Big[\Big(2 C^{V\,LL}_{ee,\,2111} + C^{V\,LR}_{ee,\,2111}\Big) C_{e\gamma,\,21}^{*} + \Big(2C^{V\,RR}_{ee,\,2111}+C^{V\,LR}_{ee,\,1121}\Big) C_{e\gamma,\,12}\Big] \Bigg]\,.
\label{eq:mu3e}
\end{align}
The ratio of branching ratios, ${\mathrm{BR}(\mu\to 3e)}/{\mathrm{BR}(\mu\to e\gamma)}$, is a very useful diagnostic of the origin of $\mu\to 3e$ transition. For instance, 
when the dipole contribution dominates, the $\mu\to 3e$ rate is directly proportional to $\mu\to e\gamma$~\citep{Kuno:1999jp},
\begin{equation}
\label{eq:mu3e:dipole}
\frac{\mathrm{BR}(\mu\to 3e)}{\mathrm{BR}(\mu\to e\gamma)}\simeq \frac{\alpha}{3\pi}\left(\ln\frac{m_\mu^2}{m_e^2}-\frac{11}{4}\right)\approx 6.6\times10^{-3}.
\end{equation}
When the contributions  from the four-fermion and dipole operators are comparable, these can interfere constructively or destructively, though the largest cancellation is only at the level of ${\mathcal O}(10\%)$. Finally, when the four-fermion operators dominate,  $\mathrm{BR}(\mu\to 3e)$ can be well above the relation in Eq.~\eqref{eq:mu3e:dipole}. 

Currently,  the bound on the dipole operator  from $\mu \to 3e$  is  weaker than the one set by $\mu\to e \gamma$. However, the expected sensitivity of Mu3e Phase-II will improve this constraint beyond the final reach of MEG~II, reaching ${\mathcal C}_{e\gamma,\,12}\,,~{\mathcal C}_{e\gamma,\,21} \lesssim 7.0\times 10^{-15}~{\rm GeV}^{-1}$, $\Lambda \gtrsim 1.6\times 10^8~{\rm GeV}$. For the contact interactions, the present bound on, e.g.,~the left-handed vector current operator results
\begin{align}
    {\mathcal C}^{V\,LL}_{ee,\,2111} \lesssim 2.3\times 10^{-11}~{\rm GeV}^{-1} \quad \Longrightarrow \quad \Lambda \gtrsim 2.1\times 10^5~{\rm GeV}\,,
\end{align}
where the limit on $\Lambda$ is obtained from Eq.~\eqref{eq:4lep} setting the quantity in parentheses equal to one.
The final sensitivity of Mu3e will improve the limit on the WC by two orders of magnitude (hence one order of magnitude for the scale $\Lambda$). 
Interestingly, in contrast to $\mu \to e\gamma$, $\mu\to 3 e$ can arise at the tree level in UV-complete BSM scenarios,\footnote{Prime examples are the doubly-charged scalar in Type~II seesaw, cf.~the bottom-mid diagram in Fig.~\ref{fig:NuMass}, and a CLFV $Z^\prime$ gauge boson as in~\cite{Smolkovic:2019jow}.} hence the above limits can be directly translated to a bound on the mass of the exchanged NP particles if their interactions with SM leptons are of $\ord{1}$.

As a summary of the above discussion on the interplay between dipole and contact interactions in CLFV muon decays, we display in Fig.~\ref{fig:mue-summary} current and future limits from $\mu\to e$ observables (including $\mu\to e$ conversion in nuclei, which is the subject of the next subsection) in terms of the SMEFT scale $\Lambda$ and the (dimensionless) dipole coefficient $\hat C_{e\gamma,\,12} $, see Eq.~\eqref{eq:dipole},
 while the WCs of the left-handed 4-fermion operators have been set equal to one, $\hat{C}_{\ell q,\,12 11}^{(1)} = \hat{C}_{\ell \ell,\,2111} =1$ (cf.~Table~\ref{tab:smeft} for the definition of the operators). The solid lines represent the current bounds reported in Table~\ref{tab:muon:limits}, the dashed lines the corresponding expected sensitivities of running/upcoming experiments.

Let us end this subsection by mentioning another purely leptonic observable: the conversion of a muonium $\rm M$ (a QED $\mu^+e^-$ bound state) into an anti-muonium $\overline{\rm M}$ ($\mu^-e^+$), a process that would imply the existence of CLFV interactions that change individual lepton numbers by two units, $|\Delta L_\mu|=|\Delta L_e|=2$. The current limit on the muonium--anti-muonium conversion probability, $P({\rm M}\to \overline{\rm M})< 8.2 \times 10^{-11}$~($90\%$~CL), was obtained by the MACS experiment at PSI about three decades ago~\citep{Willmann:1998gd}. A next-generation experiment, MACE, is being designed and developed in China with the aim of improving the current limit by two-three orders of magnitude~\citep{Bai:2024skk} --- see also \citet{Chen:2026tdg} for an overview of experimental muonium physics. The potential of these searches for testing NP has been discussed in~\cite{Conlin:2020veq,Fukuyama:2021iyw}. In particular,  ${\rm M}\to \overline{\rm M}$ would be induced by EFT operators such as $\mathcal{O}^{V\,LL}_{ee,\,2121} = (\bar \mu_L\gamma_\mu e_L)(\bar \mu_L\gamma^\mu e_L)$. The resulting constraints on the NP scale are relatively weak compared to the processes discussed above --- currently $\Lambda\gtrsim 5\,{\rm TeV}$~\citep{Conlin:2020veq}, implying a sensitivity of up to $\Lambda\approx 20-30\,{\rm TeV}$ at MACE. However, if NP preserves (at least approximately) certain flavor symmetries, $|\Delta L_\mu|=|\Delta L_e|=1$ effects could be suppressed, in which case $|\Delta L_\mu|=|\Delta L_e|=2$ operators would become the dominant source of CLFV and ${\rm M}\to \overline{\rm M}$  conversion the most sensitive probe~\citep{Heeck:2016xwg,Heeck:2024uiz,Calibbi:2025fzi}.

\subsection{$\mu\to e$ conversion}
\label{sec:mu:e:conv}

\begin{figure}[t]
    \centering
    \includegraphics[width=0.7\textwidth]{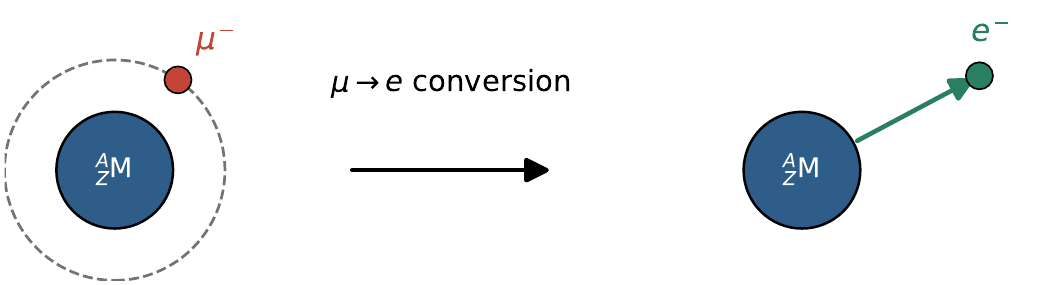}
    \caption{Schematic depiction of $\mu \to e$ conversion in nuclei.}
    \label{fig:MuToEDiagram}
\end{figure}

The process of $\mu \to e$ conversion is a hypothesized transition in which a $\mu^-$ bound to the nucleus (forming a muonic atom)
converts into an electron
without any accompanying neutrino emission~\citep{Czarnecki:1998iz, Kitano:2002mt}.
The produced electron is ultra-relativistic, and is ejected from the atom, carrying away almost all of the energy of the converted muon, while the atom absorbs the recoil momentum. 
A schematic picture is given in Fig.~\ref{fig:MuToEDiagram}.
If observed, it would be a clear signal of CLFV.
Experimentally, $\mu \to e$ conversion has a clean signature and well-understood backgrounds.
Theoretically, it is a complementary probe of CLFV
due to its sensitivity to interactions with quarks and gluons in the nucleus.
However, this also complicates the theoretical description of the process,
bridging many energy scales from the high-energy BSM physics down to the low-energy nuclear and atomic structure, cf. Fig. \ref{fig:EFTs:ladder}.
That is, in order to robustly constrain CLFV models using results of past and future searches for $\mu \to e$ conversion, a reliable theoretical description of $\mu \to e$ conversion is essential.  
In this section, we first briefly discuss experimental searches, 
and then focus on the unique set of challenges faced
when estimating $\mu \to e$ conversion rates.

The experiments searching for $\mu \to e$ conversion use a negative muon beam.
The $\mu^-$ are first directed on a target consisting of the isotope of interest, $^{A}_{Z}M$, with mass number $A$ and atomic number $Z$ (for both upcoming experiments, Mu2e and COMET, the target is $^{27}_{13}$Al).
A negatively charged muon stopped in the material  forms a muonic atom, which initially populates a high-lying excited state, and then 
relaxes to the ground state by emitting Auger electrons and $X$ rays.
Once in the $1s$ state, the muonic atom has the possibility to undergo
$\mu\to e$ conversion.
The event signature is the emission of a single mono-energetic electron
with energy
\begin{equation}
    E_{\mu e} = \frac{(m_{M} + m_\mu - B_\mu)^2 - m_{M}^2 + m_e^2}{2(m_{M} + m_\mu - B_\mu)}\,,
\end{equation}
where $m_{M}$ is  the mass of the nucleus,  $m_{\mu(e)}$ the muon (electron) mass, and $B_\mu$ the binding energy of the muon.

To definitively identify an event with an emitted electron of this energy as being due to $\mu \to e$ conversion,
we need to consider alternative outcomes for the bound muon,
some of which contribute as backgrounds to the measurement.
First, muon may be captured by the nucleus via SM weak interactions, emitting a neutrino,
\begin{equation}
     {}^A_ZM\,\mu^- \to {}^A_{Z-1}M' \,\nu_\mu\,.
\end{equation}
This process is referred to as the ordinary muon capture~\citep{Suzuki:1987jf, Measday:2001yr, King:2021jdb, Jokiniemi:2024zdl}.
Since an outgoing neutrino is not measured by the experiment, and there is no
electron with energy $E_{\mu e}$ created, 
ordinary muon capture does not contribute to the background.
Quite to the contrary, the ordinary muon capture rate is conventionally taken as the rate by
which the $\mu \to e$ conversion rate is normalized,
i.e., the normalized $\mu \to e$ conversion rate is defined as,
\begin{equation}
    \mathrm{BR}({}^A_ZM\,\mu^- \to {}^A_ZM\,e^-)\equiv \frac{\Gamma({}^A_ZM\,\mu^- \to {}^A_ZM\, e^-)}{\Gamma({}^A_ZM\,\mu^- \to {}^A_{Z-1}M'\, \nu_\mu)}\,.
\end{equation}
Second, the muon may be captured by nucleus, emitting a neutrino and a photon,
\begin{equation}
     {}^A_ZM\,\mu^- \to {}^A_{Z-1}M' \,\nu_\mu\gamma\,,
\end{equation}
a process known as the radiative muon capture~\citep{Measday:2001yr, Plestid:2020irv}.
The emitted $\gamma$ can subsequently generate an electron-positron pair;
depending on the relative energies of the neutrino and the photon,
electron's energy may equal $E_{\mu e}$, in which case this can be a source of a background for $\mu \to e $ searches. 
Third, the muon can decay while bound (i.e., ``in orbit''),
\begin{equation}
     {}^A_ZM\,\mu^- \to {}^A_{Z}M \,e^-\,\nu_\mu\,\bar{\nu}_e\,,
\end{equation}
a process known as the decay in orbit~\citep{Czarnecki:2011mx, Heeck:2021adh, Uesaka:2026tdd}.
Depending on the distribution of released energy among neutrino, antineutrino, and electron,
the electron may carry away energy $E_{\mu e}$ and be in a signal window for $\mu\to e$.
Additional sources of backgrounds are related to the muon beam and cosmic muons,
which can produce electrons that interact with the detector.
Considerable effort has been put into understanding all of these backgrounds
such that they are under control for next generation experiments~\citep{Mu2e:2014fns,COMET:2018auw}.
In particular, the next generation experiments are going to employ pulsed muon beams with a time separation much greater than the pion lifetime, so as to suppress the background from residual pions in the beam.

The current best limits for $\mu \to e$ conversion are given by the SINDRUM-II experiment~\citep{Wintz:1998rp, SINDRUMII:2006dvw}.
Specifically, the conversion rate in titanium~\citep{Wintz:1998rp} is limited to be
\begin{equation}
    \mathrm{BR}(\mathrm{Ti}\,\mu^- \to \mathrm{Ti}\,e^-)<6.1\times 10^{-13}~(95\%\,\text{CL})\,,
\end{equation}
and the conversion rate in gold~\citep{SINDRUMII:2006dvw} is bound to be below
\begin{equation}
    \mathrm{BR}(\mathrm{Au}\,\mu^- \to \mathrm{Au}\,e^-)<7\times 10^{-13}~(95\%\,\text{CL})\,.
\end{equation}
The upcoming experiments Mu2e~\citep{Mu2e:2012eea,Mu2e:2014fns} and COMET~\citep{Kuno:2013mha,COMET:2018auw}
are both searching for $\mu\to e$ conversion in aluminum.
They are expected to improve the above limits by up to four orders of magnitude,
making $\mu \to e$ conversion one of the strongest tests of CLFV, as illustrated in Fig.~\ref{fig:mue-summary}.
Constraints from searches for $\mu \to e$ conversion can be combined with limits on CLFV in other channels,
such as $\mu \to e\gamma$ or $\mu \to 3e$,
to more strongly constrain specific sources of flavor violation; see, e.g.,~\cite{Crivellin:2017rmk, Davidson:2020hkf, Davidson:2022nnl, Ardu:2023yyw, Ardu:2024bua}.

Crucially, any effort to constrain CLFV sources
relies on robust theory predictions with quantified uncertainties.
For $\mu\to e$ conversion, this requires many aspects to be considered:
\begin{enumerate}
    \item While the BSM physics that could generate CLFV is at energy scales above $1\,\TeV$,
    the relevant energy scales for $\mu \to e$ conversion are much lower.
    They are the mass of the muon $m_\mu \sim 100\,\MeV$,
    which gives the momentum of the electron;
    the ground-state energy of the nucleus $\sim8\,\MeV/\mathrm{nucleon}$;
    and the binding energy of the muon in the muonic atom $\lesssim1\,\MeV$.
    This requires a a series of effective field theory approaches in order
    to connect the BSM physics
    with the low-energy physics of the process itself~\citep{Kitano:2002mt,Cirigliano:2009bz,Petrov:2013vka,Crivellin:2013hpa,Crivellin:2014cta, Crivellin:2017rmk, Davidson:2018kud,Cirigliano:2022ekw,Hoferichter:2022mna,Rule:2021oxe,Haxton:2022piv,Noel:2024led,Noel:2024swe,Haxton:2024lyc,Borrel:2024ylg,Delzanno:2024ooj}.
    \item While the solution of the bound state of the muon is a rather routine task,
    accurately modeling the emitted electron requires
    accounting for long-range Coulomb interactions between the nucleus and the electron~\citep{Kitano:2002mt, Hoferichter:2022mna, Haxton:2022piv, Noel:2026byc}.
    \item Additionally, the structure of the nucleus itself contributes to the process~\citep{Chiang:1993xz, Siiskonen:2000nz}.
    The charge density distribution gives the potential of the nucleus,
    subtly affecting the wave functions of both the bound muon and the scattered electron~\citep{Heeck:2022wer, Noel:2024led}.
    Additionally, the $\mu\to e$ conversion process is sensitive to various nuclear structure functions, also often referred to as nuclear responses,
    notably the point-proton and point-neutron distributions~\citep{Kitano:2002mt}.
    While the point-proton density distribution may be to a good approximation
    obtained from the charge density distribution,
    which can be studied experimentally,
    the point-neutron distribution is much more difficult to measure.
    As a result, theoretical computations of the structure of the nucleus
    are required, in order to analyze properly the $\mu\to e$ conversion~\citep{Hoferichter:2022mna, Haxton:2022piv, Heinz:2024cwg}.
\end{enumerate}
While each of the above points merits a comprehensive discussion in their own right, 
we focus for brevity on the core ideas and illustrative special cases, while
providing references to more comprehensive discussion in the literature
for the interested reader.

Let us start by reviewing the calculation of  the matrix element for the $\mu\to e$ conversion process.
The incoming state $\ket{\mu(1s); M(J^\pi, M_i)}$ consists of a nucleus $M$ in a state with angular momentum and parity $J^\pi$ and angular momentum projection $M_i$
and of the muon in a bound $1s$ state.
In elastic $\mu\to e$ conversion,
the outgoing state $\bra{e; M(J^\pi, M_f)}$ consists of a nucleus in the same $J^\pi$ state, but now with angular momentum projection $M_f$,
and the emitted ultra-relativistic electron (for inelastic $\mu\to e$ conversion, where the final nucleus is in an excited state, see~\cite{Haxton:2024amf}).
Note that the wave function of the outgoing electron is distorted by the electric potential of the nucleus.
The matrix element for $\mu \to e$ conversion is given by,\footnote{Note that for this definition we absorb the delta function that enforces three-momentum conservation in the definition of matrix element $\mathcal{M}_{\mu \to e}$. This is a useful convention for $\mu\to e$ conversion since nucleus is so heavy that it can to a good approximation be taken to be at rest also after muon conversion, making momentum conservation essentially automatic. Inclusion of recoil corrections is also straightforward, as outlined in, e.g., \cite{Foldy_Recoils}.}
\begin{equation}
\label{eq:rel:M:mu:e}
    i \mathcal{M}_{\mu \to e} =  \int d^3x\, \braket{e; M(J^\pi, M_f) | \mathcal{L}^{\mu\to e}_\text{LEFT}(x) |\mu(1s);M(J^\pi, M_i)}\,,
\end{equation}
where in the LEFT Lagrangian, Eq.~\eqref{eq:left}, we kept only the operators relevant for $\mu\to e$ conversion. The degrees of freedom are the light quarks, $u$, $d$, $s$,
muons, electrons,
photons, and gluons.
Heavier degrees of freedom in the Standard Model and beyond have been integrated out,
replaced by effective short-range interactions.
We will work in the $m_e\to0$ limit, in which the left- and right-handed components of the electron spinors decouple, so that $\mathcal{M}_{\mu \to e}=\mathcal{M}_{\mu \to e,L}+\mathcal{M}_{\mu \to e,R}$ and the squared spin averaged amplitude given by $|\overline{\mathcal M}_{\mu \to e}|^2=|\overline{\mathcal M}_{\mu \to e,L}|^2+|\overline{\mathcal M}_{\mu \to e,R}|^2$, where $\mathcal{M}_{\mu \to e,L(R)}$ is the amplitude for outgoing electron with negative (positive) helicity. 

To illustrate how one calculates the nuclear matrix elements in Eq.~\eqref{eq:rel:M:mu:e}, we focus on dimension-6 scalar interaction as a representative case,
\begin{equation}
    \label{eq:ScalarLagrangianQQ}
  \mathcal{L}^{\mu\to e}_\text{LEFT} \supset  \mathcal{L}^{(6)}_S = \sum_{Y=L,R}\sum_{q=u,d,s} C^{S,q}_Y O^{S,q}_Y\,,
\end{equation}
where we shortened the notation and defined
\beq
\label{eq:ScalarOperatorQQ}
O^{S,q}_{L,R}\equiv (\bar e_{L,R} \mu_{R,L})(\bar q q).
\eeq
The complex LEFT WCs $C^{S,q}_{L,R}\propto 1/\Lambda^2$ are linear combinations of the WCs in the basis in Table \ref{tab:left}, and
are related to the higher-energy SMEFT WCs  via renormalization group.
An example of a mechanism that dominantly generates scalar CLFV operators is Higgs-mediated $\mu \to e$ conversion~\citep{Davidson:2010xv, Goudelis:2011un, Blankenburg:2012ex, Harnik:2012pb,McKeen:2012av, Crivellin:2014cta, Cirigliano:2022ekw}.

The operators in Eq.~\eqref{eq:ScalarOperatorQQ} factorize into a product of a leptonic current $(\bar e_{L,R} \mu_{R,L})$ 
and a quark current $(\bar q q)$.
For light nuclei, such as aluminum, that are of current interest for $\mu \to e$ conversion experiments, the typical Bohr radius of $1s$ muon function is much larger than the size of the nucleus so that
the leptonic and nuclear wavefunctions also factorize.
This means that we can treat the leptonic and nuclear parts of the LEFT operator in Eq.~\eqref{eq:ScalarOperatorQQ} separately.
The leptonic current is already in a form that makes it straightforward to evaluate its contribution to the matrix element in \eqref{eq:rel:M:mu:e},
once we have the muon and electron wave functions.
The quark current, on the other hand,  still requires some work;
while the LEFT Lagrangian consists of quark and gluon operators,
the nuclear wave functions are complicated bound state of nonrelativistic nucleons,
which themselves consist of quarks and gluons.

We therefore need to match LEFT onto a different effective theory, in which the degrees of freedom are not quarks and gluons but rather colorless bound states --- mesons and baryons. 
The interactions between the later
are described using chiral perturbation theory
(and more generally, for bound nucleons, using chiral effective field theory)~\citep{Bernard:1995dp, Epelbaum:2008ga, Machleidt:2011zz, Hammer:2019poc}. Chiral EFT organizes different contributions in terms of a momentum expansion, with single current interactions dominating.  
This makes the embedding of quark level operators, such as the one in Eq.~\eqref{eq:ScalarOperatorQQ}, into hadron level operators straightforward.
For instance, consider a nucleon with the incoming four-momentum $p$ and spin $s$
and outgoing momentum $p'$ and spin $s'$, so that the momentum transfer is $q=p - p'$.
The single nucleon hadronic matrix element of the $\bar{q} q$ current is given by
\begin{equation}
    \braket{N | \bar{q}q | N} = \bar{u}_N(p', s') \frac{m_N}{m_q} f^{N}_q(q) u_N(p, s)\,,
\end{equation}
where  $m_N$ is the nucleon mass,
 $m_q$ the quark mass,
and the scalar nucleon form factors $f^N_q(q)$ is a function of $q^2$. Since in $\mu\to e$ conversion the momentum transfer $q\sim {\mathcal O}(100\,\text{MeV})$ is much smaller than the typical energy at which chiral EFT breaks down, the form factors are typically expanded around $q^2=0$, 
with $f^{N}_q(0)$ the so called scalar nucleon couplings~\citep{Hoferichter:2015dsa, Durr:2015dna, Abdel-Rehim:2016won, Bali:2016lvx}, while $f^{N\prime}_q(0)$
define the scalar radii of the nucleon~\citep{Hoferichter:2016nvd}.

Here, for illustration purposes, we work in the $q \to 0$ limit, in which case the LEFT Lagrangian in \eqref{eq:ScalarLagrangianQQ} matches onto the following chiral EFT Lagrangian,
\begin{equation}
    \mathcal{L}^{\mathrm{nucleon}}_{S,N} = \sum_{Y=L,R} C^{S,S}_{Y,N} \big({\bar{e}_Y \mu}\big) \big(\bar N N\big),
    \label{eq:ScalarLagrangianNN}
\end{equation}
with the effective couplings $C^{S,S}_{Y,N}$ given by a product of hadronic form factors and quark-level WCs,
\begin{equation}
    C^{S,S}_{Y,N} = \sum_{q=u,d,s} C^{S,q}_Y \frac{m_N}{m_q} f^{N}_q(0)\,.
\end{equation}
If we had not restricted the discussion to dimension-6 scalar interaction in Eq.~\eqref{eq:ScalarLagrangianQQ},
but considered instead the full set of LEFT interactions up to and including dimension 7,
the dimension-7 operators containing the CP even gluon-gluon current, $(\bar e_Y \mu) GG$, would then also contribute to $C^{S,S}_{Y,N}$.
These contributions, though suppressed by an additional factor of $1/m_N$, can be important in more complete theories such as Higgs mediated CLFV interactions where quark currents are suppressed by small Yukawa interactions.

At this point, we require a nonrelativistic reduction of the nucleon fields, taking into account that $\mu \to e$ is a low energy process, i.e., compared to a typical momentum exchange nucleons are heavy and the total nuclear recoil is small. For this, note that acting with $\big(\bar N N\big)$ on incoming and outgoing nucleon wave functions is equivalent to acting with operator $\bar{u}_{N}' \mathds{1} u_N$, with $u_N = u_N(p,s)$ and $\bar{u}_N' = \bar{u}_N(p',s')$.
A nonrelativistic expansion of Dirac spinor $u_N(p,s)$  yields
\begin{equation}
    u_N(p,s)=\sqrt{\frac{E_N + m_N}{2 E_N}}\begingroup\setlength{\arraycolsep}{0.1em}\left(\hskip -\arraycolsep\begin{array}{c} \mathds{1} \chi_s \\ \frac{\vec{p} \cdot \vec{\sigma}}{E_N + m_N}\chi_s \end{array}\hskip -\arraycolsep\right)\endgroup = \begingroup\setlength{\arraycolsep}{0.1em}\left(\hskip -\arraycolsep\begin{array}{c} \Big(1 - \frac{\vec{p}^2}{8 m_N^2}\Big)\mathds{1} \chi_s \\ \Big(\frac{1}{2 m_N}\Big)(\vec{p} \cdot \vec{\sigma})\chi_s \end{array}\hskip -\arraycolsep\right)\endgroup + \mathcal{O}\left( \frac{\vec{p}^3}{m_N^3}\right)\,.
\end{equation}
Here, $E_N$ is the nucleon energy, $m_N$ the nucleon mass, $\vec{p}$
the 3-momentum, $\vec{\sigma}$ the 3-vector of Pauli spin matrices,
and  $\chi_s$ the two-component Pauli spinor, corresponding to spin $s$.
With some work, one can expand $\bar{u}_N' \mathds{1} u_N$ in powers of $m_N^{-1}$,
yielding several different spin and momentum structures.
This procedure is general and can be applied to other nucleon currents that would arise from different operators~\citep{Fitzpatrick:2012ix}.
The simplest term we get from $\bar{u}_N' \mathds{1} u_N$ is $\chi_{s'}^\dagger \mathds{1} \chi_s$
with a kinematic prefactor of $1 - \vec{q}^2/(8m_N^2)$ that can be absorbed via field redefinitions. 
Replacing this with nonrelativistic nucleon operators gives us the nonrelativistic Lagrangian for Eq.~\eqref{eq:ScalarLagrangianNN} (and for clarity neglecting the $\mathcal{O}(m_N^{-2})$ term in the kinematic prefactor and other more complex momentum structures):
\begin{equation}
    \mathcal{L}^{\text{non-rel.}}_{S,N} = \sum_{Y=L,R}  \mathcal{L}^{\text{non-rel.}}_{Y,N}=  \sum_{Y=L,R} C^{S,S}_{Y,N} \big({\bar{e}_Y \mu}\big)\big( N^{+} N^{-}\big).
    \label{eq:ScalarLagrangianNNNonRel}
\end{equation}
The non-relativistic nucleon operators are defined such that acting on non-relativistic nucleon wave functions one has,
\beq
    N^-(x)\ket{N(p, s)} = \chi_s e^{i\vec{p}\cdot \vec{x}}, \qquad\qquad \bra{N(p',s')}N^+(x) = \chi_{s'}^\dagger e^{-i \vec{p}\cdot \vec{x}}\,.
\eeq
The $N^\pm$ operators act on all the nucleons bound inside a nucleus, so that in the hadronic matrix element one obtains a sum of contributions from single nucleon currents. Formally, we can make this explicit already at the Lagrangian level, by introducing auxiliarly labels for each of the nucleons so that for a nucleus $M$ with $A$ nucleons, $Z$ protons,
 the effective Lagrangian is given by
\begin{equation}
    \mathcal{L}^{\mathrm{nuclear}}_{S} = \sum_{i=1}^{A}\mathcal{L}^{\text{non-rel.}}_{S,N_i}, \qquad   \mathcal{L}^{\mathrm{nuclear}}_{Y} = \sum_{i=1}^{A}\mathcal{L}^{\text{non-rel.}}_{Y,N_i},
\end{equation}
where the sum runs over all nucleons, $N_i$, that can be either a proton $p$ or a neutron $n$.

With the nuclear-level Lagrangian $\mathcal{L}^{\mathrm{nuclear}}_{S}$ in hand, we can return to the calculation of nuclear matrix elements. 
As anticipated above, the leptonic and nuclear wave functions factorize, so that the matrix element separates into leptonic and nuclear parts, giving the matrix element
\beq
\begin{split}
    i \mathcal{M}_{\mu\to e, Y} &= \int d^3x\, \braket{e; M(J^\pi, M_f) | \mathcal{L}^{\mathrm{nuclear}}_Y(x) |\mu(1s);M(J^\pi, M_i)}\,, \\
    &=\int d^3x\, \sum_{i=1}^A \braket{e; M(J^\pi, M_f) | \mathcal{L}^{\text{non-rel.}}_{Y,N_i}(x) |\mu(1s);M(J^\pi, M_i)}\,,  \\
    &=\int d^3x\, \sum_{i=1}^A  C^{S,S}_{Y,N_i} \braket{e | \bar{e}_Y \mu |\mu(1s)} \braket{ M(J^\pi, M_f) |  N_i^{+} N_i^{-} |M(J^\pi, M_i)}\,,
    \label{eq:MuToEMatrixElement1}
\end{split}
\eeq
where the electron wave function is taken to have the correct helicity such that this matrix element is nonzero. 

The leptonic and nuclear parts can now be treated independently. We begin with the leptonic current,
\begin{equation}
    \braket{e | \bar{e}_Y \mu |\mu(1s)} = \overline{\psi_{e_Y}^{\kappa}}(x)\psi_\mu^{(1s)}(x)\equiv l_Y(x)\,,
\end{equation}
where we define the leptonic current $l_Y(x)$.
$\psi_{e_Y}^{\kappa}(x)$ is the wave function of the electron in the partial wave defined by $\kappa$,
\beq
    j = |\kappa| - \frac{1}{2}, \qquad\qquad l = \begin{cases}
        \kappa & \kappa > 0, \\
        - \kappa - 1 &\kappa < 0,
    \end{cases}
\eeq
and $\psi_\mu^{(1s)}(x)$ is the wave function of the bound muon in the $1s$ partial wave
(with $\kappa = -1$).
These wave functions are obtained by solving the Dirac equation for the electron and muon in the potential of the nucleus $V(r)$, computed from the nuclear charge distribution,
\begin{equation}
    V(r)=-8\alpha \int_0^\infty dx\, x^2 \rho_\text{ch}(x) \frac{\pi}{2\max(r, x)},
\end{equation}
with the fine-structure constant $\alpha =e^2/(4\pi)$.
The finite distribution of charge in the nucleus complicates the solution of the Dirac equation.
In particular, the so-called Coulomb distortions need to be taken into account
to capture the modification of the wave function from the solution for a pure Coulomb potential.
For the bound muon, the lowest energy levels can be noticeably shifted
and for the scattered electron the asymptotic plane wave solutions are shifted by a phase
according to interactions near the nucleus.
Both can in practice be solved exactly~\citep{Kitano:2002mt, Heinz:2024cwg}, though a reasonable approximation
for Coulomb distortions in a plane-wave formalism can be obtained
by modifying the electron momentum near the nucleus to some $q_\mathrm{eff}$~\citep{Haxton:2022piv}. Note that at the current level of precision it suffices for nuclear charge distribution to be treated classically (in particular any back-reaction from muon wave function can be neglected). 

To evaluate the hadronic part of the matrix element in Eq.~\eqref{eq:MuToEMatrixElement1},
we first Fourier transform the leptonic current to momentum space,
\begin{equation}
    l_Y(x) = \int \frac{d^3 q}{(2\pi)^3} \tilde{l}_Y(\vec{q}) e^{-i \vec{q}\cdot \vec{x}}\,, 
\end{equation}
defining the Fourier-transformed current $\tilde{l}_Y(\vec{q})$.
The  matrix element is then given by
\beq
\begin{split}
    i \mathcal{M}_{\mu\to e,Y} &=  \int d^3x\, \sum_{i=1}^A C^{S,S}_{Y,N_i} l_Y(x) \braket{ M(J^\pi, M_f) |  N_i^{+} N_i^{-} |M(J^\pi, M_i)} \\
    &= \int d^3x\, \sum_{i=1}^A  C^{S,S}_{Y,N_i} \int \frac{d^3 q}{(2\pi)^3} \tilde{l}_Y(\vec{q}) e^{-i \vec{q}\cdot \vec{x}} \braket{ M(J^\pi, M_f) |  N_i^{+} N_i^{-} |M(J^\pi, M_i)} \\
    &= \int \frac{d^3 q}{(2\pi)^3} \tilde{l}_Y(\vec{q}) \int d^3x\, \sum_{i=1}^A  C^{S,S}_{Y,N_i} \braket{ M(J^\pi, M_f) |  e^{-i \vec{q}\cdot \vec{x}}  N_i^{+} N_i^{-} |M(J^\pi, M_i)}\,.
    \label{eq:MuToEMatrixElement2}
\end{split}
\eeq
In the final line we shuffled things around to focus on the nuclear structure function on the right:
\begin{equation}
    \int d^3x\, \sum_{i=1}^A  C^{S,S}_{Y,N_i} \braket{ M(J^\pi, M_f) |  e^{-i \vec{q}\cdot \vec{x}}  N_i^{+} N_i^{-} |M(J^\pi, M_i)} = \int d^3x\,e^{-i \vec{q}\cdot \vec{x}}\left( C^{S,S}_{Y,p} \rho_p(x) + C^{S,S}_{Y,n} \rho_n(x)\right)\,.
    \label{eq:NuclearStructureFunction}
\end{equation}
In this case, the nuclear matrix element simplifies to the point-proton and point-neutron densities,
$\rho_p(x)$ and $\rho_n(x)$,
which when Fourier-transformed give the point-proton and point-neutron nuclear structure functions,
$F_p(q)$ and $F_n(q)$.
These characterize the leading response of the nucleus to the scalar interactions, similar to the leading response to electromagnetic current, but now also neutrons carrying a nonzero scalar charge encoded in  the coefficient $C^{S,S}_{Y,n}$.
For electromagnetic interactions, a general framework to decompose the complete response of the nucleus in electron scattering exists,
requiring a multipole decomposition~\citep{Donnelly:1975ze, Serot:1978vj, Donnelly:1979ezn, Fitzpatrick:2012ix}.
This separates space-like and time-lime components
and decomposes the angular dependence of the operators,
with $L=0$ being the spherical component.
In general a complete description of lepton scattering off of a nucleus with a ground state spin $J$
requires the consideration of multipoles with $L\leq 2J$.

The nuclear structure function in Eq.~\eqref{eq:NuclearStructureFunction} is somewhat special.
It is commonly denoted as the $M_L$ structure function~\citep{Fitzpatrick:2012ix}.
Restricting to $L=0$
this is simply the point-proton and point-neutron scattering form factors, $F^{M_0}_p(q)$ and $F^{M_0}_n(q)$,
which then gives (in practice, we still have to average over initial spins
and sum over final spins)
\begin{align}
    i \mathcal{M}_Y &=\int \frac{d^3 q}{(2\pi)^3} \tilde{l}_Y(\vec{q}) \Big( C^{S,S}_{Y,p} F^{M_0}_p(q) + C^{S,S}_{Y,n} F^{M_0}_n(q)\Big).
    \label{eq:MuToEMatrixElement3}
\end{align}
Because the two form factors are spin independent, they are coherently enhanced,
with $F^{M_0}_p(q=0)=Z$ and $F^{M_0}_n(q=0)=A-Z$.
For this reason, in many cases the leading order description
of $\mu\to e$ conversion requires only these two nuclear structure functions.
In general, however, a number of other nuclear structure functions also contribute.
Depending on the nature of CLFV interactions, the operators leading to
spin-independent responses can have suppressed WCs, and thus the remaining nuclear response functions  could be important~\citep{Cirigliano:2017azj, Davidson:2017nrp},
motivating a complete treatment of all nuclear responses in $\mu\to e$ conversion~\citep{Hoferichter:2022mna, Haxton:2024lyc}.

If $\mu\to e $ conversion is observed, an immediate question will be how much one can learn about the UV physics. For this so called inverse problem, reliable estimates of uncertainties on theoretical predictions, especially the nonperturbative inputs such as hadronic matrix elements, and nuclear structure functions, are essential.
An important quantity that can help reduce uncertainties 
is the charge density.
Charge densities of nuclei are typically studied experimentally via electron scattering and muonic atom spectroscopy.
However, the standard compilation of model-independent parameterizations of charge densities in~\cite{DeVries:1987atn} does not provide uncertainty estimates for the truncation of the Fourier--Bessel expansion of the charge density.
This has motivated the reanalysis of electron scattering data,
along with the inclusion of constraints from muonic atoms,
to provide precise and, crucially, uncertainty-quantified charge densities for key isotopes of interest to experimental efforts~\citep{Heeck:2022wer, Noel:2024led}.
Precisely understanding uncertainties in the charge density
impacts both the calculation of the nuclear response to CLFV operators involving scalar and vector quark currents
and indirectly constrains the nuclear structure functions
through correlations between the charge density and other nuclear structure functions.

The nuclear structure functions must in general be predicted using nuclear theory calculations.
A seminal work by \cite{Kitano:2002mt} used experimental data on the charge density
and some model assumptions about the relationship between the charge density, the point-proton density, and the point-neutron density to model $\mu \to e$ conversion.
Another approach is to compute the nuclear structure functions using the nuclear shell model~\citep{Cirigliano:2022ekw, Rule:2021oxe, Hoferichter:2022mna}.
The nuclear shell model diagonalizes an effective shell-model Hamiltonian, which is empirical and adjusted to data~\citep{Caurier:2004gf}.
This approach has two challenges when it comes to uncertainty quantification.
First, the shell model makes a basic model assumption
that the essential nuclear structure of the ground state
can be modeled within the small valence space employed.
This assumption is in many cases well motivated,
but there is no way to test this as shell-model Hamiltonians are not systematically improvable.
Second, we require nuclear structure functions for both protons and neutrons,
and for neutrons experimental data (especially on the weak/neutron distribution) is extremely scarce.
So shell model Hamiltonians that may be very accurate for proton nuclear structure functions
rely on significant extrapolations to provide predictions for neutron nuclear structure functions.
Again, the uncertainty here is difficult to quantify.

Ab~initio calculations of nuclei offer a different approach,
one that is more consistent with the tower of effective field theories used here.
Such calculations start from nuclear Hamiltonians from chiral effective field theory~\citep{Epelbaum:2008ga, Machleidt:2011zz, Hammer:2019poc},
a systematically improvable expansion of quantum chromodynamics at low energies.
They then solve the many-body Schrödinger equation using either exact~\citep{Barrett:2013nh, Carlson:2014vla} or systematically improvable~\citep{Hagen:2013nca, Hergert:2015awm, Stroberg:2019mxo, Soma:2020xhv, Tichai:2020dna, Lovato:2022tjh, Lee:2025req} many-body methods.
Because both the Hamiltonians and the many-body method are systematically improvable,
all uncertainties can in principle be quantified.
Such calculations are growing more and more essential
in the interpretation of precision experiments~\citep{Miyagi:2025rvx, Noel:2026byc},
including providing constraints for BSM physics~\citep{Payne:2019wvy, Novario:2020dmr, King:2022zkz, Belley:2023lec, Longfellow:2024svx, Door:2024qqz, Heinz:2024cwg, Ng:2025hgx, Sargsyan:2026ygi, He:2026zie, Belley:2026bde, Shickele:2026fbg, Li:2026aeq}.
As far as $\mu \to e$ conversion is concerned,
the absolute uncertainties of ab~initio calculations are sizable.
However, the uncertainties are correlated in systematic ways,
e.g., due to the underlying symmetries of nuclear Hamiltonians,
the optimization to nucleon-nucleon scattering data,
the constraints from quantum many-body physics.
In particular, systematic correlations arise 
between the charge density and the proton and neutron nuclear structure functions~\citep{Heinz:2024cwg}.
This allows experimental data on the charge density to provide constraints for the more uncertain neutron nuclear structure functions.
Additionally, it allows for covariances between different nuclear structure functions to be quantified,
which is essential for any analysis seeking to disentangle multiple sources of CLFV.

The last uncertainty we will discuss comes from higher orders in the chiral EFT expansion.
At higher orders,
there are not only simple one-nucleon currents
but also two-nucleon currents,
where the $\mu\to e$ lepton current couples to a correlated pair of nucleons.
Such two-body currents arise naturally in many contexts in nuclear EFT~\citep{Cirigliano:2012pq, Klos:2013rwa, Hoferichter:2016nvd,Korber:2017ery, Gysbers:2019uyb, Hoferichter:2020osn, Miyagi:2023zvv, Chambers-Wall:2024fha, Brase:2025rus, Acharya:2023ird, Li:2025exk},
and their importance depends on the specifics of the interaction.
For scalar interactions that we discussed above,
two-body currents appear already at next-to-leading order~\citep{Bartolotta:2017mff}
and affect the amplitude for $\mu\to e$ conversion at the 10\,\% level~\citep{Cirigliano:2022ekw}.
In such cases, neglecting two-body currents is a significant uncertainty, and precision studies should include them.

\subsection{Muon transitions and light new physics}
\label{sec:mu:light:np}

So far we assumed that all new degrees of freedom are heavy, $m_X\gg m_\mu$, and can be integrated out. If the spectrum instead contains a light, feebly interacting particle $X$ with mass below $m_X<m_\mu-m_e$, the muon can decay into it. This opens up the possibility of qualitatively new exotic muon decay channels, such as $\mu \to e+\,\text{invisible}$ (with the electron being monochromatic), $\mu\to 5e$, $\mu\to 7e$, etc.
Such light states are ubiquitous in well-motivated extensions of the SM, can have flavor violating couplings (see, e.g.,~\citet{MartinCamalich:2025srw,Echenard:2026cdv,Ziegler:2026kis}, for recent reviews). Searches for light new physics particles are part of a broader ``dark sector'' search program, motivated by the existence of dark matter. Most searches focus on flavor diagonal couplings of dark sector to the SM~\citep{Agrawal:2021dbo,Antel:2023hkf,Tammaro:2025zso}. However, qualitatively new features appear if these also have flavor violating interactions.

\begin{table*}[t]
\centering
\caption{Representative SMEFT (2nd column) and $\text{SMEFT}_X$ (4th column) operators for each exotic muon-decay
signature, together with the new-physics scale $\Lambda$ probed for projected bounds in Table~\ref{tab:muon:limits} and assuming $\mathrm{BR}(\mu\to e+X)<10^{-15}$ for all others. For $\text{SMEFT}_X$ operators the required decay mode of $\mathcal{S}$ is included. 
The golden mode $\mu\to e\gamma$
has no on-shell single-mediator $\text{SMEFT}_X$ realization and is left blank. 
Notation as in Table~\ref{tab:smeft}. Adapted from \citet{Greljo:2025ljr}.
\label{tab:smeft_vs_smeftX}}
\renewcommand{\arraystretch}{2}
\setlength{\tabcolsep}{8pt}
\begin{tabular}{c  c c  c c}
\hline
\textbf{Signature} &
$\mathcal{L}_{\text{SMEFT}}\supset$ & $\Lambda\,[\text{GeV}]$ &
$\mathcal{L}_{\text{SMEFT}_X}\supset$ & $\Lambda\,[\text{GeV}]$ \\ \hline

$\mu\to e\gamma$ &
$\dfrac{y_\mu}{16\pi^2}\dfrac{1}{\Lambda^2}(\bar L_2\sigma^{\mu\nu} E_1)H B_{\mu\nu}$ &
$3\times 10^{5}$ &
--- & --- \\

$\mu\to 3e$ &
$\dfrac{1}{\Lambda^2}(\bar L_2\gamma^\mu L_1)(\bar L_1\gamma_\mu L_1)$ &
$2\times10^{6}$ &
$\dfrac{y_\mu}{\Lambda}(\bar L_2 H  E_1)\mathcal{S}\;\;[\mathcal{S}\to e^+e^-]$ &
$\mathcal{O}(10^{14})$ \\

$\mu\to 5e$ &
$\dfrac{y_\mu}{\Lambda^6}(\bar L_1\gamma^\mu L_1)^2(\bar L_2 H  E_1)$ &
$5$ &
$\dfrac{y_\mu}{\Lambda^2}(\bar L_2 H  E_1)\mathcal{S}^2\;\;[\mathcal{S}\to e^+e^-]$ &
$\mathcal{O}(10^{6})$ \\

$\mu\to 3e\gamma$ &
$\dfrac{1}{16\pi^2}\dfrac{1}{\Lambda^4}(\bar L_2\gamma^\mu L_1)(\bar L_1\gamma^\nu L_1)B_{\mu\nu}$ &
$30$ &
$\dfrac{1}{16\pi^2}\dfrac{y_\mu}{\Lambda^3}(\bar L_2\sigma^{\mu\nu} E_1)H B_{\mu\nu}\mathcal{S}\;\;[\mathcal{S}\to e^+e^-]$ &
$\mathcal{O}(10^{3})$ \\

$\mu\to e2\gamma$ &
$\dfrac{y_\mu}{16\pi^2}\dfrac{1}{\Lambda^4}(\bar L_2 H  E_1)B_{\mu\nu}B^{\mu\nu}$ &
$80$ &
$\dfrac{y_\mu}{\Lambda}(\bar L_2 H  E_1)\mathcal{S}\;\;[\mathcal{S}\to\gamma\gamma]$ &
$\mathcal{O}(10^{14})$ \\

$\mu\to e3\gamma$ &
$\dfrac{1}{16\pi^2}\dfrac{1}{\Lambda^4}(\bar L_2\gamma^\mu\overleftrightarrow{D}^\nu L_1)B_{\mu\rho}B^{\rho}{}_{\nu}$ &
$20$ &
$\dfrac{1}{16\pi^2}\dfrac{y_\mu}{\Lambda^3}(\bar L_2\sigma^{\mu\nu} E_1)H B_{\mu\nu}\mathcal{S}\;\;[\mathcal{S}\to\gamma\gamma]$ &
$\mathcal{O}(10^{3})$ \\

$\mu\to 3e2\gamma$ &
$\dfrac{1}{\Lambda^4}(\bar L_2\gamma^\mu\overleftrightarrow{D}^\nu L_1)(\bar L_1\gamma_\mu\overleftrightarrow{D}_\nu L_1)$ &
$40$ &
$\dfrac{y_\mu}{\Lambda^2}(\bar L_2 H  E_1)\mathcal{S}^2\;\;[\mathcal{S}\to e^+e^-,\gamma\gamma]$ &
$\mathcal{O}(10^{6})$ \\ \hline
\end{tabular}
\end{table*}

\begin{table}[t]
\renewcommand{\arraystretch}{1.3}
\TBL{\caption{The landscape of exotic muon decays $\mu\to(m\,e)(n\,\gamma)$, organized by the number of electrons and positrons $m$ (rows) and photons $n$ (columns) in the final state. The three golden transitions, $\mu\to e$ ($\mu\to e$ conversion), $\mu\to e\gamma$, and $\mu\to 3e$, occupy the upper-left corner. Each entry may in addition carry missing energy, and the tracks may be prompt or displaced. Adapted from \citet{Greljo:2025ljr}.}\label{tab:exotic:grid}}
{\begin{tabular*}{\textwidth}{@{\extracolsep{\fill}}@{}l|ccccc@{}}
\toprule
$m\,\backslash\,n$ & $0$ & $1$ & $2$ & $3$ & $4$\\
\colrule
$1$ & $e$   & $e\gamma$   & $e2\gamma$  & $e3\gamma$  & $e4\gamma$\\
$3$ & $3e$  & $3e\gamma$  & $3e2\gamma$ & $3e3\gamma$ & $\dots$\\
$5$ & $5e$  & $5e\gamma$  & $5e2\gamma$ & $\dots$     & \\
$7$ & $7e$  & $7e\gamma$  & $\dots$     &             & \\
$9$ & $9e$  & $\dots$     &             &             & \\
\botrule
\end{tabular*}}{}
\end{table}

The CLFV interactions involving light NP states are due to higher dimensional operators, described by $\text{SMEFT}_X$ --- an effective field theory obtained by extending SMEFT with an additional light field, $X$, a singlet of the SM gauge group (one could also entertain several light new fields). Several examples of such operators are listed in Table~\ref{tab:smeft_vs_smeftX}, with $X$ taken to be a light scalar ${\cal S}$.  Since in $\text{SMEFT}_X$ the muon decays proceed through operators of lower dimension than in SMEFT, the exotic decays now probe much higher effective scales~\citep{Greljo:2025ljr,Hostert:2023gpk}, as can be seen from Table~\ref{tab:smeft_vs_smeftX}. Furthermore, without any additional light new states, an observable signal in high-multiplicity muon decays, such as $\mu\to 5e, \mu\to 3e\gamma,...$ is possible only, if the suppression scale $\Lambda$ in SMEFT is below the electroweak scale. Such an EFT is not internally consistent, which means that an observable high-multiplicity FCNC muon decay signal requires the existence of light new physics states. 

The possible final states are conveniently organized in terms of a small set of building blocks added to the leading $\mu\to e$ transition: additional $e^+e^-$ pairs, additional photons, or missing energy, with the visible tracks either prompt or displaced~\citep{Greljo:2025ljr}. The resulting landscape of exotic modes, $\mu\to (m\,e)(n\,\gamma)\,(+\,\mathrm{inv})$, is summarized in Table~\ref{tab:exotic:grid}. The three golden channels discussed in the previous subsections occupy its upper-left corner, while the higher-multiplicity entries: $\mu\to e\gamma\gamma$, $\mu\to 5e$, $\mu\to 7e$, and others, remain largely unexplored, both experimentally and theoretically. The exception is $\mu\to e 2\gamma$, which was searched for by Crystal Box in 1988 \citep{Bolton:1988af}, and also, more recently, by MEG, however, in this case assuming a particular decay chain, $\mu\to e a$ with an axion-like particle (ALP) $a$ having a mass in the range 20-45 MeV and decaying via $a\to \gamma\gamma$ \citep{MEG:2020zxk}. The high-multiplicity decays without missing energy are also subject to the bound from a more than 5 decade old inclusive search for exotic decays, $\text{BR}(\mu\to (m e) (n\gamma))< 4 \times 10^{-6}$~\citep{Poutissou:1974ic}. Using dedicated searches at Mu3e one should be able to place significantly more stringent bounds, at the level of $10^{-13}$ or better. In the SM these types of decays arise only from radiative corrections to the Michel decay, $\mu\to e\nu\bar\nu$ followed by photon emissions, with some photons Dalitz converted to $e^+e^-$ pairs, and always include missing energy due to neutrinos in the final state. The rates also fall steeply with the multiplicity, e.g.~$\mathrm{BR}(\mu\to 3e2\nu)\sim10^{-5}$ and $\mathrm{BR}(\mu\to 5e2\nu)\sim10^{-10}$ \citep{Hostert:2023gpk}; precise predictions of these irreducible backgrounds are available \citep{Fael:2016yle,Pruna:2016spf,Banerjee:2020rww,Colangelo:2021xix,Banerjee:2022nbr}. A NP signal can therefore be sought in an essentially background-free environment, especially for higher multiplicities and when there is no missing energy mimicking the SM neutrinos.

\begin{figure}[t]
\centering
\includegraphics[width=0.6\textwidth]{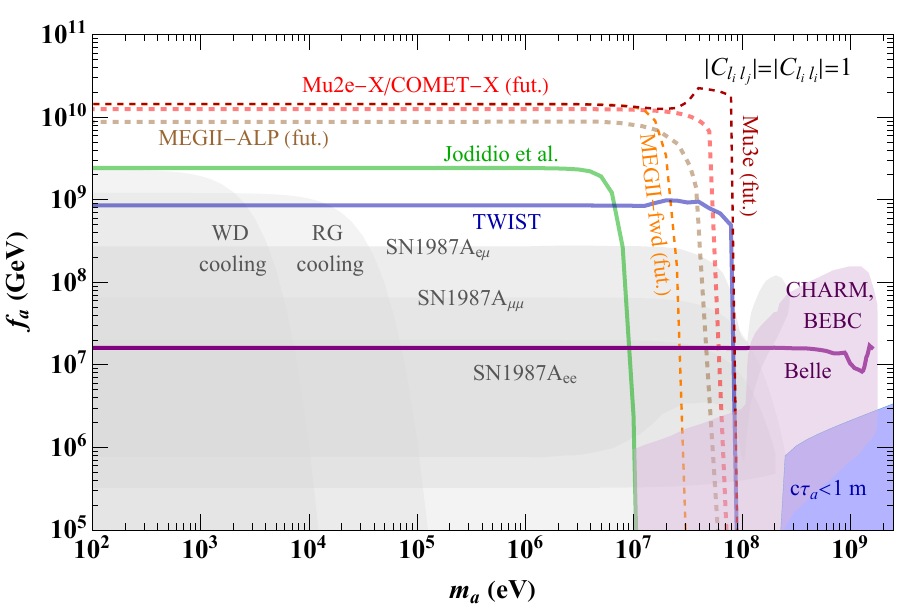}% 
\caption{Reach of searches for the two-body decay $\mu\to e\,a$ (with $a$ invisible/long-lived) and its radiative counterpart $\mu\to e\,a\,\gamma$ (MEGII-ALP) in terms of CLFV ALPs with anarchic couplings to all three generations, displayed in the plane of the ALP mass $m_a$ and the symmetry-breaking scale $f_a$ (or coupling $1/f_a$), compared with astrophysical bounds (grey shaded region).
Present exclusions are denoted with solid lines, assuming isotropic ALP production with axial couplings (that is, $C^A_{ij}=1$, $C^V_{ij}=0$), while dashed lines correspond to projections of proposed searches. The purple line shows the exclusion from analogous searches for $\tau \to \ell\,a$ at Belle \citep{Belle:2025bpu},
while  beam dump constraints are shown as purple shaded region \citep{Ema:2025bww}, see Sec.~\ref{sec:tau:light:np}.
 The figure has been adapted and updated from~\citet{Calibbi:2020jvd,Hill:2023dym} and includes in addition projections from~\cite{Jho:2022snj}, SN constraints from \cite{Li:2025beu}, updated Belle constraints and beam dump bounds. 
}
\label{fig:alp:reach}
\end{figure}

The simplest light-NP channel is the two-body decay $\mu\to eX$ with $X$ escaping the detector. The experimental signature is a monochromatic positron at $E_e\simeq m_\mu/2$ (for massless $X$, appropriately shifted if $X$ is massive) sitting on the Michel spectrum. The prototypical $X$ is a pseudo-Nambu-Goldstone boson (pNGB), or axion-like particle (ALP), $a$, of a spontaneously broken approximate global symmetry (e.g.,~a flavor symmetry~\citep{Ema:2016ops,Calibbi:2016hwq}), coupling derivatively through the dimension-5  interaction~\citep{Feng:1997tn,Cornella:2019uxs,MartinCamalich:2020dfe,Calibbi:2020jvd,Bauer:2020jbp,Bauer:2021mvw,Galda:2021hbr,Greljo:2024evt,Alda:2025uwo}
\beq
\label{eq:ALP:muon}
{\cal L}_\text{ALP}\supset \frac{\partial_\mu a}{2 f_a}\,\bar e\,\gamma^\mu(C^V_{e\mu}+C^A_{e\mu}\gamma_5)\,\mu\,,
\eeq
where $f_a$ is a NP scale (often referred to as ALP decay constant) related to the breaking of the underlying symmetry.
Concrete realizations of this ALP Lagrangian include a CLFV QCD axion, an axiflavon, a leptonic familon, and a majoron \citep{Ema:2016ops,Calibbi:2016hwq,Calibbi:2020jvd,Linster:2018avp,Bonnefoy:2019lsn}. As a pNGB, $a$ is protected by a shift symmetry and can be naturally light, in contrast to a generic light scalar, which suffers from a hierarchy problem \citep{Greljo:2025ljr}.  Because $\mathrm{BR}(\mu\to ea)\propto(m_W^2/m_\mu f_a)^2$, even a weak bound on the BR translates into an enormous reach on the symmetry-breaking scale: the current limits on $\mathrm{BR}(\mu^+\to e^+X)$ at the level of few$\times10^{-6}$, depending on ALP mass and chirality of couplings \citep{Jodidio:1986mz,TWIST:2014ovy} (see~\cite{Calibbi:2020jvd} for a systematic discussion), already imply $f_a\gtrsim10^{9}\,$GeV, which could be improved by another order of magnitude at future facilities. 
Dedicated projections for $\mu\to ea$ searches (and the associated radiative mode $\mu\to ea\gamma$) at Mu3e, MEG~II, Mu2e/COMET, including the effects of muon polarization, have been worked out in \citet{Perrevoort:2018ttp,Calibbi:2020jvd,Jho:2022snj,Xing:2022rob,Knapen:2023zgi,Hill:2023dym,Knapen:2024fvh,Bigaran:2025uzn}, with a reach complementary to astrophysical and beam-dump constraints. The reach of some of these proposals in the $m_a-f_a$ plane is illustrated in Fig.~\ref{fig:alp:reach}. 
Note that the outstanding sensitivity of searches for $\mu\to e a$ to high-energy scales is due to the fact that the operators in Eq.~\eqref{eq:ALP:muon} are of dimension-five, hence the decay rate is inversely proportional to $f_a^2$. This is to be compared with the processes discussed in the previous subsections, which are mediated by dimension-six operators so that those rates are suppressed by the fourth power of the NP scale.

In addition, virtual ALP exchange can mediate ordinary $\mu\to e$ transitions~\citep{Cornella:2019uxs,Bauer:2021mvw,Fuyuto:2023ogq}, although typically leading to sub-leading constraints in the parameter space where $\mu\to ea$ is kinematically allowed. Finally, the additional requirement that $\mu\to ea$ be the very process responsible for production of ALP dark matter in the early universe was investigated in~\cite{Panci:2022wlc}. 

If a light NP particle $X$ decays back into visible states within the detector, the muon decay cascades into exotic high-multiplicity final states, with the pattern dictated by the spin and quantum numbers of $X$. A classification of various $\text{SMEFT}_X$ operators and induced high-multiplicity decays for a single light scalar $\mathcal S$, pseudoscalar $\mathcal P$, fermion $N$, or vector $V$, was performed in \cite{Greljo:2025ljr}. A light scalar or pseudoscalar emitted in $\mu\to e+n\,X$ and decaying via $X\to e^+e^-$ or $X\to\gamma\gamma$ yields $\mu\to(2n{+}1)e$ or $\mu\to e\,n\gamma$; a light vector decaying to $e^+e^-$ gives $\mu\to eV\to3e$ or $\mu\to eVV\to5e$; a pair of light fermions gives $\mu\to eNN$ with missing energy, or additional visible tracks if $N$ itself decays. Multi-electron muon decays of exactly this type were first analyzed in detail in \citet{Hostert:2023gpk}, while the emission of light states from the external legs of the ordinary Michel decay, $\mu\to n\,X+e\nu\bar\nu$, was studied in \citet{Knapen:2023iwg,Knapen:2024fvh}; earlier proposals for visibly decaying light particles in muon decays appear in \citet{Echenard:2014lma,Heeck:2017xmg}. Because the SM background falls so rapidly with multiplicity, these channels can probe effective scales ranging from $\mathcal O(1)\,$TeV up to $\sim10^{7}\,$GeV depending on the operator.

\begin{figure}[t]
\centering
\includegraphics[width=0.6\textwidth]{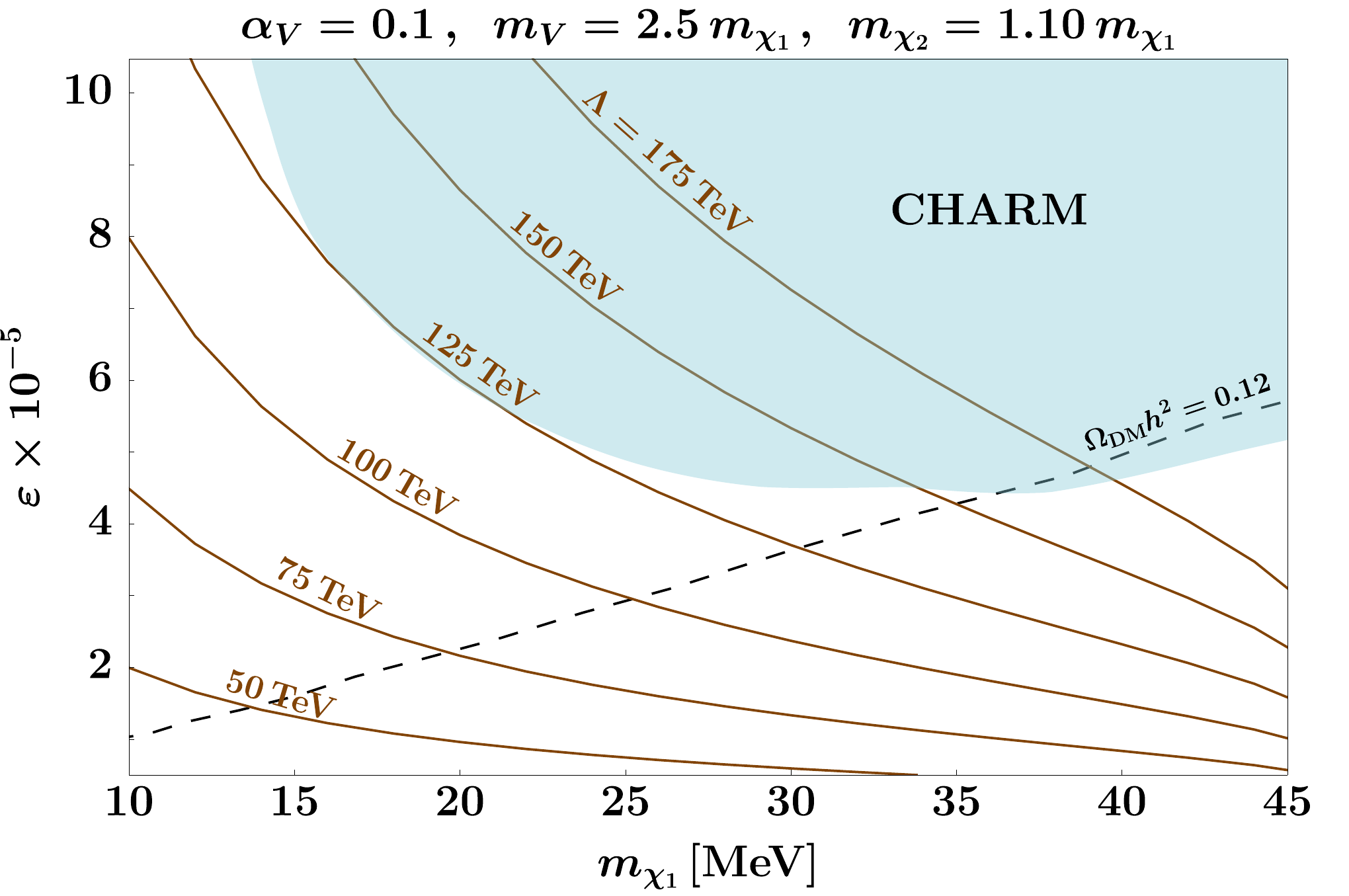}% 
\caption{Sensitivity to the CLFV scale $\Lambda$ of searches for the exotic muon decay $\mu\to 3e+\text{inv}$ in the thermal inelastic dark-matter model of~\citet{Duerr:2019dmv}, displayed on the plane of the dark-matter mass $m_{\chi_1}$ and the kinetic-mixing parameter $\varepsilon$. Also shown are the relic-abundance contour (dahsed line) and the exclusion of the CHARM beam-dump experiment. Figure reproduced from~\citet{Greljo:2025ljr}. \label{fig:dm:reach}}
\end{figure}

Several explicit, renormalisable models illustrate that such exotic decays can be the leading signature of an otherwise viable NP scenario \citep{Greljo:2025ljr}. Examples include: \emph{(i)} A flavor-protected scalar $\mathcal S$, coupled through a Yukawa-like operator with a Froggatt--Nielsen suppression, drives the chain $\mu\to e\,3\mathcal S$ with $\mathcal S\to2e$, i.e.\ $\mu\to7e$. \emph{(ii)} A flavor-violating dark photon, the gauge boson $V$ of a spontaneously broken $U(1)_d$ kinetically mixed with hypercharge, mediates the cascade $\mu\to e\varphi\to e\,(VV)\to e\,(e^+e^-)(e^+e^-)$, i.e.,\ $\mu\to5e$, a variant of the model first presented in \citet{Hostert:2023gpk}; gauging a flavor-non-universal $U(1)$ such as $U(1)_{\mu-e}$ gives closely related signatures \citep{Greljo:2021npi,Greljo:2022dwn,Allanach:2020zna,Costa:2020dph,Smolkovic:2019jow}. \emph{(iii)} A thermal inelastic dark-matter sector, in which the dark photon couples off-diagonally to two Majorana states $\chi_{1,2}$, yields $\mu\to3e+\mathrm{inv}$ or $\mu\to5e+\mathrm{inv}$ cascades. Remarkably, over part of the parameter space that reproduces the observed relic abundance the exotic muon decay is the most sensitive probe, surpassing beam-dump limits such as those from CHARM \citep{Duerr:2019dmv,Izaguirre:2015zva,Tucker-Smith:2001myb,Tsai:2019buq}, see Fig.~\ref{fig:dm:reach}. \emph{(iv)} An on- or off-shell ALP produced in $\mu\to e a^{(*)}$ with $a\to\gamma\gamma$ realizes $\mu\to e\gamma\gamma$. As mentioned above, the on-shell window is already part of the MEG~II search program \citep{MEG:2020zxk}, while heavier, off-shell ALPs evade astrophysical and beam-dump bounds at the price of a broad, rather than peaked, di-photon spectrum. {\em (v)} A rather exotic possibility is specific to $\mu\to e$ conversion experiments: these could see signs of $\mu$-proton annihilation into dark sector states; for operators involving right-handed SM fields this would be the discovery channel, surpassing even the proton bounds for  dark states in a somewhat narrow mass range \citep{Fox:2024kda}.

\section{CLFV with taus} 
\label{sec:tau}

Being much heavier and shorter-lived than muons ($m_\tau \simeq 1777\,{\rm MeV}$, $\tau_\tau \simeq 2.9\times 10^{-13}\,{\rm s}$), tau leptons cannot be produced as beams, stopped in a target, and studied at rest, as is the case with muons. The most efficient production mode of $\tau^+\tau^-$ pairs currently relies on $e^+e^-$ or $pp$ collisions at accelerator facilities such as SuperKEKB (where Belle~II is operated), BEPC~II (which hosts the BESIII detector), and the LHC. For these reasons, the limits on CLFV tau decays are several orders of magnitude weaker than the limits on the CLFV muon transitions, discussed in the previous section, and are currently spanning the $10^{-8}-10^{-7}$ range in the branching ratios. On the other hand, unlike muons, taus are heavy enough for hadronic decay modes to be kinematically allowed. Consequently, around 40 hadronic CLFV modes have been searched for in the past, mainly at the two $B$ factories BaBar and Belle. These hadronic decay modes are in addition to all possible flavor combinations of purely leptonic and radiative decays~\citep{Banerjee:2022xuw}.
This makes the tau CLFV program extremely rich and enables us to test many different CLFV LEFT operators, including operators with little or no constraint from muon CLFV. 
Improved sensitivities to $\tau$ processes are expected to be achieved by the currently running Belle~II experiment~\citep{Belle-II:2018jsg,Banerjee:2022vdd} and at future or proposed facilities, such as the High-Luminosity LHC (HL-LHC)~\citep{LHCb:2018roe,ATLAS:2018owa}, the Super Tau-Charm Factory (STCF)~\citep{Acharya:2023ird}, and high-energy $e^+e^-$ Higgs/Tera-Z factories (FCC-ee and CEPC)~\citep{Dam:2018rfz,Ai:2024nmn}.\footnote{Interestingly, complementary probes of lepton-quark CLFV interactions involving tau leptons can also be obtained at neutrino telescopes through searches for cosmic-ray muons transitions into taus~\citep{Maitra:2026qhi}.}
Given the unknown nature and flavor structure of NP, it is certainly possible that tau CLFV is discovered before muon CLFV, especially if certain non-trivial flavor symmetries are approximately preserved in the charged lepton sector, see, e.g.,~\cite{Calibbi:2025fzi}. This makes searches for tau CLFV extremely important and complementary to the searches for muonic CLFV. Systematic overviews on tau CLFV can be found in~\cite{Celis:2013xja,Celis:2014asa, Husek:2020fru, Cirigliano:2021yto, Banerjee:2022xuw}.

\subsection{Purely leptonic and radiative tau CLFV decays}

%------------------- Table ----------------------
\begin{table}[t]
\renewcommand{\arraystretch}{1.1}
\TBL{\caption{Current 90\%~CL limits and projected future sensitivities for radiative and purely leptonic CLFV $\tau$ decays. Expected future limits are collected from \cite{Banerjee:2022vdd,Achasov:2023gey,Dam:2018rfz,Ai:2024nmn}.}
\label{tab:taulimits}}
{\begin{tabular*}{\textwidth}{@{\extracolsep{\fill}}@{}lllll@{}}
\toprule
Process & Current limit & Experiment & Future reach & Experiment\\
\colrule
\multirow{3}{*}{$\tau^- \to e^-  \gamma$}
    & \multirow{3}{*}{$3.3\times10^{-8}$} & \multirow{3}{*}{BaBar~\citep{BaBar:2009hkt}}
    & $9.0\times 10^{-9}$  & Belle~II~($50\,{\rm ab}^{-1}$)\\
    &   &   & $\ord{10^{-9}}$   & STCF~($10\,{\rm ab}^{-1}$) \\ 
    &   &   &  $\ord{10^{-10}}$ & FCC-ee/CEPC \\[1mm] 
\multirow{3}{*}{$\tau^- \to \mu^-  \gamma$}
    & \multirow{3}{*}{$4.2\times10^{-8}$} & \multirow{3}{*}{Belle~\citep{Belle:2021ysv}}
    & $6.9\times 10^{-9}$  & Belle~II~($50\,{\rm ab}^{-1}$) \\
    &   &   & $5.7\times 10^{-9}$  & STCF~($10\,{\rm ab}^{-1}$) \\ 
    &   &   & $\ord{10^{-10}}$  & FCC-ee/CEPC \\%[1mm]
\midrule
\multirow{3}{*}{$\tau^- \to e^-  e^+e^-$}
    & \multirow{3}{*}{$2.5\times10^{-8}$} & \multirow{3}{*}{Belle~II~\citep{Belle-II:2025urb}}
    & $4.7\times 10^{-10}$  & Belle~II~($50\,{\rm ab}^{-1}$)\\
    &   &   & $1.9\times 10^{-10}$   & STCF~($10\,{\rm ab}^{-1}$) \\ 
    &   &   &  $\ord{10^{-10}}$ & FCC-ee/CEPC \\[1mm] 
\multirow{3}{*}{$\tau^- \to \mu^-  \mu^+\mu^-$}
    & \multirow{3}{*}{$1.9\times10^{-8}$} & \multirow{3}{*}{Belle~II~\citep{Belle-II:2024sce}}
    & $3.6\times 10^{-10}$  & Belle~II~($50\,{\rm ab}^{-1}$)\\
    &   &   & $1.9\times 10^{-10}$   & STCF~($10\,{\rm ab}^{-1}$) \\ 
    &   &   &  $\ord{10^{-11}}$ & FCC-ee/CEPC \\[1mm] 
\multirow{3}{*}{$\tau^- \to e^-  \mu^+\mu^-$}
    & \multirow{3}{*}{$2.4\times10^{-8}$} & \multirow{3}{*}{Belle~II~\citep{Belle-II:2025urb}}
    & $4.5\times 10^{-10}$  & Belle~II~($50\,{\rm ab}^{-1}$)\\
    &   &   & $1.9\times 10^{-10}$   & STCF~($10\,{\rm ab}^{-1}$) \\ 
    &   &   &  $\ord{10^{-10}}$ & FCC-ee/CEPC \\[1mm] 
\multirow{3}{*}{$\tau^- \to \mu^-  e^+e^-$}
    & \multirow{3}{*}{$1.6\times10^{-8}$} & \multirow{3}{*}{Belle~II~\citep{Belle-II:2025urb}}
    & $2.9\times 10^{-10}$  & Belle~II~($50\,{\rm ab}^{-1}$)\\
    &   &   & $1.9\times 10^{-10}$   & STCF~($10\,{\rm ab}^{-1}$) \\ 
    &   &   &  $\ord{10^{-10}}$ & FCC-ee/CEPC \\[1mm] 
\multirow{3}{*}{$\tau^- \to \mu^+  e^-e^-$}
    & \multirow{3}{*}{$1.5\times10^{-8}$} & \multirow{3}{*}{Belle~\citep{Hayasaka:2010np}}
    & $2.3\times 10^{-10}$  & Belle~II~($50\,{\rm ab}^{-1}$)\\
    &   &   & $1.9\times 10^{-10}$   & STCF~($10\,{\rm ab}^{-1}$) \\ 
    &   &   &  $\ord{10^{-10}}$ & FCC-ee/CEPC \\[1mm] 
\multirow{3}{*}{$\tau^- \to e^+  \mu^-\mu^-$}
    & \multirow{3}{*}{$2.6\times10^{-8}$} & \multirow{3}{*}{Belle~II~\citep{Belle-II:2025urb}}
    & $4.5\times 10^{-10}$  & Belle~II~($50\,{\rm ab}^{-1}$)\\
    &   &   & $1.9\times 10^{-10}$   & STCF~($10\,{\rm ab}^{-1}$) \\ 
    &   &   &  $\ord{10^{-10}}$ & FCC-ee/CEPC \\
\botrule
\end{tabular*}}{}
\end{table}
%-----------------------------------------------

Current and expected limits on leptonic and radiative tau CLFV decays are displayed in Table~\ref{tab:taulimits}. The best sensitivity to these modes is reached at $e^+e^-$ colliders. In general, searches for CLFV $\tau$ decays in $e^+e^-$ collisions exploit the clean topology of $e^+e^-\to\tau^+\tau^-$ events, in which the two $\tau$ leptons are produced nearly back-to-back in the center-of-mass (c.o.m.)~frame. Events are partitioned into signal and tag hemispheres using a plane perpendicular to the thrust axis, defined as the direction that maximizes the sum of the projections of the momenta of all reconstructed particles along a common axis. The signal hemisphere contains the candidate CLFV decay (e.g.,~$\tau\to\mu\gamma$ or $\tau\to3\mu$), while the opposite hemisphere is required to be consistent with a SM $\tau$ decay, providing powerful background suppression. Traditionally, the tag side is reconstructed in exclusive one-prong $\tau$ decay modes, corresponding to $\tau$ decays with a single charged particle in the final state, although recent Belle~II analyses have increasingly adopted inclusive $\tau$-tagging algorithms based on boosted decision trees (BDTs), which improve the overall signal efficiency~\citep{Belle-II:2024sce,Belle-II:2025urb}. 
Candidate events are selected using stringent particle-identification and vertex requirements together with the precisely known initial-state kinematics. In particular, the known beam energy and the absence of missing energy on the signal side provide powerful kinematic constraints. The invariant mass of the visible signal decay products is required to be consistent with the $\tau$ mass, while their total energy in the c.o.m.~frame must be close to half of the collision energy, $\sqrt{s}/2$, so that signal events populate a narrow region around $(m_\tau,\Delta E=0)$ in the $(m_{\rm inv},\Delta E)$ plane, where $\Delta E=E_{\rm sig}-\sqrt{s}/2$. This strategy has enabled the $B$ factories to perform searches for $\tau\to3\ell$ with very low, and in some cases essentially negligible, background. In contrast, searches for $\tau\to\ell\gamma$ at the $\Upsilon(4S)$ resonance, with c.o.m.~energy $\sqrt{s}\simeq10.6,\mathrm{GeV}$, are affected by a sizeable irreducible background arising from initial-state radiation (ISR), which ultimately limits the sensitivity achievable at Belle~II. This background is substantially reduced at lower c.o.m.~energies, such as the range of $\sqrt{s}$ foreseen at the STCF, as well as at higher energies, such as the proposed Tera-$Z$ factories.

In terms of the coefficients of the LEFT operators in Table~\ref{tab:left}, we can write expressions for $\tau$ decays that are analogous to the muon processes discussed in Sec.~\ref{sec:mu:e:gamma}.
Ignoring the final state lepton masses, the BRs of the radiative decays are given by
\begin{align}
    \mathrm{BR}(\tau\to \ell_i\gamma) = \frac{m_\tau^3}{4\pi\,\Gamma_\tau} \left(|{\mathcal C}_{e\gamma,\,i3}|^2+|{\mathcal C}_{e\gamma,\,3i}|^2 \right)\,,
    \label{eq:tau-ell-gamma}
\end{align}
where $\Gamma_\tau = 1/\tau_\tau \simeq 2.3\times 10^{-12}\,{\rm GeV}$.
The limits reported in Table~\ref{tab:taulimits} then imply
\begin{align}
    {\mathcal C}_{e\gamma,\,13}\,,~{\mathcal C}_{e\gamma,\,31} \lesssim 4.1\times 10^{-10}~{\rm GeV}^{-1} \quad & \Longrightarrow \quad \Lambda \gtrsim 6.5\times 10^5~{\rm GeV}\,, \\
     {\mathcal C}_{e\gamma,\,23}\,,~{\mathcal C}_{e\gamma,\,32}  \lesssim 4.6\times 10^{-10}~{\rm GeV}^{-1} \quad & \Longrightarrow \quad \Lambda \gtrsim 6.1\times 10^5~{\rm GeV}\,,
\end{align}
where the same assumptions as in Sec.~\ref{sec:mu:e:gamma} were made for the SMEFT-LEFT matching.

For the $\tau\to eee$ and $\tau\to\mu\mu\mu$ decays, one obtains (see, e.g.,~\cite{Calibbi:2021pyh})
\beq
\begin{split}
{\rm BR}(\tau\to \ell_j \ell_j \bar \ell_j) =~ &
\frac{m_{\tau}^5}{3 (16\pi)^3 \Gamma_{\tau}} 
\bigg[16\big|C^{V\,LL}_{ee,\,j3 jj}\big|^2+ 16 \big|C^{V\,RR}_{ee,\,j3 jj}\big|^2 + 8 \big|C^{V\, LR}_{ee,\,j3 jj}\big|^2 + 8\big|C^{V\,RL}_{ee,\,j3 jj}\big|^2 
+ \big|C^{S\,RR}_{ee,\,j3 jj}\big|^2 + \big|C^{S\,RR}_{ee,\,3j jj}\big|^2 
\\ 
+ & \frac{256 e^2}{m_{\tau}^2} \left(\log\frac{m_{\tau}^2}{m_{\ell_j}^2}-\frac{11}4\right)
\Big(\big|C_{e\gamma,\,j3}\big|^2 + \big|C_{e\gamma,\,3 j}\big|^2\Big)
- \frac{64 e}{m_{\tau}}\, Re\Big[\Big(2 C^{V\,LL}_{ee,\,j3 jj} + C^{V\,LR}_{ee,\, jjj3}\Big) C_{e\gamma,\,j3 }^{*} + \Big(2C^{V\,RR}_{ee,\,j3\, jj}+C^{V\,LR}_{ee.\,jjj3}\Big) C_{e\gamma,\,3 j}\Big] \bigg]\,.
\end{split}
\eeq
Similarly, for the $\tau\to e\mu\bar\mu$ and $\tau\to \mu e\bar e$ decays one has 
\beq
\begin{split}
{\rm BR}(\tau\to \ell_j \ell_k \bar \ell_k) 
=~ &\frac{m_{\tau}^5}{3 (16\pi)^3 \Gamma_{\tau}} 
\bigg[8\big|C^{V\,LL}_{ee,\,j3 kk}\big|^2+ 8 \big|C^{V\,RR}_{ee,\,j3 kk}\big|^2 + 8 \big|C^{V\,LR}_{ee,\,j3 kk}\big|^2 + 8\big|C^{V\,LR}_{ee,\,kkj3}\big|^2 
+ 2\big|C^{S\,RR}_{ee,\,j3 kk}\big|^2 + 2\big|C^{S\,RR}_{ee,\,3j kk}\big|^2 
\\
+& 
\frac{256 e^2}{m_{\tau}^2} \left(\log\frac{m_{\tau}^2}{m_{\ell_k}^2}-3\right)
\Big(\big|C_{e\gamma,\,j3}\big|^2 + \big|C_{e\gamma,\,3 j}\big|^2\Big)
- \frac{64 e}{m_{\tau}}\, Re\Big[\Big(C^{V\,LL}_{ee,\,j3 kk} + C^{V\,LR}_{ee,\,j3 kk}\Big) C_{e\gamma,\,j3}^{*} + \Big(C^{V\,RR}_{ee,\,j3 kk}+C^{V\,LR}_{ee,\,kkj3}\Big) C_{e\gamma,\,3 j}\Big] \bigg]\,.
\end{split}
\eeq
Finally, for $\tau\to e e \bar\mu$ and $\tau\to \mu\mu\bar e$ we have
\begin{align}
{\rm BR}(\tau\to \ell_k \ell_k \bar \ell_j) 
=\frac{m_{\tau}^5}{3 (16\pi)^3 \Gamma_{\tau}} 
\bigg[16\big|C^{V\,LL}_{ee,\,k3 jk}\big|^2+ 16 \big|C^{V\,RR}_{ee,\,k3 jk}\big|^2 + 8 \big|C^{V\,LR}_{ee,\,k3 jk}\big|^2 + 8\big|C^{V\,LR}_{ee,\,jkk3 }\big|^2
+ \big|C^{S\,RR}_{ee,\,k3 jk}\big|^2 + \big|C^{S\,RR}_{ee,\,3k kj}\big|^2 \bigg]\,.
\end{align}
Notice that these two processes (unlike the previous ones) require the three lepton family numbers to all be violated. For example, $\tau\to e e \bar\mu$ would arise if a mediator particle has couplings of both $\tau-e$ and $\mu-e$ types.

The current limits on these three-body decays shown in Table~\ref{tab:taulimits} imply a bound on $\Lambda$ around $10\,{\rm TeV}$. For instance, $\tau \to 3\mu$ 
gives the following limit on the relevant purely left-handed contact interaction:
\begin{align}
    {\mathcal C}^{V\,LL}_{ee,\,2322} \lesssim 7.6\times 10^{-9}~{\rm GeV}^{-1} \quad \Longrightarrow \quad \Lambda \gtrsim 11~{\rm TeV}\,.
\end{align}
The improvement of about two orders of magnitude in the sensitivity on such branching ratios is expected at Belle~II and other future facilities, which will then test scales as high as $30\,{\rm TeV}$.

\subsection{ Tau CLFV involving hadrons}
\label{sec:tau:hadronic}

As mentioned above, unlike muons, tau leptons can also decay into
hadronic final states due to their relatively large mass. Table \ref{tab:tauhadroniclimits} presents current limits and prospective sensitivities for representative hadronic CLFV $\tau$ decays. Most of the current limits are of order $10^{-8}$, while future sensitivities are expected to be around $10^{-9}-10^{-10}$ assuming $50\,{\rm ab}^{-1}$ of integrated luminosity accumulated at Belle~II and $1\,{\rm ab}^{-1}$ at STCF. These decay modes can arise from operators involving different quark flavors. For instance, quark-flavor conserving operators with $q=u,d$ contribute to $\tau^-\to \ell^- \pi^0$, whereas quark-flavor violating operators with $q=s,d$ contribute to $\tau^-\to \ell^- K^0_S$. Searches in different decay channels are therefore complementary and play an important role in probing the underlying origin of CLFV. On the theory side, predictions of decay rates of hadronic channels require evaluating their hadronic matrix elements, which can be determined either from fits to experimental data, or using Chiral Perturbation Theory (ChPT), and Lattice QCD calculations.

%------------------- Table ----------------------
{\renewcommand{\arraystretch}{1.1}
\begin{table}[t!]
\TBL{\caption{Current 90\%~CL limits and projected future sensitivities for representative hadronic CLFV $\tau$ decays. Expected future limits are collected from \cite{Belle-II:2022cgf} for Belle~II and \cite{Zhou2020Snowmass} for STCF.
}
\label{tab:tauhadroniclimits}}
{\begin{tabular*}{\textwidth}{@{\extracolsep{\fill}}@{}lllll@{}}
\toprule
Process & Current limit & Experiment & Future reach & Experiment\\
\colrule

{$\tau^- \to e^- \pi^0$}
    & $8.0\times10^{-8}$ & Belle \citep{Belle:2007cio}
    & $7.3\times 10^{-10}$
    & Belle~II\\[1mm] 

{$\tau^- \to \mu^- \pi^0$}
    & $1.1\times 10^{-7}$ & BaBar \citep{BaBar:2006jhm}
    & $7.1\times 10^{-10}$
    & Belle~II\\[1mm] 

{$\tau^- \to e^- K_S^0$}
    & $8\times10^{-9}$ & Belle \citep{Belle:2025iff}
    & $4.0\times 10^{-10}$
    & Belle~II\\[1mm] 

{$\tau^- \to \mu^- K_S^0$}
    & $1.2\times10^{-8}$ & Belle \citep{Belle:2025iff}
    & $4.0\times 10^{-10}$
    & Belle~II\\[1mm] 

$\tau^- \to e^- \eta$
    & $9.2\times10^{-8}$ & Belle \citep{Belle:2007cio}
    & $1.2\times 10^{-9}$
    & Belle~II\\[1mm] 

$\tau^- \to \mu^- \eta$
    & $6.5\times10^{-8}$ & Belle \citep{Belle:2007cio}
    & $8.0\times 10^{-10}$
    & Belle~II\\[1mm] 

$\tau^- \to e^- \rho^0$
    & $1.8\times10^{-8}$ & Belle \citep{Belle:2011ogy}
    & $3.5\times 10^{-10}$
    & Belle~II\\[1mm] 

$\tau^- \to \mu^- \rho^0$
    & $1.2\times10^{-8}$ & Belle \citep{Belle:2011ogy}
    & $5.5\times 10^{-10}$
    & Belle~II\\%[1mm] 

\midrule

$\tau^- \to e^- \pi^+\pi^-$
    & $2.3\times10^{-8}$ & Belle \citep{Belle:2012unr}
    & $5.8\times 10^{-10}$
    & {Belle~II}\\[1mm] 

\multirow{2}{*}{$\tau^- \to \mu^- \pi^+\pi^-$}
    & \multirow{2}{*}{$2.1\times10^{-8}$} & \multirow{2}{*}{Belle \citep{Belle:2012unr}}
    & $5.6\times 10^{-10}$
    & Belle~II\\
    & & & ${\cal O}(10^{-9})$ & STCF\\[1mm] 

{$\tau^- \to e^- \pi^+ K^-$}
    & {$3.7\times10^{-8}$} & {Belle \citep{Belle:2012unr}}
    & $7.1\times 10^{-10}$
    & Belle~II\\[1mm]

\multirow{2}{*}{$\tau^- \to \mu^- \pi^+ K^-$}
    & \multirow{2}{*}{$8.6\times10^{-8}$} & \multirow{2}{*}{Belle \citep{Belle:2012unr}}
    & $1.2\times 10^{-9}$
    & Belle~II\\
    & & & ${\cal O}(10^{-9})$ & STCF\\[1mm] 

$\tau^- \to e^- K^+ K^-$
    & $3.4\times10^{-8}$ & {Belle \citep{Belle:2012unr}}
    & $6.5\times 10^{-10}$
    & Belle~II\\[1mm] 

\multirow{2}{*}{$\tau^- \to \mu^- K^+ K^-$}
    & \multirow{2}{*}{$4.4\times10^{-8}$} & \multirow{2}{*}{Belle \citep{Belle:2012unr}}
    & $1.1\times 10^{-9}$
    & Belle~II\\
    & & & ${\cal O}(10^{-9})$ & STCF\\

\botrule
\end{tabular*}}{}
\end{table}
%-----------------------------------------------

Let us start with two-body $\tau$ decays. Neglecting lepton mass in the final states, the partial decay widths for $\tau^-\to \ell^- P$ decays are given by
\begin{align}
    {\Gamma}\left(\tau^- \to \ell^- \,P \right)=\frac{m^3_{\tau}}{32\pi}\left(1-\frac{m^2_P}{m^2_{\tau}} \right)^2G_F^2\,\Bigg[\left|\,A^P_L \,\right|^2+ \left|\,A^P_R\, \right|^2\Bigg]\,,
\end{align}
where $P=\pi^0,\,K_S^0,\,\eta,\,\eta^{\prime}$, and $G_F$ is the Fermi constant. The coefficients
$A_{L,R}^P$ for $P=\pi^0$ and $K_S^0$ are given by the following combinations of the LEFT WCs (corresponding to operators listed in Table~\ref{tab:left})~\citep{Cirigliano:2021img}\footnote{Note that the normalization of the dim-6 LEFT operators used here differs by a factor of $4G_F/\sqrt{2}$ from that in \cite{Cirigliano:2021img}.}  
\begin{align}
    A_{L}^{\pi^0} = &~\frac{\sqrt{2}\,f_{\pi}}{4G_F}\,\Bigg[(C_{ue}^{V\,RL}-C_{ue}^{V\,LL})_{11\ell 3}-(u\to d)+\frac{m^2_{\pi}}{m_{\tau}(m_u+m_d)}\,\Bigg\{(C_{ue}^{S\,RR*})_{113\ell}-(C_{ue}^{S\,RL})_{11\ell3}-(u\to d)\Bigg\}\Bigg]\,,\\[0.2cm]
    A_{R}^{\pi^0} = &~\frac{\sqrt{2}\,f_{\pi}}{4G_F}\,\Bigg[(C_{ue}^{V\,LR}-C_{ue}^{V\,RL})_{11\ell3}-(u\to d)+\frac{m^2_{\pi}}{m_{\tau}(m_u+m_d)}\,\Bigg\{(C_{ue}^{S\,RR})_{11\ell 3}-(C_{ue}^{S\,RL})_{113\ell}-(u\to d)\Bigg\}\Bigg]\,,
\end{align}
and
\begin{align}
    A_{L}^{K} = &~\frac{\sqrt{2}\,f_K}{4G_F}\,\Bigg[(C_{de}^{V\,LR}-C_{de}^{V\,LL})_{12\ell3} +\frac{m^2_K}{m_{\tau}(m_d+m_s)}\Bigg\{(C_{de}^{ S\,RR*})_{213\ell}-(C_{de}^{S\,RL})_{12\ell3}\Bigg\}-\left(d\leftrightarrow s \right)\Bigg]\,,\\[0.2cm]
    A_{R}^{K} = &~\frac{\sqrt{2}\,f_K}{4G_F}\,\Bigg[(C_{de}^{V\,LR}-C_{de}^{\rm VRR})_{12\ell3} +\frac{m^2_K}{m_{\tau}(m_d+m_s)}\Bigg\{(C_{de}^{S\,RR})_{12\ell 3}-(C_{de}^{S\,RL*})_{213\ell}\Bigg\}-\left(d\leftrightarrow s \right)\Bigg]\,.
\end{align}
The pseudoscalar meson decay constants are defined by $\langle P(p) |\,\bar{q}\gamma^{\mu}\gamma_5q\,| 0\rangle = ip^{\mu}f_P$, where Lattice QCD calculations give $f_{\pi} = 130.2~(1.2)~{\rm MeV}$ and $f_{K} = 155.7~(3)~{\rm MeV}$~\citep{PDG2026}.
The $\tau^-\to \ell^- \eta$ decays, on the other hand, can also be mediated by CLFV gluon and strange-quark operators, as well as the up- and down-quark operators, giving
\begin{align}
A_L^{\eta} & \supset \frac{\sqrt{2}\,a_{\eta}}{m_{\tau}v}~(C_{G\tilde{G}})_{3\ell}^*+\frac{1}{2G_F}~\Bigg[\,f^s_{\eta}\,(C_{de}^{V\,RL}-C_{de}^{V\,LL})_{22\ell3}+\frac{h_{\eta}^s}{2m_{\tau}m_s}\,\Bigg\{(C_{de}^{S\,RR*})_{223\ell}-(C_{de}^{S\,RL})_{22\ell3}\Bigg\} \Bigg]\,,\\[0.2cm]
A_R^{\eta} & \supset \frac{\sqrt{2}\,a_{\eta}}{m_{\tau}v}~(C_{G\tilde{G}})_{\ell3}+\frac{1}{2G_F}~\Bigg[\,f^s_{\eta}\,(C_{de}^{V\,LR}-C_{de}^{\rm VRR})_{22\ell3} +\frac{h_{\eta}^s}{2m_{\tau}m_s}\,\Bigg\{(C_{de}^{S\,RR})_{22\ell3}-(C_{de}^{S\,RL *})_{223\ell}\Bigg\} \Bigg]\,.
\end{align}
To shorten the expressions the contributions from up- and down-quark operators were omitted above.\footnote{The CLFV gluon operator is of dimension 7 and is given by  ${\cal L}_\text{LEFT}\supset C_{G\tilde G}\,\alpha_s/(4\pi v^3)\,G\tilde{G}\,\bar{e}_Le_R$. The complete expressions for $A_{L,R}^{\eta}$ can be found in \cite{Cirigliano:2021img}.} The expressions for $\tau^-\to \ell^- \eta^{\prime}$ can be obtained by replacing $\eta$ with $\eta^{\prime}$. Using the Feldmann-Kroll-Stech scheme, which assumes a single $\eta-\eta'$  mixing phase \citep{Feldmann:1998vh, Feldmann:1998sh, Feldmann:1999uf}, the decay constants $a_{\eta,\eta'},\, f^s_{\eta,\eta'}$ and $h^s_{\eta,\eta'}$ can be extracted from fits to experimental data and are given by \citep{Beneke:2002jn}
\begin{align}
    a_{\eta}&=0.022\pm0.002~{\rm GeV}^3\,, \hspace{1cm} f^s_{\eta}=-0.11\pm 0.01~{\rm GeV}\,, \hspace{1cm} h^s_{\eta}=-0.055\pm 0.003~{\rm GeV}^3\,,\\%[0.1cm]
    a_{\eta^{\prime}}&=0.056\pm0.002~{\rm GeV}^3\,, \hspace{0.92cm} f^s_{\eta^{\prime}}=0.135\pm 0.006~{\rm GeV}\,, \hspace{0.83cm} h^s_{\eta^{\prime}}=0.068\pm 0.005~{\rm GeV}^3\,.
\end{align}
For a recent study of the decay constants based on the Lattice QCD calculations, see \cite{Ottnad:2025zxq}. We can interpret the experimental limits on BR($\tau^-\to \ell^- \eta$) in terms of the vector and scalar operators as $v^2_\text{EW} C \lesssim \mathcal{O}(10^{-3} - 10^{-4})$ for the strange-quark operators, 
while the limits on the gluon operators are ${\cal O}(1)$.

The determination of hadronic matrix elements needed for three-body decays is more complicated than for two-body decays, since these now depend on the invariant mass $s$ of the hadronic system, which can vary over a relatively large range, $2m_\pi\leq \sqrt{s}\leq m_{\tau}-m_{e/\mu}$. For $\tau^- \to \ell^- \pi^+\pi^-$, there are four relevant matrix elements
\begin{subequations}
\begin{align}
    \langle\pi^+(p_{\pi^+})\,\pi^- (p_{\pi^-})|\,\tfrac{1}{2}\left(\bar{u}\gamma^{\mu}u-\bar{d}\gamma^{\mu}d \right)\,|0 \rangle &=\left(p_{\pi^+}-p_{\pi^-} \right)^{\mu}\,F_V(s)\,,\\%[0.2cm]
    \langle\pi^+(p_{\pi^+})\,\pi^- (p_{\pi^-})|\,m_u\bar{u}u+m_d\bar{d}d\,|0 \rangle &=\Gamma_{\pi}\, (s)\,,\\%[0.2cm]
    \langle\pi^+(p_{\pi^+})\,\pi^- (p_{\pi^-})|\,m_s\bar{s}s \,|0\rangle &=\Delta_{\pi} \,(s)\,,\\%[0.2cm]
    \langle\pi^+(p_{\pi^+})\,\pi^- (p_{\pi^-})|\,\theta^{\mu}_{\mu} \,|0\rangle &=\theta_{\pi} \,(s)\,,
\end{align}
\end{subequations}
where $s=(p_{\pi^+}+p_{\pi^-})^2$ and the matrix element of the gluon operator can be reduced to that of the trace of the energy-momentum tensor $\theta^{\mu}_{\mu}$ via the relation $\theta_{\mu}^{\mu}=-9\alpha_s/(8\pi)\,G^a_{\mu\nu}G^{a\mu\nu}$. The vector form factor $F_V(s)$ is known directly from measurements of $e^+e^-\to \pi^+\pi^-$  \citep{NA7:1986vav, KLOE:2004lnj, CMD-2:2006gxt, BaBar:2009wpw, KLOE:2010qei} and $\tau\to \pi^-\pi^0 \nu_{\tau}$ \citep{CLEO:1999dln, Belle:2008xpe}. The form factors for scalar quark current and gluon operators have been studied using ChPT \citep{Donoghue:1990xh, Arganda:2008jj}; however, it was pointed out that the ChPT prediction is valid only for very low energies \citep{Celis:2013xja}. These form factors have been properly derived in \cite{Daub:2012mu, Celis:2013xja} within the BSM frameworks of R-parity violating supersymmetric models and extended Higgs models, and have subsequently been used in a comprehensive model-independent analysis \citep{Celis:2014asa}. Below, we will employ the form factors derived in \cite{Celis:2013xja} and follow the analysis in \cite{Celis:2014asa}.

%---------------- Figures ---------------------
\begin{figure}[t]
  \centering
  \includegraphics[width=7cm]{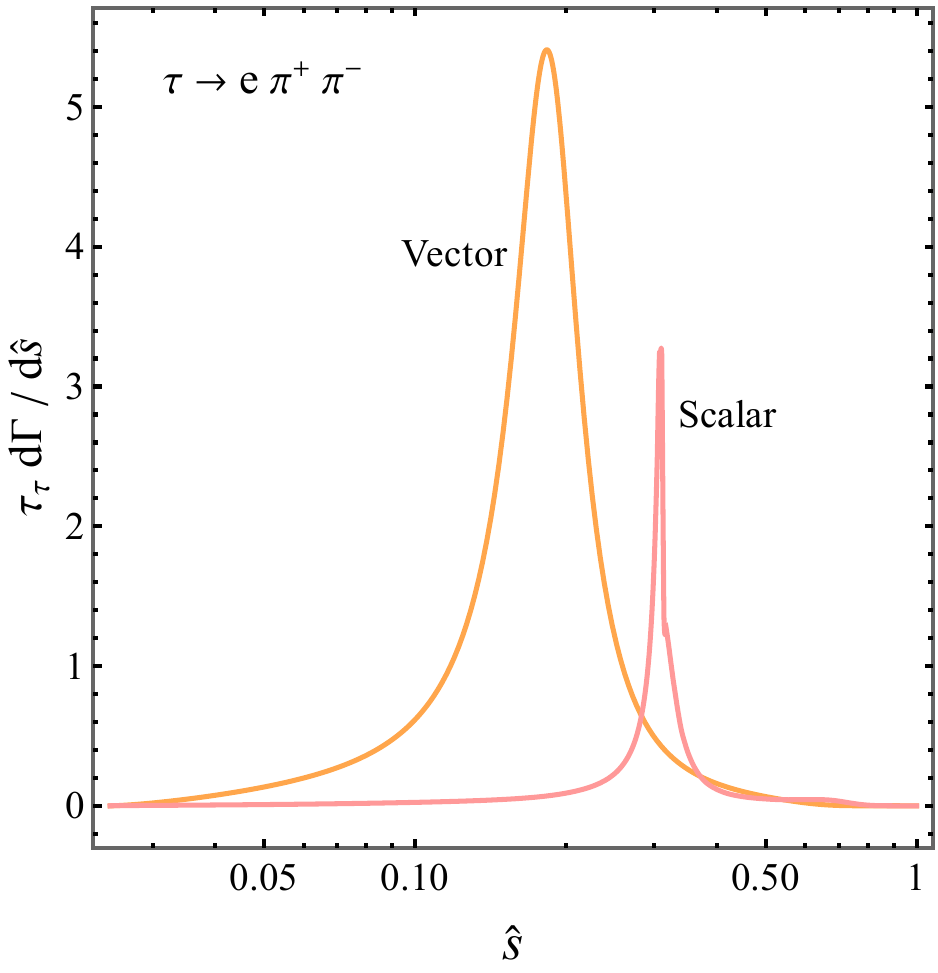}
  \vspace{0.3cm}
  \caption{Resonance contributions to $\tau \to e \pi^+\pi^-$ around $\hat{s}\simeq 0.2$ ($\rho(770)$) and $\hat{s}\simeq 0.3$ ($f_0(980)$). Figure adapted from \cite{Cirigliano:2021img}.}
  \label{fig:FF_tauePP}
\end{figure}
%------------------------------------------------

With the above form factors, the differential decay width for $\tau \to \ell \pi^+\pi^-$ is given by
\begin{align}
\tau_\tau \frac{d\Gamma}{d\hat s}
&=
\frac{1}{40\,\hat{\Gamma}_\tau}
\left(1-\frac{\rho_\pi}{\hat s}\right)^{1/2}
(1-\hat s)^2
\Bigg\{
6m_\tau^2\,
\left(\,
|Q_L|^2+|Q_R|^2\,
\right)
+8\left(1-\frac{\rho_\pi}{\hat s}\right)\,
|F_V(s)|^2\,
\left[\,
\frac{2+\hat s}{4\hat s}
\left(
|A_L|^2+|A_R|^2
\right)
+
(B_L+B_R)\,
\right]
\Bigg\}\,,
\end{align}
where the normalized dimensionless quantities are $\hat{s}=s/m^2_{\tau}$ and $\rho_{\pi}=4m^2_{\pi}/m^2_{\tau}$. In this notation, the kinematically allowed region is $\rho_{\pi}<\hat{s}<1$. The $Q_{L,R}$ coefficients contain contributions from gluon and scalar operators, and are given by,
\begin{align}
Q_L
&=
\frac{2}{9v}
\Bigl(
\theta_\pi(s)-\Gamma_\pi(s)-\Delta_\pi(s)
\Bigr)
(C_{GG})_{\ell 3}
+\frac{\Delta_\pi}{m_s}\frac{\sqrt{2}}{4G_F}
\left\{
(C_{de}^{S\,RR})_{22\ell3}+(C_{de}^{S\,RL *})_{223\ell}
\right\}
+\frac12\,\Gamma_\pi(s)
\left[
\frac{1}{\hat m}
\left\{
(C_{ue}^{S\,RR})_{11\ell3}+(C_{ue}^{S\,RL*})_{113\ell}
\right\}
+
(u\to d)
\right]\,,\\[0.2cm]
Q_R
&=
\frac{2}{9v}
\Bigl(
\theta_\pi(s)-\Gamma_\pi(s)-\Delta_\pi(s)
\Bigr)
(C_{GG})^{*}_{3\ell}
+\frac{\Delta_\pi}{m_s}\frac{\sqrt{2}}{4G_F}
\left\{
(C_{de}^{S\,RR*})_{223\ell}+(C_{de}^{S\,RL})_{22\ell3}
\right\}
+\frac12\,\Gamma_\pi(s)
\left[
\frac{1}{\hat m}
\left\{
(C_{ue}^{S\,RR*})_{113\ell}+(C_{ue}^{S\,RL})_{11\ell3}
\right\}
+
(u\to d)
\right]\,.
\end{align}
The $A_{L,R}$ coefficients encode contributions from dipole and tensor operators,
\begin{align}
A_L
&=
2e\,
\frac{v^2}{m_\tau}
(C_{e\gamma})_{\ell 3}
+
\frac{\sqrt{2}}{4G_F}\hat s\,\sqrt{\rho_\pi}\,
B^{T}_{\pi,u}(0)
\left[
(C_{de}^{T\,RR})_{11\ell 3}
-
(C_{ue}^{T\,RR})_{11\ell 3}
\right],\\[0.2cm]
A_R
&=
2e\,
\frac{v^2}{m_\tau}
(C_{e\gamma})^{*}_{3 \ell}
+
\frac{\sqrt{2}}{4G_F}\,\hat s\,\sqrt{\rho_\pi}\,
B^{T}_{\pi,u}(0)
\left[
(C_{de}^{T\,RR})^{*}_{113\ell}
-
(C_{ue}^{T\,RR})^{*}_{113\ell}
\right]\,,
\end{align}
while $B_{L,R}$ coefficients describe contributions from vector operators,
\begin{align}
B_L
&=\frac{\sqrt{2}}{4G_F}\,
\Big[
(C_{de}^{V\,LL}+C_{de}^{V\,RL})_{11\ell 3}
-
(d\to u)
\Big]
\Big[
3A_L
+
\frac{\sqrt{2}}{4G_F}(2\hat s+1)\,
\Big\{
(C_{de}^{V\,LL}+C_{de}^{V\,RL})_{11\ell 3}
-
(d\to u)
\Big\}
\Big]\,,\\[0.2cm]
B_R
&=\frac{\sqrt{2}}{4G_F}\,
\Big[
(C_{de}^{\rm VRR}
+
C_{de}^{V\,LR})_{11\ell 3}
-
(d\to u)
\Big]\,
\Big[
3A_R
+\frac{\sqrt{2}}{4G_F}
(2\hat s+1)\,
\Big\{
(C_{de}^{\rm VRR}
+
C_{de}^{V\,LR})_{11\ell 3}
-(d\to u)
\Big\}
\Big]\,.
\end{align}
Here, $B^T_{\pi, u}$ represents the pion matrix element of the tensor current, where its zero momentum transfer value $B^T_{\pi, u}(0)=0.195$ was obtained in \cite{Hoferichter:2018zwu}. It should be noted that two resonances contribute to the $\tau \to \ell 2\pi$ decay, namely $\rho(770)$ ($J^{PC}=1^{--}$) and $f_0(980)$ ($J^{PC}=0^{++}$). 
Figure \ref{fig:FF_tauePP} illustrates the resonance contributions to $\tau^- \to e^- \pi^+\pi^-$. The orange contour corresponds to the vector contribution assuming $(C_{V\,LL}^{ed})_{13 11}=1$, while the pink one shows the scalar contribution with $(C_{S\,RR}^{eu})_{13 qq} \propto Y_q$. The two peaks are located at around $\hat{s} \simeq 0.2$, for the $\rho(770)$ resonance, and at $\hat{s} \simeq 0.3$, for the $f_0(980)$ resonance. The latest bound on the $\pi\pi$ decay channel translates to the following bound on vector and scalar operators, $v^2_\text{EW}\,C\lesssim{\cal O}(10^{-4})$.% with $v\simeq 246$ GeV.

A similar resonant effect can arise from $K^*(892)$ and $\phi(1020)$ in $\tau^-\to e^-\pi^{\pm}K^{\mp}$ and $\tau^-\to e^-K^+K^-$, respectively. For $\tau\to e\pi K$, the vector and scalar form factors have been obtained in \cite{Antonelli:2013usa}, and the lattice QCD results for the normalization, $f_{+,0}^{K\pi}(0)$, are available in \cite{FlavourLatticeAveragingGroupFLAG:2024oxs}. The tensor form factor was derived using the elastic unitarity
relation in \cite{Hoferichter:2018zwu, Cirigliano:2017tqn}. It results in stronger constraints on $ds$ and $sd$ components of vector operators than $\tau^-\to e^-K^0_S$ \citep{Cirigliano:2021img}. The estimates for form factors entering the $\tau \to e K^+K^-$ predictions have also been developed in recent years. The vector form factors were extracted from $e^+e^-$ data by \cite{Beloborodov:2019uql}, while the tensor form factor was described using a dispersive approach by \cite{Gonzalez-Solis:2019lze}. More recently, the scalar and tensor form factors have also been investigated within the framework of U(3) ChPT in \cite{Hao:2026vrw}.
It should be emphasized that the $KK$ mode is sensitive not only to isovector but also to isoscalar contributions. Therefore, it plays an essential role in lifting a flat direction among WCs, which remains unconstrained by the $\pi\pi$ mode.

\subsection{Tau decays and light new physics}
\label{sec:tau:light:np}

The light-NP models that lead to exotic muon decays, which we discussed in Sec.~\ref{sec:mu:light:np}, have a natural and richer analogue in the tau sector. Three features distinguish the tau case. First, the larger tau mass, $m_\tau/m_\mu\simeq17$, opens a considerably wider kinematic window: any light state $X$ with $m_X<m_\tau-m_\ell$ can be produced on-shell in the two-body decay $\tau\to\ell X$, including states heavy enough that the analogous muon decay is kinematically forbidden. Second, due to higher allowed mass of $X$, the final state can now involve decays into hadronic modes. For instance, if  $X$ decays to two dark vectors $V$ and $V$ decays to SM particles, this can results not only in multi-lepton final state such as in the cascade $\tau\to\ell X\to\ell(VV)\to\ell(\ell^+\ell^-)(\ell^+\ell^-)$, but also in 
$\tau\to3\mu\,2\pi$ and $\tau\to\mu\,4\pi$, etc.,~depending on the decay modes of the vector boson $V$. If $V$ is a dark photon, the rates to different hadronic states are  controlled by the 
$R$-ratio, with ${\rm BR}(V\to\pi\pi)/{\rm BR}(V\to\mu\mu)\sim10$ near the $\rho$ resonance, because the couplings to the SM only arise from mixing with the ordinary photon. Third, the $\tau$ samples are typically much smaller than the  muon samples. For instance, Belle~II is expected to reach $4.5\times 10^{10}$ $\tau^+\tau^-$ pairs (see~\cite{Ema:2025bww} for other facilities), to be compared with, e.g.,~a number of stopped muons of the order of $10^{16}$ at Mu3e. Finally, unlike the muon case, the tau rest frame cannot be fully reconstructed event by event at $B$-factories or hadron colliders, because the associated tau in $e^+e^-\to\tau^+\tau^-$ also produces neutrinos. This makes the two-body invisible modes $\tau\to\ell a$,  with $a$ escaping the detector, experimentally more challenging than $\mu\to ea$; a monoenergetic lepton signal in the tau rest frame translates to a smooth distribution of charged lepton energies in the lab frame, or any partially reconstructed psuedo-rest-frame. The signal therefore overlaps more closely with the irreducible three-body SM background $\tau\to\ell\nu\bar\nu$, making the search more challenging. For these reasons, experimental limits are weaker than the corresponding muon decays. For example, in the case of an invisible $a$, the current limits on ${\rm BR}(\tau\to \ell a)$ are $\ord{10^{-4}}$ depending on the ALP mass~\citep{Belle:2025bpu}.

\begin{figure}
    \centering       
     \includegraphics[width=0.59\textwidth]{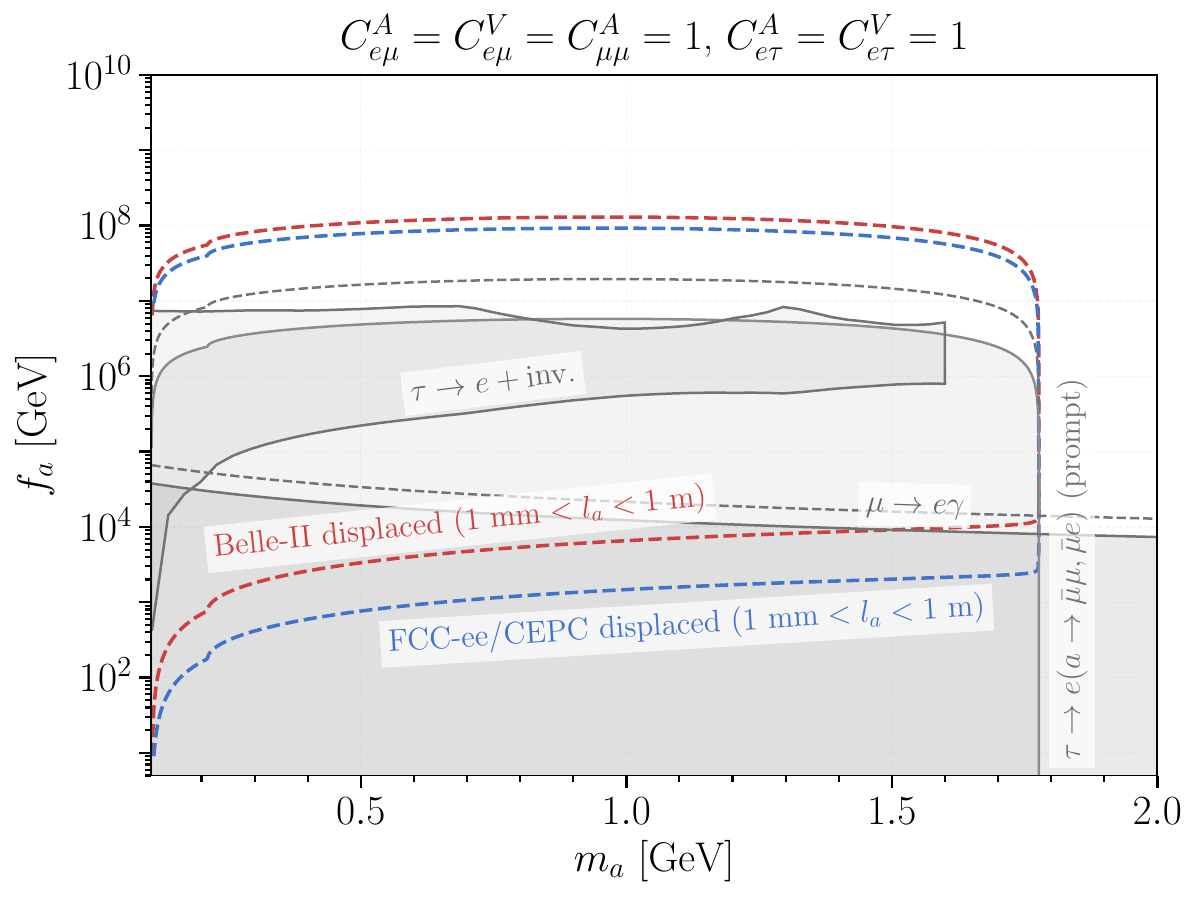}
     \includegraphics[width=0.4\textwidth]{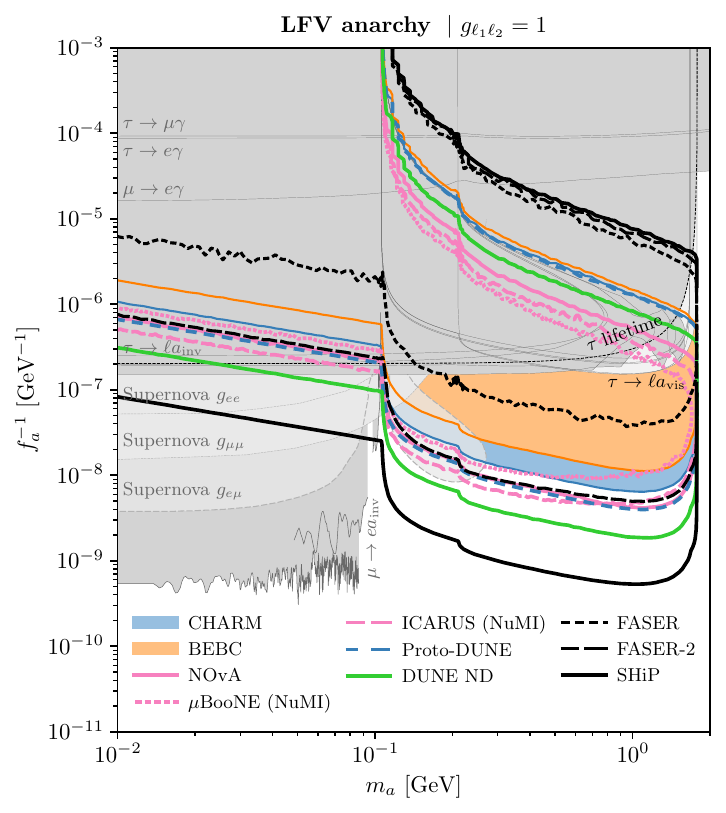}
    \caption{{\em Left:} Prospects at Belle~II (red dashed line) and CEPC/FCC-ee (blue dashed line) sensitivity to the ALP scale $f_a$ from searches for $\tau \to e a$ with the ALP decaying as $a \to \mu \mu~\text{or}~\mu e$ with a displacement $1\,\text{cm} <l_a <1\,\text{m}$ ($l_a$ being the ALP decay length in the laboratory frame), assuming that a limit similar to the prompt searches will be achieved, i.e.,~$\text{BR}(\tau \to e a\, (\to \mu \mu,\,e\mu)_\text{displ}) \lesssim 10^{-10}$. These displaced searches have the potential to surpass in sensitivity both the prompt $\tau\to e\ell \ell^{(\prime)}$ and $\tau \to \mu$ + inv, shown in gray. Reproduced from~\citep{Ardu:2026vsr}. {\em Right.} Sensitivities of current (colored regions are excluded) and projected (solid and dashed lines) beam dump experiments to CLFV ALP with anarchic flavor couplings, compared to existing constraints (grey shaded regions). Reproduced from \citep{Ema:2025bww}.
    \label{fig:tau-displ}}
\end{figure}

The simplest example of CLFV decays involving light NP states is the two-body mode $\tau\to\ell a$ with $a$ an ALP, a pNGB of a spontaneously broken global symmetry, coupling derivatively as 
\beq
\label{eq:ALP:tau}
{\cal L}_\text{ALP}\supset \frac{\partial_\mu a}{2 f_a}\,\bar\ell \,\gamma^\mu(C^V_{\ell\tau}+C^A_{\ell\tau}\gamma_5)\,\tau\,.
\eeq  
As in the case of muon decays, Eq.~\eqref{eq:ALP:muon}, concrete realizations include CLFV QCD axions, axiflavons, leptonic familons, and majorons~\citep{Calibbi:2020jvd}.
The current best 90\% CL~limits, ${\rm BR}(\tau\to\mu(e)\,a)<2.8\,(4.5)\times10^{-4}$ (for $m_a \approx 0$) from Belle~\citep{Belle:2025bpu} 
already constrain NP scales up to $f_a\gtrsim\mathcal{O}(10^7)\,$GeV. The reach in $f_a$ is parametrically the same as in the muon sector, $\mathrm{BR}(\tau\to\ell\,a)\propto(m_W^2/m_\tau f_a)^2$, but the absolute sensitivity is presently far weaker than $f_a\gtrsim\mathcal{O}(10^9)\,$GeV for $\mu\to ea$ \citep{Jodidio:1986mz,TWIST:2014ovy}. Here, one order of magnitude difference is explained by the $m_\tau/m_\mu$ ratio (with lepton masses picked up by the derivative in Eqs.~\eqref{eq:ALP:muon} and~\eqref{eq:ALP:tau}), while the remainder is due to much weaker experimental bounds because of the experimental challenges described above. With $50\,\text{ab}^{-1}$ at Belle~II and $\mathcal{O}(10^{11})$ tau pairs at CEPC/FCC-ee, significant improvements are expected. A Belle~II search with the first $63\,{\rm fb}^{-1}$ was published in~\citet{Belle-II:2022heu}, extrapolating which we can expect an ultimate Belle~II limit in the order of ${\rm BR}(\tau \to \ell a) \lesssim 2-3 \times 10^{-5}$ for a practically massless ALP. For what concerns Tera-Z factories, a preliminary CEPC study estimated a potential improvement of one additional order of magnitude, that is, ${\rm BR}(\tau \to \ell a) \lesssim 3 \times 10^{-6}$~\citep{Ai:2024nmn}.
Furthermore, proposals exploiting the threshold production mode at the STCF, where the tau rest frame is fully reconstructable, could bring the sensitivity to ${\rm BR}(\tau\to\ell\,a)$ below current limits \citep{Bondar:private}. 
If ALP decays into leptons are kinematically open, there exists an interesting interplay between searches for an invisible ALP and searches for $\tau \to \ell a\,(\to \mu\mu,\,\mu e)$ with the ALP decay being either prompt or displaced. For prompt decays, the bounds shown in Table~\ref{tab:taulimits} already apply. However, searches for displaced decays inside the inner detectors of Belle~II (or FCC-ee/CEPC) would be sensitive to higher values of $f_a$, up to $\mathcal{O}{(10^8)}\,{\rm GeV}$~\citep{Ardu:2026vsr}, as illustrated in Fig.~\ref{fig:tau-displ} (left). Complementary searches can be performed at beam dump experiments, which are sensitive to the part of parameter space that leads to highly displaced ALP decays, see Fig.~\ref{fig:tau-displ} (right).

More generally, if a NP particle $X$ decays back to visible SM states within the detector, the initial $\tau\to\ell\,X$ two-body decay can generate exotic high-multiplicity final states through dark cascade decays. The landscape of such modes has been systematically surveyed for the tau sector in \citet{Ema:2026dao}, covering five representative UV models: a kinetically mixed dark photon, gauged $U(1)_{L_\alpha-L_\beta}$ models, a gauged $U(1)_{L_e+L_\mu-2L_\tau}$,  chiral $U(1)'$ extensions, and a flavor-protected scalar. When the CLFV coupling arises at dimension~5 in $\text{SMEFT}_X$, the Goldstone boson equivalence theorem forces ${\rm BR}(\tau\to\ell\phi)/{\rm BR}(\tau\to\ell V)\simeq1$, with $\phi$ denoting a dark Higgs and $V$ a dark vector boson. As a consequence, direct three-body decays $\tau\to\ell V$ $(V\to \ell\ell)$ compete with the five-body cascades (from $\tau\to\ell\phi\, (\phi\to VV)$) at the same order; gauged lepton-number models in this class additionally predict large missing-energy fractions from $V\to\nu\bar\nu$ decays. When the coupling is instead dimension~6, as in the kinetically mixed dark photon model, all cascade final states can be fully visible. If the CLFV interaction of the light scalar $\phi$ with SM charged leptons arises from a dimension-5 operator, then a bound ${\rm BR}(\tau\to\ell\phi)\lesssim10^{-8}$ probes a NP scale $\Lambda_5\gtrsim10^{11}\,$GeV, far above the direct collider reach and orders of magnitude beyond what heavy-NP SMEFT analyses of the classic CLFV tau channels can access~\citep{Ema:2026dao}.

An atlas of predicted decay modes is given in Table~1 of \citet{Ema:2026dao}, which catalogues three-, five-, and seven-body leptonic modes, as well as mixed leptonic-hadronic modes such as $\tau\to3\mu\,2\pi$ and $\tau\to\mu\,4\pi$, and their missing-energy variants. Of the eighteen distinct five-body and eight seven-body channels predicted across the five benchmark models, so far  only $\tau\to5\mu$ has received even just a preliminary experimental sensitivity estimate, ${\rm BR}(\tau\to5\mu)\lesssim4\times10^{-8}$ at LHCb \citep{Overhoff:2020}; every remaining channel, including the hadronic modes and $\tau\to7\mu$, is experimentally unexplored. As for the high-multiplicity muon decays, the irreducible SM backgrounds for the  fully visible high-multiplicity tau decays also drop quickly with increasing multiplicity, so these searches can in principle be carried out close to background-free. Measuring ratios of branching fractions across channels provides discrimination between NP models: a large $\tau\to3\mu\,2\pi$ rate relative to $\tau\to5\mu$ points to kinetic mixing; comparable rates for $\tau\to5\mu$ and $\tau\to3\mu\,2\nu$ indicate a gauged $L_\mu-L_\tau$ symmetry; and $\tau\to5e$ without a corresponding $\tau\to5\mu$ would signal a chiral model coupling exclusively to right-handed electrons. Belle~II, with its low backgrounds, full detector hermeticity, would at 50~ab$^{-1}$ integrated luminosity be sensitive to branching ratios in the ${\rm BR}\sim\mathcal{O}(10^{-9})$--$\mathcal{O}(10^{-10})$ range, while FCC-ee/CEPC with $\sim10^{11}$ tau pairs could extend the reach to ${\rm BR}\sim\mathcal{O}(10^{-11})$, probing NP scales in the $10^{12}$~GeV range for dimension-5 CLFV operators.

\section{Collider searches for CLFV}
\label{sec:colliders}

\subsection{Probing contact interactions at high-energy colliders: LHC, EIC, $\mu$TRISTAN, etc.} 

CLFV has been extensively searched for at the LHC through different-flavor dilepton final states with high invariant mass. The CMS collaboration searched for heavy resonances decaying into $e\mu,e\tau,$ and $\mu\tau$ final states using the full Run-2 dataset corresponding to an integrated luminosity of $138~{\rm fb}^{-1}$ at $\sqrt{s}=13$ TeV~\citep{CMS:2022fsw}, while the ATLAS collaboration performed a complementary search using $139~{\rm fb}^{-1}$ of Run-2 data \citep{ATLAS:2020tre}. Both analyses are based on the invariant-mass spectrum of different-flavor dilepton pairs, where heavy CLFV resonances would appear as an excess at large invariant masses. So far, no significant deviations from the SM prediction have been observed, and stringent limits have been set on benchmark models, including CLFV $Z^{\prime}$ bosons, R-parity-violating supersymmetric models, and quantum black holes. 
If the particles responsible for CLFV are much heavier than the energy scale directly accessible at the LHC, their effects can be described by EFT operators discussed in Sec.~\ref{sec:eft}. In this case, the absence of a signal in the high-mass dilepton searches can also be interpreted as constraining the corresponding CLFV EFT operators, thereby probing NP scales in the multi-TeV regime for WCs of order unity. The interpretation of these searches within the EFT framework has been discussed in \cite{Han:2010sa, Greljo:2017vvb, Angelescu:2020uug, Cirigliano:2021img, Allwicher:2022gkm, Delzanno:2024ooj}.

The LHC has also been searching for CLFV interactions involving the top quark. An early ATLAS search analyzed $t\to \ell\ell^{\prime}q~(q=u,c)$ within an EFT framework using Run-2 data with $79.8~{\rm fb}^{-1}$ in top-quark pair events and presented ${\rm BR}(t\to e\mu q)<1.86\times 10^{-5}$ at 95$\%$ CL \citep{ATLAS:2018avw}, where only electrons and muons were considered as reconstructed leptons in the analysis. More recently, CMS searched for CLFV top-quark production and decays in the $e\mu$ channel using the full Run-2 dataset ($138~{\rm fb}^{-1}$), obtaining 95\% CL upper limits of
\begin{align}
    {\rm BR}(t\to u e\mu)&<\left(0.13,\hspace{0.3cm}0.07,\hspace{0.3cm}0.25\right)\times 10^{-6},\\[0.1cm]
    {\rm BR}(t\to c e\mu)&<\left(1.31,\hspace{0.3cm}0.89,\hspace{0.3cm}2.59\right)\times 10^{-6},
\end{align}
respectively, for vector, scalar, and tensor operators \citep{CMS:2022ztx}. Dedicated searches for the $\mu\tau$ channel have subsequently been performed by both collaborations. Using the full Run-2 dataset, ATLAS ($140~{\rm fb}^{-1}$) obtained BR$(t\to q\mu^{\pm}\tau^{\mp})<8.7\times 10^{-7}$ \citep{ATLAS:2024njy}, while CMS ($138~{\rm fb}^{-1}$) reported limits of \citep{CMS:2025xnf}
\begin{align}
    {\rm BR}(t\to u \mu\tau)&<\left(0.08,\hspace{0.3cm}0.04,\hspace{0.3cm}0.12\right)\times 10^{-6},\\[0.1cm]
    {\rm BR}(t\to c \mu\tau)&<\left(1.71,\hspace{0.3cm}0.81,\hspace{0.3cm}2.05\right)\times 10^{-6}.
\end{align}
Assuming WCs of order unity, the results can be translated into lower bounds on the scale of new physics, $\Lambda \gtrsim \mathcal{O}\,(1~\mathrm{TeV})$. For theoretical studies and EFT interpretations of these searches, see~\cite{Davidson:2015zza, Cirigliano:2021img, Delzanno:2024ooj}.

Deep inelastic scattering serves as an alternative search channel for CLFV, offering insights complementary to traditional low-energy searches. A prominent implementation of this approach was carried out at the Hadron–Electron Ring Accelerator (HERA), where high-energy electron-proton collisions were successfully utilized to perform direct searches for CLFV processes mediated by hypothetical leptoquarks \citep{ZEUS:2005nsy, H1:2007dum, H1:2011rlk}. While the HERA set initial constraints, the future Electron-Ion Collider (EIC) will take these searches to the next level employing an unprecedented luminosity. The EIC is a state-of-the-art, high-luminosity particle accelerator planned for construction at the Brookhaven National Laboratory, which will collide high-energy electrons with protons and heavy ions, with the primary scientific goal of exploring the inner structures of nucleons and atomic nuclei. The collision c.o.m.~energies will range from 20 to 100 GeV (upgradable to 140 GeV), and the instantaneous luminosity in the  $10^{33}-10^{34}~{\rm cm}^{-2}\,{\rm s}^{-1}$ range, with physics operations scheduled to begin in the early to mid-2030s. The outstanding capability of the EIC opens new avenues to search for CLFV processes via $e^-\,p\to \ell\,(=\tau~{\rm or}~\mu)~X$, where $X$ denotes all possible final states. Based on the SMEFT framework, theoretical studies have been conducted in \cite{Cirigliano:2021img} for $e\to \tau$ and \cite{Delzanno:2024ooj} for $e\to \mu$ transitions.\footnote{Recent work has extended this framework to explore flavorful lepton number violation (LNV) at the EIC, capturing the interplay between both CLFV and LNV effects \citep{UrrutiaQuiroga:2026yee}.} These studies analyzed the discovery potential of the EIC in the CLFV transitions and compared it with the existing bounds on SMEFT operators from low-energy and high-energy searches. The analysis of low-energy decays also constrained operators involving heavy-quarks using renormalization group equations, which then induce low energy CLFV observables starting at one loop order. 
The study shows that, while low-energy experiments impose highly stringent constraints on $e-\mu$ interactions, the EIC has the potential to compete with current searches, particularly for operators involving bottom and charm quarks.\footnote{2-quark 2-lepton CLFV operators involving  either flavor-conserving or flavor-violating 2nd- and 3rd-generation quark currents can also be tested in hadron decays, such as $B\to K \ell_i\ell_j$, $J/\psi\to \ell_i\ell_j$, $B_s \to \ell_i\ell_j$ etc.,~which are searched for at flavor factories (Belle~II, BESIII, LHCb, NA62 etc.). However, the typical sensitivities are often inferior to indirect limits from muon and tau decays and the collider probes discussed here~\citep{Angelescu:2020uug,Calibbi:2022ddo,Becirevic:2024vwy,Chattopadhyay:2025air}, with the noteworthy exception of CLFV kaon decays such as $K^+\to\pi^+ \mu^+ e^-$~\citep{Sher:2005sp}.}  This is because the CLFV signals are heavily suppressed either by a single electroweak loop in low-energy decays or by double heavy-quark parton distribution functions at the LHC. Crucially, the EIC projections rely on imposing simple kinematical cuts on final state particles, identifying $\tau \to \mu \bar{\nu}_{\mu}\nu_{\tau}$ as a promising channel for reconstruction. More dedicated analyses incorporating $\tau$ hadronic decay channels, which feature the largest branching ratio, alongside heavy flavor tagging, would be able to explore the full potential of the EIC in searching for CLFV processes.

Leptonic colliders, such as $e^+e^-$ Higgs/Tera-Z factories, will also provide plenty of opportunities to test CLFV interactions, as illustrated in the next subsection. In this subsection, we focus instead on the potential of proposed facilities involving high-energy muon beams. 
One of such facilities is the $\mu$TRISTAN project~\citep{Hamada:2022mua}, which proposes to cool and focus $\mu^+$ beams employing the techniques that are being developed by the J-PARC muon $g-2$ experiment~\citep{Abe:2019thb}. Subsequently, the $\mu^+$ beam could be accelerated to 1~TeV and collided with the 30~GeV $e^-$ beam from the TRISTAN ring at KEK, yielding a $\mu^+e^-$ Higgs factory~\citep{Lu:2020dkx} with c.o.m.~energy $\sqrt{s}\simeq 346$~GeV and an expected integrated luminosity of $\mathcal{L}=1~\mathrm{ab}^{-1}$. The facility could also operate as a $\mu^+\mu^+$ collider at $\sqrt{s}=2$~TeV, providing a high-energy leptonic discovery machine based on existing technology. The latter operation mode would provide an ideal environment for studying CLFV processes mediated by doubly-charged scalars (such as one of the states in the Type~II seesaw triplet)~\citep{Fridell:2023gjx,Lichtenstein:2023iut}, as well as other CLFV interactions that violate individual lepton flavor numbers by two units~\citep{Heeck:2024uiz}, e.g.~processes such as $\mu^+\mu^+ \to \tau^+\tau^+$ corresponding to virtually unconstrained NP operators, to which $\mu$TRISTAN could be sensitive up to $\Lambda \approx 30\,{\rm TeV}$~\citep{Heeck:2024uiz,Calibbi:2025fzi}. The $\mu^+e^-$ mode could be employed to test CLFV processes such as $\mu^+e^- \to \mu^-e^+$ and $\mu^+e^- \to \mu^+\mu^+$~\citep{Bossi:2020yne,Dev:2023nha,Ding:2024zaj,Calibbi:2024rcm}, whose underlying NP is already constrained by, respectively, muonium to anti-muonium conversion and (indirectly) $\mu\to e \gamma$. However, in the presence of NP mediators with purely flavor-violating couplings (that is, suppressed flavor-conserving interactions), $\mu\to e \gamma$ is suppressed and $\mu$TRISTAN would provide a valuable probe~\citep{Calibbi:2024rcm}. In addition, $\mu^+e^-$ scattering into final states involving tau leptons can in some cases offer a sensitivity comparable and complementary to Belle~II searches for the leptonic CLFV $\tau$ decays~\citep{Lichtenstein:2023iut}. 

More ambitious (and futuristic) proposals envision collisions of high-energy $\mu^+$ and $\mu^-$ beams. The exceptional potential of high-energy $\mu^+\mu^-$ colliders to discover NP particles and perform precision measurements of Higgs boson couplings has been extensively studied in recent years; see~\cite{Accettura:2023ked,InternationalMuonCollider:2025sys} for overviews. A high-energy $\mu^+\mu^-$ collider would also directly probe CLFV four-lepton contact interactions through processes such as $\mu^+\mu^- \to \mu^\pm e^\mp$, $\mu^+\mu^- \to \tau^\pm e^\mp$, and $\mu^+\mu^- \to \tau^\pm \mu^\mp$. For instance, a machine operating at $\sqrt{s}=10\,\mathrm{TeV}$ could be sensitive to NP scales up to $\Lambda \simeq 100-300\,\mathrm{TeV}$~\citep{Glioti:2025zpn,Asadi:2025dii}.
The collinear emission of neutral gauge bosons from one of the beams could also be used to probe operators that do not involve a flavor-conserving muon current, for example through the $2\to 3$ scattering process $\mu V \to \tau\tau\tau$ ($V=\gamma,\,Z$). The corresponding sensitivity could reach scales of $20-30\,\mathrm{TeV}$, surpassing indirect bounds from low-energy observables such as the CLFV $\tau$ decays~\citep{Asadi:2025dii}.

\subsection{Heavy bosons CLFV and Tera-Z factories} 

Future colliders offer unique opportunities to search for CLFV in heavy boson decays, $Z\to \ell_i\ell_j$ and $h\to \ell_i\ell_j$. 
The prospects for CLFV $Z$ and Higgs decays are summarized in Table~\ref{tab:ZLFV}. While the HL-LHC is expected to improve current limits by roughly a factor 5, further progress is hindered by irreducible backgrounds, especially from $Z \to\tau\tau$ and $h\to \tau\tau$ followed by leptonic tau decays.
 In contrast, a high-energy electron-positron collider would benefit from a much cleaner environment, excellent $\tau$ identification, precise knowledge of the initial-state kinematics, and superior momentum and particle-identification performance. These features will enable Tera-Z factories to improve present sensitivities on the $Z$ decays by several orders of magnitude, down to ${\rm BR}(Z\to\tau\ell)\approx 10^{-9}$ and ${\rm BR}(Z\to\mu e)\approx 10^{-10}-10^{-8}$, with the latter estimate depending on how well muon misidentification as electrons in the ECAL can be kept under control~\citep{Dam:2018rfz}. Sensitivities to Higgs decays at future $e^+e^-$ colliders will also exceed the HL-LHC prospects but, ultimately, they will be statistically limited because of the sample size --- no more than $\mathcal{O}(10^6)$ Higgs bosons are expected to be produced in the Higgs-factory runs at these machines. 

\begin{table}[t]
\TBL{\caption{Current 95\%~CL limits on CLFV in $Z$ boson and Higgs decays and projected sensitivities for HL-LHC and the proposed Higgs/Tera-Z factories FCC-ee and CEPC~\citep{Dam:2018rfz,Ai:2024nmn,Qin:2017aju}. For an approximate HL-LHC projection, current limits (obtained by ATLAS and CMS based on $\approx138\,{\rm fb}^{-1}$) were scaled to the target luminosity of $3~\text{ab}^{-1}$.}\label{tab:ZLFV}}
{\begin{tabular*}{\textwidth}
{@{\extracolsep{\fill}}@{}lllll@{}}
\toprule
Process & Current limit & Experiment & Future reach & Experiment\\
\colrule
\multirow{2}{*}{$Z\to e\mu$}  & \multirow{2}{*}{$1.9\times 10^{-7}$} & \multirow{2}{*}{CMS~\citep{CMS:2025wqy}} &  $4\times 10^{-8}$ &  HL-LHC~($3\,{\rm ab}^{-1}$) \\
 & & & $10^{-8}-10^{-10}$ & FCC-ee/CEPC \\[1mm] 
 \multirow{2}{*}{$Z\to e\tau$}  & \multirow{2}{*}{$5.0\times 10^{-6}$} & \multirow{2}{*}{ATLAS~\citep{ATLAS:2021bdj}} &  $10^{-6}$ &  HL-LHC~($3\,{\rm ab}^{-1}$) \\
 & & & $10^{-9}$ & FCC-ee/CEPC \\[1mm] 
\multirow{2}{*}{$Z\to \mu\tau$}  & \multirow{2}{*}{$6.5\times 10^{-6}$} & \multirow{2}{*}{ATLAS~\citep{ATLAS:2021bdj}} &  $10^{-6}$ &  HL-LHC~($3\,{\rm ab}^{-1}$) \\
 & & & $10^{-9}$ & FCC-ee/CEPC \\
\midrule
\multirow{2}{*}{$h\to e\mu$}  & \multirow{2}{*}{$4.4\times 10^{-5}$} & \multirow{2}{*}{CMS~\citep{CMS:2023pte}} &  $9\times 10^{-6}$ &  HL-LHC~($3\,{\rm ab}^{-1}$) \\
 & & & $4 \times 10^{-6}$ & FCC-ee/CEPC \\[1mm] 
 \multirow{2}{*}{$h\to e\tau$}  & \multirow{2}{*}{$2.0\times 10^{-3}$} & \multirow{2}{*}{ATLAS~\citep{ATLAS:2023mvd}} &  $4 \times 10^{-4}$ &  HL-LHC~($3\,{\rm ab}^{-1}$) \\
 & & & $8\times 10^{-5}$ & FCC-ee/CEPC \\[1mm] 
\multirow{2}{*}{$h\to \mu\tau$}  & \multirow{2}{*}{$1.5\times 10^{-3}$} & \multirow{2}{*}{CMS~\citep{CMS:2021rsq}} & $3 \times 10^{-4}$ &  HL-LHC~($3\,{\rm ab}^{-1}$) \\
 & & & $7\times 10^{-5}$ & FCC-ee/CEPC \\
\botrule
\end{tabular*}}{}
\end{table}

The key observation is that any new physics inducing CLFV in $Z$ or $h$ decays interactions will also contribute to CLFV $\mu$ and $\tau$ processes: in fact, effective flavor-violating $Z$ or $h$ interactions, together with the standard couplings of the heavy bosons, unavoidably generate diagrams contributing to low-energy CLFV processes.\footnote{This fact is exemplified by Eq.~\eqref{eq:4lep}, where Higgs-lepton operators match onto a LEFT 4-lepton operator after  the $Z$ boson is integrated out.}
In order to quantify the resulting indirect bounds on $Z\to \ell_i\ell_j$ and $h\to \ell_i\ell_j$, one can consider the SMEFT operators shown in Table~\ref{tab:smeft}.
CLFV $Z$ decays are induced by three independent combinations of Higgs-lepton operators, $Q_{H e}$, $Q_{H \ell } \equiv Q_{H \ell }^{(1)}+ Q_{H \ell }^{(3)}$, and $Q_{eZ} \equiv \left(\sin\theta_W Q_{eB} + \cos\theta_W Q_{eW}\right)$:
\begin{equation}
\mathrm{BR}(Z \to \ell_i \ell_j ) =
\frac{m_Z^3}{12\pi\, \Gamma_Z} \frac{v_\text{EW}^2}{\Lambda^4}\Bigg[\left(\left| \hat{\mathcal{C}}_{H\ell,\,ij} \right|^2 + \left|\hat{\mathcal{C}}_{H e,\,ij} \right|^2  \right) + \frac14 \left(\left| \hat{\mathcal{C}}_{eZ,\,ij} \right|^2 + \left| \hat{\mathcal{C}}_{eZ,\,ji} \right|^2\right) \Bigg]\,,
\end{equation}
where $\Gamma_Z \simeq 2.5\,{\rm GeV}$, $i\neq j$ and we summed over the two possible combinations of electric charges.
On the other hand, the only contribution to CLFV Higgs decays at dimension six is provided by $Q_{eH3}$,
\begin{equation}
\mathrm{BR}(h \to \ell_i \ell_j ) = \frac{m_h}{16\pi\, \Gamma_h} \frac{v_\text{EW}^4}{\Lambda^4} \left(\left| \hat{\mathcal{C}}_{eH3,\,ij} \right|^2 +\left| \hat{\mathcal{C}}_{eH3,\,ji} \right|^2\right)\,,
\end{equation}
where $\Gamma_h \simeq 4.1\,{\rm MeV}$, $i\neq j$ and, again, we summed over the possible final-state charges.

\begin{figure}[t]
    \centering    
    \includegraphics[width=.8\textwidth]{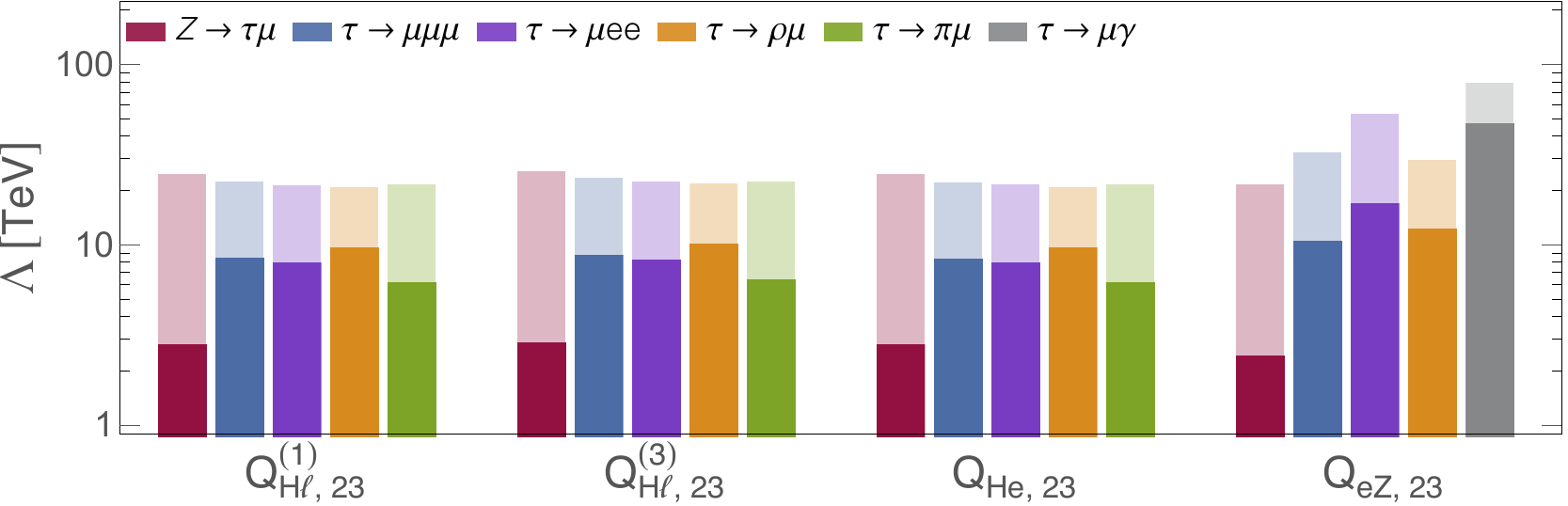}
    \vspace{0.1cm}
    \caption{Sensitivity to the NP scale $\Lambda$ associated with SMEFT operators inducing $Z\to \tau \mu$, obtained by setting the indicated WC equal to one (and all the others to zero). Darker colors correspond to present limits, lighter one to the future expected sensitivities of Tera-Z factories for the $Z$ decays and Belle~II ($50\,{\rm ab}^{-1}$) for the $\tau$ decays. Adapted and updated from~\cite{Calibbi:2021pyh}. \label{fig:Ztaumu}}
\end{figure}

As we mentioned, the WCs of the operators that give rise to these decay modes are already constrained by the $\mu$ and $\tau$ observables discussed in Secs.~\ref{sec:rare:muon:trans} and~\ref{sec:tau}.
In particular, barring accidental cancelations among different contributions, searches for CLFV in muon decays imply ${\rm BR}(Z\to\mu e)\lesssim10^{-12}$~\citep{Calibbi:2021pyh}, ${\rm BR}(h\to\mu e)\lesssim10^{-8}$~\citep{Harnik:2012pb}, beyond the reach of future experiments.
On the other hand, the channels $Z\to\tau\ell$ and $h\to \tau\ell$ are less constrained by low energy observables. For instance, the bounds on $h\to \tau\ell$ set the most stringent constraints on $Q_{eH3,\,i3}$ operator. Hence, a discovery of $h\to \tau\ell$ already at the (HL-)LHC is possible~\citep{Harnik:2012pb}. The present LHC limits translate to:
\begin{align}
    \Lambda \gtrsim 5.8\,{\rm TeV}~~(h\to \tau e)\,,\quad\quad
    \Lambda \gtrsim 6.2\,{\rm TeV}~~(h\to \tau\mu)\,,
\end{align}
where, for illustration, we set $\hat{\mathcal{C}}_{eH3,\,13} = \hat{\mathcal{C}}_{eH3,\,23} =1$, 
$\hat{\mathcal{C}}_{eH3,\,31} = \hat{\mathcal{C}}_{eH3,\,32} =0$.

For  $Z\to\tau\ell$, the branching ratios can still reach $\mathcal{O}(10^{-7})$ without conflicting with existing bounds from CLFV $\tau$ decays~\citep{Calibbi:2021pyh}, making them accessible at Tera-$Z$ factories (but not at HL-LHC). In particular, NP scales up to roughly $\Lambda \approx 20-30\,{\rm TeV}$ can be probed through $Z\to\tau\ell$, a reach comparable to that expected from $\tau$ decays at Belle~II with $50~\mathrm{ab}^{-1}$, as illustrated in Fig.~\ref{fig:Ztaumu} for the $Z\to\tau\mu$ case (the results for $Z\to\tau e$ are comparable). 
This demonstrates the remarkable complementarity of searches for CLFV at high and low energies. 

Beyond resonant $Z$ decays, future $e^+e^-$ colliders can also probe CLFV through non-resonant $e^+e^-\to\tau\mu$ production~\citep{Altmannshofer:2023tsa}. The energy dependence of this process provides valuable information on the underlying operator structure. Contributions from Higgs-lepton operators are enhanced at the $Z$ pole. Away from the resonance, dipole operators lead to cross sections that approach a constant at large values of the c.o.m.~energy squared $s$, whereas Higgs-current operators result in a $1/s$ behaviour. By contrast, four-fermion contact interactions generate cross sections that grow linearly with $s$. Combining measurements at different c.o.m.~energies would therefore provide a powerful way to disentangle the origin of a possible CLFV signal while probing NP scales of order $10$~TeV.

\section{Conclusions}
\label{sec:conclusions}
Searching for CLFV processes represents one of the cleanest and most powerful probes of BSM physics. Because the minimal scenario, where the SM is supplement only by either Dirac or Majorana neutrino mass terms (the so called $\nu$SM), predicts negligibly small rates for CLFV processes, any experimental observation of CLFV would immediately be a clear signal of new physics. This chapter reviewed the theoretical aspects of CLFV processes, with primary focus on low-energy muon and tau transitions, while also providing an overview of complementary searches at high-energy colliders. We reviewed the EFT framework, which provides a robust, model-independent language that systematically connects low- and high-energy CLFV physics. The chapter also addressed recent developments in the theoretical analysis of nuclear structure functions for $\mu\to e$ conversion, studies of CLFV scenarios involving light BSM particles that lie beyond the scope of the conventional EFT framework, and advances in the theory of hadronic CLFV $\tau$ decays. 
Fig.~\ref{fig:LFVsummary} summarizes the sensitivity to the NP scale of some representative processes discussed in the previous pages.
For flavor anarchic CLFV heavy new physics, rare muon transitions  set the most stringent limits, currently probing NP scales around $10^{3}-10^4$ TeV, with the next-generation experiments (MEG II, Mu3e, Mu2e, COMET) expected to  improve this reach significantly. For CLFV that couples more strongly to third generation, on the other hand, CLFV tau decays offer the most sensitive probes. Furthermore, high-energy collider searches provide an essential complement to low-energy CLFV probes. They offer a competitive window into the flavor structure of BSM physics, particularly in the $\tau\ell$ sector, highlighting the synergy between low- and high-energy searches. Such multi-channel approach, encompassing both low- and high-energy probes, will thus ensure that the searches for CLFV processes will remain at the frontier of BSM discovery in the coming decade.

\begin{figure}[t]
    \centering        \includegraphics[width=0.7\textwidth]{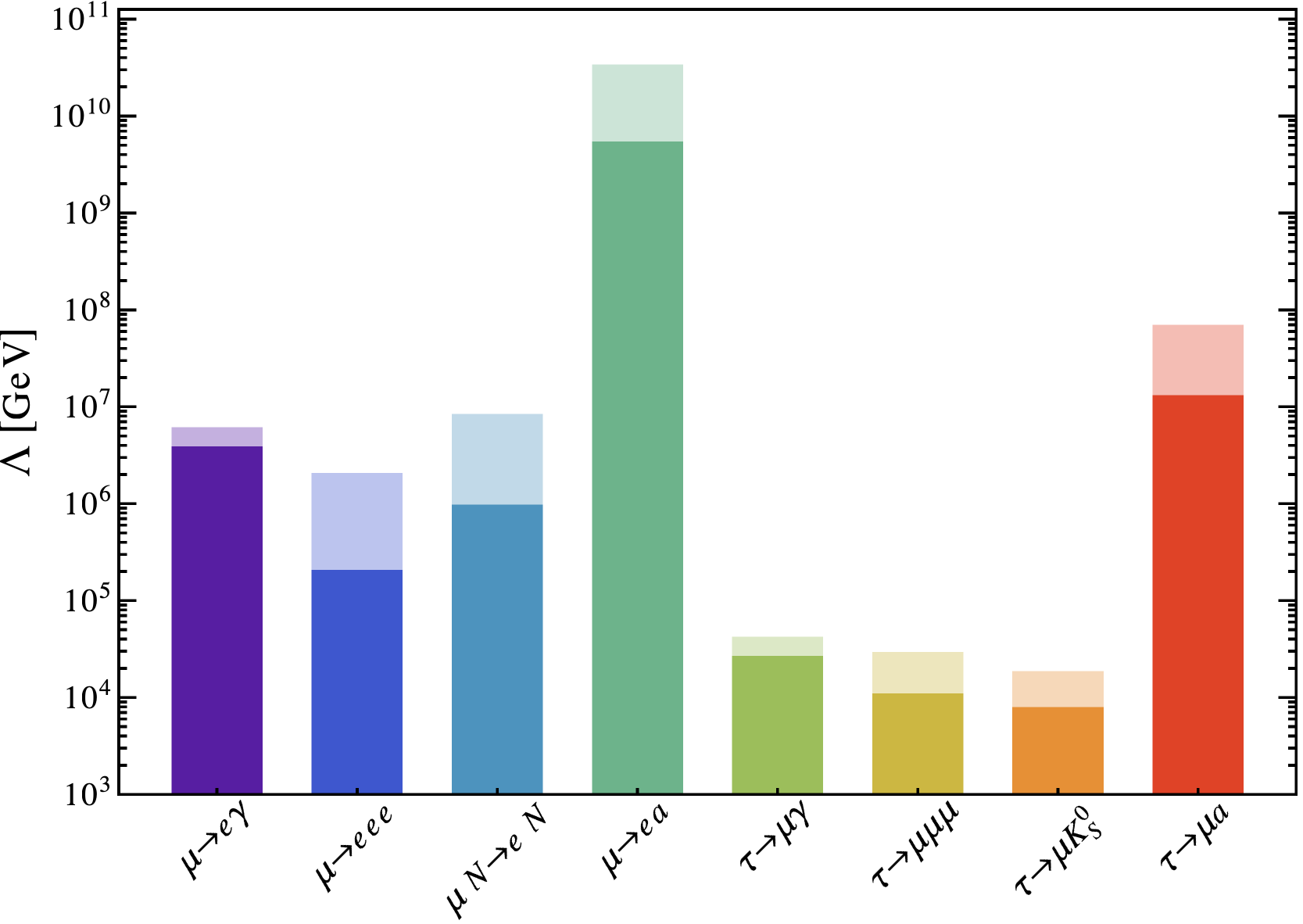}
    \vspace{0.1cm}
    \caption{Sensitivity to the NP scale $\Lambda$ for some of the key observables discussed in this review, obtained by switching on a single operator (usually the contact interaction that is most constrained in each case) and setting its WC equal to one, with the exception of the photon dipole, for which we include the loop factor and electromagnetic coupling. 
    Explicitly, the WCs coefficients of SMEFT operators (cf.~Table~\ref{tab:smeft}) are set to: 
    $\hat{\mathcal{C}}_{e\gamma,\,12}= e/(4\pi)^2$ (for $\mu\to e \gamma$); $\hat{\mathcal{C}}_{\ell \ell,\,2111}= 1$ (for $\mu\to 3e$); $\hat{\mathcal{C}}^{(1)}_{\ell q,\,1211}= 1$ (for $\mu\to e$ conversion);  $\hat{\mathcal{C}}_{e\gamma,\,32}= e/(4\pi)^2$ (for $\tau\to \mu \gamma$); $\hat{\mathcal{C}}_{\ell \ell,\,3222}= 1$ (for $\tau\to 3\mu$); $\hat{\mathcal{C}}^{(1)}_{\ell q,\,32 12}= 1$ (for $\tau\to \mu K^0_S$). 
  The labels  $\mu\to ea$ and $\tau\to \mu a$ refer to searches for CLFV decays into a (practically) massless and invisible ALP, for which we defined $\Lambda \equiv 2 f_a$ and set $C^A_{ij}=1$, $C^V_{ij}=0$ --- cf.~Eqs.~\eqref{eq:ALP:muon} and~\eqref{eq:ALP:tau} for the definition of these parameters. Darker colors show present bounds, lighter colors denote the expected final sensitivities of running or upcoming experiments: MEG~II, Mu3e, COMET, Mu2e, and Belle~II.
    \label{fig:LFVsummary}}
\end{figure}

\begin{ack}[Acknowledgments]
{}
KF acknowledges support in part by RIKEN iTHEMS SUURI-COOL (KEK) program.
JZ acknowledges support in part by DOE grant DE-SC0011784, DE-SC0026301, and by NSF grants OAC-2103889, OAC-2411215, and OAC-2417682.
MH acknowledges support in part by the Laboratory Directed Research and Development Program of Oak Ridge National Laboratory, managed by UT-Battelle, LLC, for the U.S.\ Department of Energy
and by the U.S.~Department of Energy, Office of Science, Office of Advanced Scientific Computing Research and Office of Nuclear Physics, Scientific Discovery through Advanced Computing (SciDAC) program (SciDAC-5 NUCLEI).
This research used resources of the Oak Ridge Leadership Computing Facility at the Oak Ridge National Laboratory, which is supported by the Advanced Scientific Computing Research programs in the Office of Science of the U.S. Department of Energy under Contract No. DE-AC05-00OR22725. LC and JZ acknowledge support by the Munich Institute for Astro-, Particle and BioPhysics (MIAPbP) which is funded by the Deutsche Forschungsgemeinschaft (DFG, German Research Foundation) under Germany's Excellence Strategy – EXC-2094 – 390783311, and thank the organizers of the MIAPbP ``Precision Lepton Physics'' workshop for creating a stimulating discussion environment. We also thank S. Gonz{\'a}lez-Sol{\'i}s for the useful comments on form factors relevant for $\tau$ hadronic decay channels.
\end{ack}

\bibliographystyle{Harvard}
\bibliography{reference}

\end{document}